\documentclass[pra,showkeys,superscriptaddress,twocolumn,10pt]{revtex4-2}
\usepackage{booktabs} 
\usepackage{amsmath, amssymb, amsfonts,mathtools}
\usepackage{physics}
\usepackage{overpic}
\usepackage{amsmath}
\usepackage{amssymb}
\usepackage{graphicx}
\usepackage{bm,subfigure,braket,float,pstricks}
\usepackage{lmodern}
\usepackage{enumitem}
\usepackage[utf8]{inputenc}
\usepackage[T1]{fontenc}
\usepackage{hyperref}
\hypersetup{
colorlinks=true,
linkcolor=blue,
filecolor=magenta,
urlcolor=cyan,
citecolor=blue,
}
\usepackage{cleveref} 
\usepackage[paperwidth=8.3in, paperheight=11in, left=0.5in, right=0.5in, top=0.6in, bottom=0.6in]{geometry}
\usepackage{pstricks}
\usepackage{orcidlink}
\usepackage{xcolor}
\newcommand{\GP}{\text{GP}}
\newcommand{\SGP}{\text{SGP}}
\newcommand{\OGP}{\text{OGP}}
\newcommand{\LGP}{\text{LGP}}
\newcommand{\HGP}{\text{HGP}}
\newcommand{\AGP}{\text{AGP}}
\newcommand{\PSI}{$\ket{\psi}^+$}
\newcommand{\PHI}{$\ket{\Phi}^+$}
\newcommand{\PP}{$\ket{++}$}
\newcommand{\PZ}{$\ket{+0}$}
\newcommand{\ZP}{$\ket{0+}$}
\newcommand{\ZZ}{$\ket{00}$}
\begin{document}
\title{The Quantum Control Hierarchy: When Physics-Informed Design Meets Machine Learning}

\author{Atta ur Rahman}
\affiliation{School of Physical Sciences, 
University of Chinese Academy of Science,
Yuquan Road 19A, Beijing 100049, China}

\author{M. Y. Abd-Rabbou} 
\affiliation{School of Physical Sciences, 
University of Chinese Academy of Science,
Yuquan Road 19A, Beijing 100049, China}
\affiliation{Department of Mathematics and Computer Science, Faculty of Science, Al-Azhar University Nasr City, Cairo, 11884, Egypt}
\author{Cong-feng Qiao}
\email{Corresponding Author: qioacf@ucas.ac.cn}
\affiliation{School of Physical Sciences, 
University of Chinese Academy of Science,
Yuquan Road 19A, Beijing 100049, China}
\affiliation{Key Laboratory of Vacuum Physics, University of Chinese Academy of Sciences, Beijing 100049, China}
\begin{abstract}
We address a wide spectrum of quantum control strategies, including various open-loop protocols and advanced adaptive methods. These methodologies apply to few-qubit scenarios and naturally scale to larger N-qubit systems. We benchmark them across fundamental quantum tasks: entanglement preservation/generation, and directed quantum transport in a disordered quantum walk. All simulations are performed in a challenging environment featuring non-Markov colored noise, imperfections, and the Markov Lindblad equation. With a complex task-dependent performance hierarchy, our deterministic protocols proved highly effective for entanglement generation/preservation, and in specific pulse configurations, they even outperformed the RL-optimization. In contrast, more advanced methods demonstrate a marked specialization. For entanglement preservation, a physics-informed hybrid Quantum Error Correction and Dynamical Decoupling (QEC-DD) protocol provides the most stable and effective solution, outperforming all other approaches. Conversely, for dynamic tasks requiring the discovery of non-trivial control sequences, such as DD, Floquet engineering, and rapid entanglement generation or coherent transport, the model-free Reinforcement Learning (RL) agents consistently find superior solutions. We further demonstrate that the control pulse envelope is a non-trivial factor that actively shapes the control landscape, which determines the difficulty for all protocols and highlights the adaptability of the RL agent. We conclude that no single strategy is universally dominant. A clear picture emerges: the future of high-fidelity quantum control lies in a synthesis of physics-informed design, as exemplified by robust hybrid methods, and the specialized, high-performance optimization power of adaptive machine learning.
\end{abstract}
\maketitle

\section{Introduction}
Recent technological advancements are rapidly transforming the study of open quantum systems from a purely theoretical pursuit into an engineering discipline \cite{a1, a2, a3}. In the realm of superconducting circuits \cite{a4}, unprecedented control over microwave environments now allows researchers to design and implement specific decoherence channels on demand \cite{a5}, effectively creating quantum simulators for environmental noise \cite{a6}. This has enabled the direct observation of non-Markovian dynamics, the engineering of dissipative processes to stabilize quantum states, and the development of error syndromes for quantum error correction codes. Similarly, in trapped-ion and neutral-atom platforms, precise laser-driven control makes it possible to introduce localized, tunable coupling to engineered reservoirs, facilitating the study of heat transport in quantum networks and the exploration of many-body open systems \cite{a7, a8}. Furthermore, the integration of photonic circuits with solid-state emitters has paved the way for creating structured photonic baths. Thus providing a powerful toolkit to manipulate decoherence pathways, enhance emitter coherence, and explore collective effects like superradiance in a highly controlled, scalable manner \cite{a9}. These platforms are collectively shifting the paradigm from passively observing decoherence to actively engineering the system-environment interaction as a resource for quantum control and information processing.

A parallel and equally transformative advancement lies in the precise temporal control of quantum systems using tailored electromagnetic pulses \cite{a10, a11}. By moving beyond simple on/off switching to dynamically modulating Hamiltonian parameters with shaped pulses, such as Gaussian, DRAG (Derivative Removal by Adiabatic Gate), or optimal control-derived waveforms, researchers can now steer quantum evolution with remarkable fidelity \cite{a12, a13, a14}. These techniques are crucial for engineering effective quantum dynamics and achieving successful protocols in the presence of imperfections and noise. For instance, carefully shaped microwave or laser pulses can dynamically tune the coupling strengths between qubits \cite{a15}, effectively decoupling them from unwanted environmental interactions or selectively activating desired information-exchange pathways \cite{a16}. This temporal control allows for the suppression of leakage to non-computational states, the execution of high-fidelity quantum gates in shorter times than the system's decoherence timescale, and the robust preparation of complex entangled states. By engineering the pulse shape, one can effectively create time-dependent Hamiltonians that guide the system along a desired trajectory in Hilbert space, yielding protocols for quantum state transfer, entanglement generation, and computation that are resilient to both control errors and environmental noise \cite{a17}.

In the current era of Noisy Intermediate-Scale Quantum (NISQ) technology, the significance of improving quantum protocol efficacy and designing novel, resource-efficient protocols cannot be overstated. While hardware advancements are steadily increasing qubit counts and coherence times, existing quantum devices remain highly susceptible to errors and environmental noise \cite{a18, a19}. Therefore, the practical utility of these machines hinges directly on the efficiency and robustness of the quantum algorithms and protocols they execute. Improving a protocol's efficacy for instance, by reducing the number of required gates, shortening its execution time, or increasing its resilience to a specific noise channel, can mean the difference between an output dominated by noise and a result that demonstrates a genuine quantum advantage \cite{a20, a21, a22}. The design of new protocols tailored to the specific connectivity and noise characteristics of a given hardware platform is equally critical \cite{a23}. This co-design approach, where software and hardware development are mutually informed, is essential for extracting maximum value from today's imperfect quantum processors therefore, improving for fault-tolerant quantum computation. Ultimately, every incremental improvement in protocol performance and every new protocol designed for near-term hardware represents an important step toward solving meaningful problems in fields like materials science \cite{a24}, drug discovery \cite{a25}, and optimization \cite{a26}, making this area of research a central pillar of the global quantum technology effort.

Further amplifying these control capabilities, a paradigm-shifting approach has emerged through the integration of machine learning, particularly RL \cite{a27}. In this model, an autonomous RL agent, a classical optimization algorithm, interacts directly with the quantum hardware. It proposes control strategies (e.g., pulse shapes or gate sequences) as actions and receives feedback from experimental outcomes, such as a measured state fidelity, as a reward. The profound advantage of this method is its ability to operate effectively without a perfect analytical model of the system's complex, noisy Hamiltonian. By systematically exploring the vast parameter space through trial and error, the RL agent can discover non-intuitive and highly robust control solutions that implicitly compensate for unknown noise sources and hardware imperfections. This has already led to significant breakthroughs in discovering high-fidelity quantum gates \cite{a28}, designing robust state-preparation protocols \cite{a29}, and learning optimal strategies for quantum error correction \cite{a30}. By automating the discovery of quantum protocols that can outperform human-designed counterparts, RL is becoming an indispensable tool for pushing the performance limits of NISQ hardware and accelerating the path toward demonstrating a tangible quantum advantage.

Achieving high-fidelity control over quantum systems is a central challenge in the development of quantum technologies, with a primary tool being the precise temporal shaping of electromagnetic fields to steer quantum evolution. This principle of physics-informed pulse engineering has been successfully applied across diverse platforms. In superconducting circuits, for instance, the optimization of Gaussian and DRAG pulse shapes has become a standard technique for boosting qubit gate fidelities by mitigating leakage and coherent errors \cite{Wiening2025, Krantz2019, Astapenko2022}. This control extends beyond computation, with Gaussian-modulated laser pulses being crucial for enabling high-speed continuous-variable quantum key distribution \cite{Madsen2023}, and for enhancing sensitivity in quantum sensing with nitrogen-vacancy centers by extending coherence times \cite{Wang2023}. While analytical pulse design is remarkably successful, its reliance on accurate system models has motivated a parallel trend in adaptive machine learning. RL has emerged as a particularly powerful model-free paradigm, leading to significant breakthroughs such as the discovery of novel, error-robust entangling gates for transmon qubits that outperform their human-designed counterparts \cite{Romesh2024}. This synergy is also seen in quantum annealing, where machine learning methods optimize pulse sequences to correct for hardware biases and reduce error rates \cite{Yamamoto2023}. These two powerful approaches—physics-informed analytical design and model-free RL optimization—have often been advanced in parallel, with techniques like dynamical decoupling benefiting from both tailored pulse shaping in ion traps \cite{Hill2024} and adaptive learning. However, a comprehensive, direct comparison between these two control philosophies across a wide range of quantum protocols remains a critical area of investigation. This work addresses this gap by systematically benchmarking the performance of a hierarchy of tailored pulse protocols against RL-optimized strategies for several key tasks in open quantum systems, aiming to elucidate the conditions under which each approach provides a tangible advantage.

We investigate the design and implementation of robust protocols for entanglement engineering in open quantum systems by dynamically controlling the system Hamiltonian. To compare control strategies, we modulate the system's driving fields and coupling strengths using both conventional, analytically-defined pulse sequences (comprising established protocols as well as our own designed) and machine-optimized pulses. These optimized pulses were either autonomously discovered by RL or developed by refining a pre-selected pulse within the RL framework. Protocol success is quantified by the resulting entanglement, measured via  entanglement or relative measure. This work provides a systematic evaluation of the trade-offs between physics-informed control and model-free machine learning optimization for achieving high-fidelity quantum operations in realistic, noisy environments.

We organize this work as: Sec. \ref{sec:physical_model}, we introduce the theoretical model for our assumed open quantum system dynamical setup. along with introducing the protocols and relative RL-employment. The results are presented in Sec. \ref{results}, while we summarize the work in Sec. \ref{sec:conclusion} while appendix is provided in Sec. \ref{appendix}.

\section{System Model and Theoretical Framework}
\label{sec:physical_model}
\begin{figure*}
\includegraphics[scale=0.8]{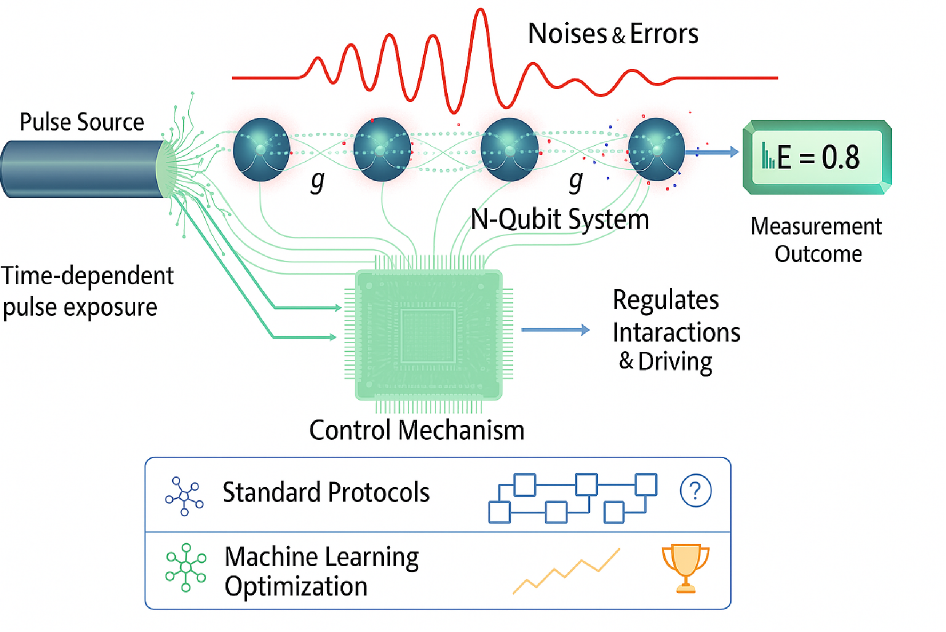}
\caption{Schematic of our study representing a quantum control schematic for an N-qubit system, highlighting the crucial role of the Control Mechanism. This mechanism dictates the pulses from the Pulse Source, which in turn shape the external driving and interactions between qubits. Despite the presence of environmental Noises and inherent coupling (g), the system's evolution is carefully managed to optimize the final Measurement Outcome. The control strategies can be either Standard Protocols or advanced Machine Learning Optimization techniques, which work to improve the system's performance.}\label{t-model}
\end{figure*}

We consider a system of $N$ coupled qubits $\{q_k\}_{k=1}^N$ defined on the Hilbert space $\mathcal{H} = \bigotimes_{k=1}^N \mathbb{C}^2$. The system Hamiltonian is considered with a time-dependent control term, $\hat{H}_c(t)$, which consists of driving and interaction components. The core of our study focuses on employing specially designed pulse envelopes to manipulate $\hat{H}_c(t)$ (illustrated by Control Mechanism in Fig. \ref{t-model}), with the explicit goal of generating or maintaining initial entanglement or other  relative quantum features. This approach is intended to build resilience against realistic experimental imperfections, including decoherence, detuning errors, pulse errors, and initial state preparation inaccuracies etc (also illustrated in Fig. \ref{t-model}). The total Hamiltonian $\hat{H}(t)$ is decomposed as:
\begin{equation}
 \hat{H}(t) = \hat{H}_s(t) + \hat{H}_c(t),\label{total Ham}
\end{equation}

where $\hat{H}_s(t)$ contains the intrinsic, time-independent energy terms of the qubits plus a fluctuating classical noise component, and $\hat{H}_c(t)$ encapsulates all externally applied, time-dependent control fields. The control Hamiltonian itself consists of inter-qubit interactions and external drives, and is given by \cite{a31}:
\begin{equation}
 \hat{H}_c(t) = \hat{H}_i(t) + \hat{H}_d(t). \label{control ham}
\end{equation}
To define these terms, we denote the standard Pauli operators for a single qubit as $\hat{\sigma}_x, \hat{\sigma}_y, \hat{\sigma}_z$ and the identity as $\hat{I}$. An operator $\hat{O}$ acting on the $k$-th qubit is implicitly embedded in the N-qubit Hilbert space as $\hat{O}_k \equiv \hat{I}^{\otimes(k-1)} \otimes \hat{O} \otimes \hat{I}^{\otimes(N-k)}$, with ladder operators defined as $\hat{\sigma}_{\pm,k} = (\hat{\sigma}_{x,k} \pm i\hat{\sigma}_{y,k})/2$.

The system Hamiltonian describes the intrinsic energy of each qubit and  classical noise affecting the configuration \cite{Nielsen2010}. Generalizing an asymmetric two-qubit model, we consider each qubit $k$ to have a unique Zeeman splitting \cite{a32} and a time-dependent noise term \cite{a33}, leading to the form:
\begin{equation}
\hat{H}_s(t) = \sum_{k=1}^{N} \left( \frac{\omega_{q,k}}{2} + \delta\omega_{z,k}(t) \right) \hat{\sigma}_{z,k},
\label{eq:H_static_N}
\end{equation}
where for each qubit $k$, $\omega_{q,k}$ is its transition frequency or Zeeman splitting, and $\delta\omega_{z,k}(t)$ represents time-dependent classical noise. This noise is modeled as a stochastic process with a Lorentzian power spectral density (PSD) \cite{Clerk2010}. Specifically, we consider a fluctuating detuning $\delta\omega_{z,A}(t)$ for a designated qubit A, synthesized from a sum of weighted cosine functions:
\begin{equation}
\delta\omega_{z,A}(t) = \sum_{j=1}^{M} \sqrt{2 S(\Omega_j) \Delta \Omega} \cos(\Omega_j t + \phi_j),\label{lorent}
\end{equation}
where $M$ is the number of frequency components, $\Omega_j$ are frequencies sampled uniformly from a range $[\Omega_{\text{min}}, \Omega_{\text{max}}]$, $\Delta \Omega = (\Omega_{\text{max}} - \Omega_{\text{min}})/M$ is the frequency spacing, and $\phi_j$ are random phases drawn from $\mathcal{U}[0, 2\pi]$. The Lorentzian PSD is given by \cite{a36}: $S(\Omega) = S_0/ (1 + \Upsilon^2\Big)$ with $\Upsilon=\left(\frac{\Omega - \Omega_c}{\Gamma}\right)$. Since the qualitative impact of such noise sources is well-studied \cite{Paladino2014}. Therefore, we do no intend to investigate the impact of each related noise parameter, instead for simplification, we fix the parameters as: $S_0$ is the noise power scale, $\Omega_c$ is the center frequency, and $\Gamma$ is the half-width at half-maximum (HWHM). For the current simulations, noise is applied only to qubit A, i.e., $\delta\omega_{z,k}(t) = 0$ for $k \neq A$.

The inter-qubit coupling is modeled as a nearest-neighbor XY interaction \cite{a34}. This model describes an interaction where a qubit only couples with its immediate neighbors. The XY part specifies that this interaction is governed by the qubits' quantum states along the x and y axes of their Bloch sphere representation. The XY interaction is a powerful mechanism for creating entangled states, which are essential for most quantum algorithms. Therefore, we are interested in amplifying the role of this interaction using pulses, which would be a plus point for quantum resource preservation. Hence, we define the bare interaction Hamiltonian as:
\begin{equation}
\hat{H}_{i,base} = \sum_{k=1}^{N-1} g_{0,k} \left( \hat{\sigma}_{+,k} \hat{\sigma}_{-,k+1} + \hat{\sigma}_{-,k} \hat{\sigma}_{+,k+1} \right),
\label{eq:H_int_base_N}
\end{equation}
where $g_{0,k}$ is the intrinsic coupling strength between qubits $k$ and $k+1$ \cite{Blais2021}.

Next, in the case of control, we consider the driving part $\hat{H}_d(t)$, which is important time-dependent term of the Hamiltonian, along with the ability to tune its amplitude, frequency, and phase, allowing us to precisely control qubit rotations \cite{x-1}. While the XY interaction models how qubits inherently interact, the external drive here is used to activate and precisely manage these interactions. In addition, the accompanied time-dependent pulse in $\hat{H}_d(t)$  can be shaped to turn-on the coupling between two qubits for a specific duration, allowing them to become entangled, and then turn-off to preserve the state. Our major goal would be to stabilize the energy and speed of these pulses enabling quantum operations to be performed much faster than the rate at which environmental noise can destroy the qubit's state. In this regard, the general base form of the external drive Hamiltonian is defined as:
\begin{equation}
\hat{H}_{d,base}(t) = \sum_{k=1}^{N} \frac{\lambda_{d,k}}{2} \cos(\omega_{d,k} t + \phi_{d,k})\hat{\sigma}_{x,k},
\label{eq:H_drive_base_N}
\end{equation}
where each qubit $k$ is driven with an amplitude $\lambda_{d,k}$, frequency $\omega_{d,k}$, and phase $\phi_{d,k}$. The time-dependent pulse envelopes that shape these interactions and drives are detailed next.

\subsection{Control Pulse Envelopes} \label{pulses}
The coherent control of the system, encapsulated in $\hat{H}_c(t)$, is achieved by dynamically shaping the interaction and drive amplitudes using time-dependent pulse envelopes. This is of critical importance because it allows for precise, time-dependent manipulation of individual qubits, which is essential for performing quantum operations and mitigating the effects of decoherence.

Although there are various methods to regulate the resourcefulness of quantum resources, we present a general non-complex method of accompanying the pulses with the control part of the Hamiltonian in various ways. We define a set of common pulse shapes, $f(\mathcal{C})$, which are dimensionless functions of the normalized time $\mathcal{C}=(t-t_c)/T$, where $t_c$ is the pulse center and $T$ is its characteristic duration. The pulses used in this work are based on Gaussian functions and their derivatives, chosen for their well-defined spectral properties \cite{a37,a38}. The mathematical forms are given by:

\begin{align}
f^{\GP}(\mathcal{C})   &= \frac{1}{\sqrt{\pi}} e^{-\mathcal{C}^2} \label{eq:pulse_gp} \\
f^{\SGP}(\mathcal{C})  &= e^{-\mathcal{C}^n}, \quad (\text{with}~n > 2) \label{eq:pulse_sgp} \\
f^{\OGP}(\mathcal{C})  &= \frac{1}{\sqrt{\pi}} e^{-\mathcal{C}^2} \cos (5 \pi\mathcal{C}) \label{eq:pulse_ogp} \\
f^{\LGP}(\mathcal{C})  &= |\mathcal{C}| e^{-\mathcal{C}^2}, \quad (\text{with }~p=0, l=1) \label{eq:pulse_lgp} \\
f^{\HGP}(\mathcal{C})  &= \sqrt{\frac{2}{\pi}} H_n(\sqrt{2}\mathcal{C}) e^{-\mathcal{C}^2}, \quad (\text{with}~ n=1) \label{eq:pulse_hgp} \\
f^{\AGP}(\mathcal{C})  &= \frac{1}{2} e^{-\mathcal{C}^2} \left[1 + \mathrm{erf}\left(\frac{\mathcal{C}T}{\mathrm{t}_{as}}\right)\right] \label{eq:pulse_agp}
\end{align}
where $H_n(x)$ is the Hermite polynomial of order $n$, and $\mathrm{erf}(x)$ is the error function. In Eq.~\eqref{eq:pulse_agp}, $\mathrm{t}_{as}$ is the asymmetry time scale.

The pulses are scaled to ensure consistent amplitude application in simulations. For modulated Hamiltonians, the envelopes are applied as $\hat{H}_d(t) = f(\mathcal{C})\hat{H}_{d,base}(t)$. The analytical time derivatives $\dot{f}(t)$ are provided in Appendix \ref{appendix}. Note that we replace $\mathcal{C}$ by relative appropriate symbolic forms depending upon the notion and requirement of the protocol in the later section.

\subsection{Open System Dynamics and Decoherence}
To realistically model the system's interaction with its environment, we employ the Lindblad master equation under the standard Markovian approximation \cite{a39}.  This framework is essential because it moves beyond the idealized assumption of a perfectly isolated quantum system. It realistically accounts for how the environment introduces noise, causing qubits to lose their delicate quantum properties.  This framework describes the irreversible evolution of the system's density matrix $\rho$ due to quantum noise channels. The master equation is:
\begin{equation}
\frac{d \rho}{dt} = -i [\hat{H}(t), \rho] + \sum_{k=1}^N \sum_{j} \mathcal{D}[L_{j,k}]\rho,
\label{eq:lindblad_compact}
\end{equation}
where $\mathcal{D}[L]\rho = L\rho L^\dagger - \frac{1}{2}\{L^\dagger L, \rho\}$ is the Lindblad dissipator for the collapse operator $L$. We incorporate two primary decoherence channels for each qubit $k$: amplitude damping and dephasing. The first process models the spontaneous emission of energy from the qubit to its environment, causing decay from the excited state $\ket{1}$ to the ground state $\ket{0}$. It is described by the collapse operator $L_{1,k} = \sqrt{\gamma_{1,k}}\hat{\sigma}_{-,k}$, where the rate $\gamma_{1,k} = 1/T_1^{(k)}$ is the inverse of the $T_1$ relaxation time. The second process, dephasing, models the loss of phase coherence without energy exchange, arising from low-frequency fluctuations in the qubit's transition frequency. It is described by the operator $L_{\phi,k} = \sqrt{\gamma_{\phi,k}}\hat{\sigma}_{z,k}$, where the dephasing rate is related to the coherence time $T_2^*$ by $\gamma_{\phi,k} = 1/T_2^{*(k)} - \gamma_{1,k}/2$.

Note that the stochastic noise implication like the $\delta \omega$ term is a realistic model for the imperfections in quantum control hardware. It's an explicit, non-Markovian term that captures how real-world control signals fluctuate and can have a memory of past events. The master equation models the interaction with the quantum environment. It's a key approximation that assumes the bath's influence is instantaneous and has no memory. This is a common and valid simplification for many physical systems and makes the problem computationally solvable.

\section{Deterministic versus Machine-Learned Protocols}\label{sec:control_protoocls}
This section compares quantum control protocols for N-qubit systems, crucial for achieving fault-tolerant quantum computing. We detail deterministic methods using shaped Gaussian pulses before introducing RL to dynamically optimize control strategies for tasks like entanglement generation. This comparison explores Artificial Intelligence's potential to overcome the limitations of traditional, hand-tuned protocols, which are increasingly challenging to design (see Fig. \ref{t-model}).

The comparison between the RL agent and the benchmark protocols is a  methodological comparison  of two distinct control philosophies. Traditional benchmarks are  static, open-loop strategies  with pre-programmed actions, making them inflexible to dynamic, real-world noise. In contrast, the RL agent uses a  dynamic, closed-loop strategy, learning an optimal policy by interacting with the environment. This gives it the freedom to select continuous pulse amplitudes, phases, and timings, a core design feature that enables it to discover more robust and effective protocols than those designed by humans. The resulting performance gap demonstrates the powerful potential of RL to achieve superior quantum control in noisy environments.

In addition, we also explore the role of pulsed dynamics by examining specific active duration windows during which pulses influence the system. This approach helps us assess each pulse's contribution to entanglement generation and preservation. We indicate these active durations with lightly shaded regions in relevant plots for clarity.

\subsection{Single-Pulse Envelope Protocol}
First, let us consider a control protocol based on a single pulse envelope applied to the dynamical map of the system. The system's dynamics are governed by a pulse-controlled Hamiltonian for a specific time window, outside of that it undergoes its free evolution. In this regard, the control Hamiltonian is the sum of the inter-qubit coupling and external drive terms, $\hat{H}_c(t) = \hat{H}_{\text{int}}(t) + \hat{H}_d(t)$ \cite{a38}. A general laboratory-frame drive can be written as:
\begin{equation}
\hat{H}_{d, \text{lab}}(t) = \sum_{k=1}^N \lambda_{d,k} E_k(t) \cos(\omega_{d,k} t + \phi_{d,k}) \hat{\sigma}_{x,k},
\label{eq:drive_lab_N}
\end{equation}
where for each qubit $k$, $\lambda_{d,k}$ is the peak drive amplitude, $\omega_{d,k}$ is the drive frequency, and $E_k(t)$ is a time-dependent pulse envelope. Moreover, we allow both the inter-qubit interaction and the external drive terms to be modulated. The time-independent portion of the interaction, $\hat{H}_{i, \text{base}}$ (e.g., the nearest-neighbor coupling from Eq.~\eqref{eq:H_int_nearest_neighbor}), becomes time-dependent when modulated by a pulse envelope $f(\mathcal{C})$ \cite{a38} as:
\begin{equation}
\hat{H}_{\text{int}}(t) = f(\mathcal{C}) \hat{H}_{i, \text{base}}.
\end{equation}

To improve control fidelity, the drive Hamiltonian used in our numerical implementation incorporates a corrective principle \cite{a40}. The key  shift is from a simple drive term to a more sophisticated one. A simple drive, like $\hat{H}_{d, \text{lab}}(t)$, is a general representation but ignores the subtle errors that arise from a pulse's changing amplitude over time. To counteract these effects, the more advanced drive protocol adds a corrective component. For each qubit $k$, the expression includes a $\hat{\sigma}_{y,k}$ component that is proportional to the time derivative of the primary $\hat{\sigma}_{x,k}$ drive envelope:
\begin{equation}
\begin{aligned}
\hat{H}_{d}(t) = &\sum_{k=1}^N \frac{\lambda_{d,k}}{2} \Big[ f_k(\mathcal{C}_k) \cos(\omega_{d,k} t + \phi_{d,k}) \hat{\sigma}_{x,k} \\
& + \frac{1}{\omega_{d,k}} \dot{f}_k(\mathcal{C}_k) \cos(\omega_{d,k} t + \phi_{d,k}) \hat{\sigma}_{y,k} \Big],
\label{eq:drive_code_form_N_qubit}
\end{aligned}
\end{equation}
where $\mathcal{C}_k = (t-t_{c,k})/T_k$ is the normalized time for the pulse on qubit $k$. The function $f_k(\mathcal{C}_k)$ is the pulse envelope, and $\dot{f}_k(\mathcal{C}_k)$ is its analytical derivative with respect to its normalized argument $\mathcal{C}_k$. The parameters $t_{c,k}$ and $T_k$ are the center time and characteristic duration for the pulse on qubit $k$, respectively. This corrective drive is crucial because the added term counteracts unwanted off-resonant effects and phase shifts that occur when a pulse has a non-zero time derivative, leading to a more accurate and robust control protocol.

\subsubsection{RL Framework for Single-Pulse  Protocol}
\label{sec:general_rl_framework}

\begin{figure}[h]
\centering
\includegraphics[scale=0.5]{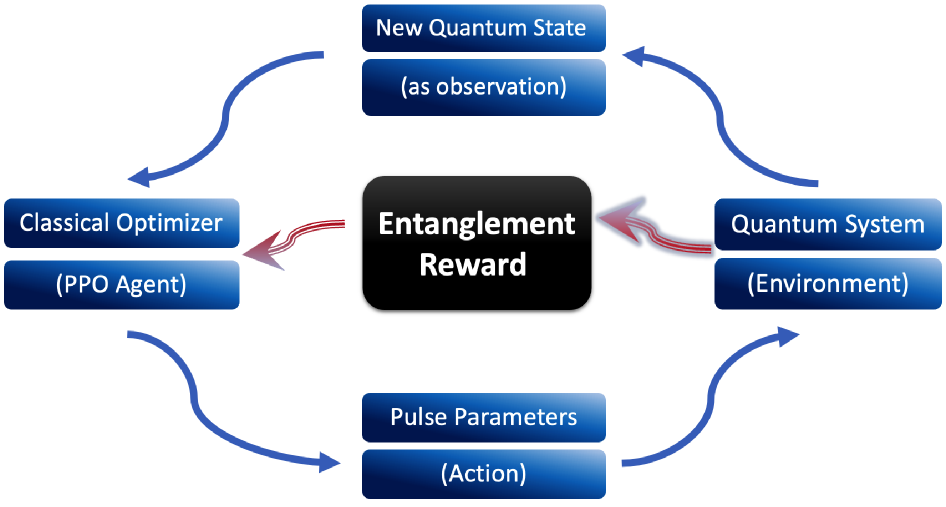}
\caption{RL framework for quantum control. The quantum system represents the environment. The PPO agent's action is a set of control parameters for the external pulses. The observation is the measured quantum state. The reward is a function quantifying the success of the operation (e.g., entanglement fidelity). This feedback loop allows the agent to discover high-performance control policies without prior analytical knowledge.} \label{fig00}
\end{figure}

To optimize the control protocols, we employ a model-free RL approach. The control problem is formulated as a Markov Decision Process (MDP), where an agent learns to interact with the quantum system to achieve a desired objective \cite{a41}. The general RL framework is illustrated in Fig.~\ref{fig00}, and its core components are defined as follows:

\paragraph{Environment:} The N-qubit quantum system, whose dynamics is simulated using the Lindblad master equation (Eq.~\eqref{eq:lindblad_compact}), which describes the time evolution of the density matrix $\hat{\rho}(t)$.

\paragraph{State:} At each discrete time step $t$, the observation provided to the agent is the system's state, $s_t = \hat{\rho}(t)$. This complex-valued matrix is flattened into a real-valued vector, which serve as  input to the agent's policy network.

\paragraph{Action:} At each step $t$, the agent selects an action $\mathbf{a}_t$ from a predefined action space $\mathcal{A}$. This action consists of a vector of control parameters (e.g., pulse amplitudes, phases) that directly parameterize the control Hamiltonian $\hat{H}_c(t)$ for the subsequent time interval.

\paragraph{Reward:} The agent's learning is guided by a reward function designed to maximize a chosen N-qubit entanglement measure, denoted by $C_t$. The reward formulation is adapted to the specific task. The reason behind the task-based reward function implication, is that it is difficult or quite challenging to design a single RL reward both encountering preservation and generation. In our case,  however, as we are utilizing six various initials states, therefore, the RL was completely failing either in preservation or generation with a single RL reward function. Hence, we designed separate RL rewards for preservation and generation of the entanglement, causing a one step down performance than the deterministic protocols with Gaussian pulses.

For entanglement preservation, the reward penalizes deviations from the initial entanglement level and instabilities. The reward at step $t$ is given by:
\begin{equation}
\begin{aligned}
R_{t_p} ={}& -w_{dev} (C_{\text{initial}} - C_{t+1})^2 - w_{stab} (C_{t+1} - C_t)^2 \\
        & - w_{cost} \|\mathbf{a}_t\|_1 - w_{sm} \|\mathbf{a}_t - \mathbf{a}_{t-1}\|_1.
\label{eq:reward_preservation}
\end{aligned}
\end{equation}

 For entanglement generation, the reward guides the agent towards a highly entangled state by combining a shaping term, an integral-like term, and cost penalties:
\begin{equation}
\begin{aligned}
R_{t_g} ={}& w_{shape}(C_{t+1}-C_t)(1+2C_{t+1}^2) + w_{\text{int}} C_{t+1}^2 \\
        & - w_{cost}\|\mathbf{a}_t\|_1 - w_{sm}\|\mathbf{a}_t - \mathbf{a}_{t-1}\|_1.
\label{eq:reward_generation}
\end{aligned}
\end{equation}
At the end of the episode, an additional terminal reward is provided based on the final entanglement level, $C_{\text{final}}$:
\begin{equation}
R_{\text{terminal}} =
\begin{cases} w_{\text{bonus}} (C_{\text{final}})^4 & \text{if } C_{\text{final}} > 0.95, \\
-w_{\text{penalty}} & \text{if } C_{\text{final}} < 0.2, \\
0 & \text{otherwise}.
 \end{cases}
 \end{equation}
In these equations, $C_t$ is the entanglement measure at step $t$, $\mathbf{a}_t$ is the action vector, and the $w$ terms are weight hyperparameters.

For entanglement preservation, the deviation weight $w_{\text{dev}}=75.0$ and the stability weight $w_{\text{stab}}=10.0$. For the entanglement generation task, the shaping weight $w_{\text{shape}}= 15.0$ and the integral weight $w_{\text{int}}= 0.5$. A large bonus of 75.0 is awarded if the  $C_{\text{final}}$ is above a threshold of 0.95, while a penalty of -30.0 is applied for low final entanglement measure. Additionally, for both tasks, a cost penalty $w_{\text{cost}}= 0.05$ is applied to the magnitude of the control action ($\|\mathbf{a}_t\|_1$), and a smoothness penalty $w_{\text{sm}}$= 0.02 is used to discourage abrupt changes ($\|\mathbf{a}_t - \mathbf{a}_{t-1}\|_1$).

\paragraph{Policy:} We use the PPO algorithm to learn the parameters $\theta$ of a stochastic policy $\pi_\theta(\mathbf{a}_t|s_t)$. The agent's goal is to discover an optimal policy that maximizes the total expected future reward, or return, defined as:
\begin{equation}
J(\pi_\theta) = \mathbb{E}_{\tau \sim \pi_\theta} \left[ G_t \right] = \mathbb{E}_{\tau \sim \pi_\theta} \left[ \sum_{k=0}^{T} \gamma^k R_{t+k+1} \right],
\label{eq:objective_function}
\end{equation}
where $\tau = (s_0, \mathbf{a}_0, s_1, \dots)$ is a trajectory, $T$ is the episode length, and $\gamma \in [0, 1]$ is a discount factor.

\subsection{Multi-Segment Control Protocol}
\label{subsec:multi_pulse_superposition}
Sophisticated quantum control can be achieved by constructing complex pulse shapes from multiple time-dependent segments. This approach allows for fine-tuned control over the frequency spectrum of the drive and can lead to solutions that are more robust to noise \cite{a42}. Here, we extend the single-pulse exposure to a multi-pulse case by using periodic windows. One powerful method involves creating a control envelope from a superposition of multiple, potentially overlapping, simpler shaped pulses. In this framework, the total control Hamiltonian, $\hat{H}_c(t) = \hat{H}_{\text{int}}(t) + \hat{H}_d(t)$, is modulated by such composite envelopes \cite{a38}. The interaction envelope, which controls the inter-qubit coupling, is constructed as a superposition of $M_{\text{int}}$ individual pulse shapes:
\begin{equation}
\mathcal{E}_{\text{int}}(t) = \sum_{j=1}^{M_{\text{int}}} f_j\left(\frac{t - t_{\text{int},j}}{T_j}\right).
\label{eq:H_int_envelope_new}
\end{equation}
Similarly, the drive Hamiltonian applies external controls to each of the $N$ qubits. The envelope for each qubit $k$, denoted $\mathcal{E}_{d,k}(t)$, is also a superposition of multiple pulses. The general form of the drive Hamiltonian with I/Q (in-phase and quadrature) control is given by:
\begin{equation}
\begin{aligned}
\hat{H}_d(t) = &\frac{1}{2} \sum_{k=1}^{N} \lambda_{d,k} \mathcal{E}_{d,k}(t) \Big[ \cos(\omega_{d,k} t + \phi_{d,k}) \hat{\sigma}_{x,k} \\
& + \sin(\omega_{d,k} t + \phi_{d,k}) \hat{\sigma}_{y,k} \Big],
\label{eq:H_drive_multi_combined_new}
\end{aligned}
\end{equation}
where the composite drive envelope for each qubit $k$ is defined as:
\begin{equation}
\mathcal{E}_{d,k}(t) = \sum_{j=1}^{M} f_j\left(\frac{t - t_{k,j}}{T_j}\right).
\label{eq:total_envelope_drive_new}
\end{equation}
The total interaction Hamiltonian models the coherent coupling between adjacent qubits in a chain, driven by the shared envelope $\mathcal{E}_{\text{int}}(t)$:
\begin{equation}
\hat{H}_{\text{int}}(t) = g_{\text{int}} \mathcal{E}_{\text{int}}(t) \sum_{i=1}^{N-1} (\hat{\sigma}^{+}_i \hat{\sigma}^{-}_{i+1} + \hat{\sigma}^{-}_i \hat{\sigma}^{+}_{i+1}).
 \label{eq:H_int_nearest_neighbor}
\end{equation}
This expression specifically models a coherent exchange of energy between only nearest-neighbor qubits.
\subsubsection{RL for Sequential Control Protocol}
\label{sec:sequential_control_revised}
We employ a model-free RL approach to find optimal sequential control protocols for a system of N-qubits. The control problem is formulated as a MDP where an agent learns to interact with the quantum system to achieve a desired objective. The framework's core components are defined as follows:

\paragraph{Action:} The control protocol is a piecewise-constant (PWC) pulse sequence. At each discrete time step $t$, the agent's action is a vector of $M$ amplitudes, $\mathbf{a}_t = [a_{t,1}, a_{t,2}, \dots, a_{t,M}]^\intercal$, drawn from a continuous action space $\mathcal{A} = [-1, 1]^M$. This vector defines the amplitudes for $M$ corresponding time windows that comprise the total pulse duration. The policy $\pi_\theta(\mathbf{a}_t | s_t)$ learns to map the system's current state $s_t$ (described by its $2^N \times 2^N$ density matrix) to an optimal action vector.

\paragraph{Reward:} The agent receives an instantaneous reward at each time step, guiding its learning. This reward is based on the change in a chosen N-qubit entanglement measure $C_t$.

For entanglement preservation, the instantaneous reward at step $t$ is:
\begin{equation}
\begin{aligned}
R_{t_p} ={}& -w_{dev} (C_{\text{initial}} - C_{t+1}) - w_{stab} (C_{t+1} - C_t)^2 \\
       & - w_{cost} \|\mathbf{a}_t\|_1 - w_{sm} \|\mathbf{a}_t - \mathbf{a}_{t-1}\|_1.
\label{eq:reward_pwc_preservation_new}
\end{aligned}
\end{equation}
The objective is to minimize deviations from the initial entanglement level and maintain stability. The cost and smoothness penalties are calculated on the full action vector $\mathbf{a}_t$, and the $w$ terms are weight hyperparameters.

 For entanglement generation, the instantaneous reward at step $t$ is:
\begin{equation}
\begin{aligned}
R_{t_g} ={}& w_{shape}(C_{t+1}-C_t)(1+2C_{t+1}^2) + w_{\text{int}} C_{t+1}^2 \\
       & - w_{cost} \|\mathbf{a}_t\|_1 - w_{sm} \|\mathbf{a}_t - \mathbf{a}_{t-1}\|_1. \label{eq:reward_pwc_generation_new}
\end{aligned}
\end{equation}
This reward encourages progress towards a highly entangled state. Additionally, a large terminal bonus or penalty is applied based on the final entanglement level, $C_{\text{final}}$:
\begin{equation}
R_{\text{terminal}} =
\begin{cases} w_{\text{bonus}} (C_{\text{final}})^4 & \text{if } C_{\text{final}} > 0.95, \\
-w_{\text{penalty}} & \text{if } C_{\text{final}} < 0.2, \\
0 & \text{otherwise}.
 \end{cases}
\end{equation}
The total return for an episode is the sum of all instantaneous rewards plus the terminal reward. This MDP formulation allows the agent to learn a policy that generates a sequence of optimal action vectors to maximize the cumulative reward.

In the current case, for the multi-segment control protocol, the reward function hyperparameters are defined with the following values. For the entanglement preservation task, the $w_{\text{dev}}= 75.0$ and the $w_{\text{stab}}=10.0$. For the entanglement generation task, $w_{\text{shape}}=50.0$ and the integral weight $w_{\text{int}}=1.5$. A terminal bonus weight $w_{\text{bonus}}=75.0$ is applied if $C_{\text{final}}>0.95$, while $w_{\text{penalty}}=30.0$ is applied if the $C_{\text{final}}< 0.2$. Common to both tasks $w_{\text{cost}}=0.01$ and a $w_{\text{sm}}=0.005$.
\subsection{Multi Segment Sequential Protocols Utilizing Polarized Pulses}
\label{subsec:multi_pulse_sequential_new}

Many quantum algorithms are constructed from a fixed sequence of operations, such as a series of temporally distinct pulses that enable discrete gates. This strategy allows for simultaneous gating of the inter-qubit interaction and the application of local single-qubit drives in a structured manner \cite{a43, a44}. Here, we extend the case of multi-pulse case in Sec. \ref{subsec:multi_pulse_superposition}, to be employed in a sequential case, along with certain polarization. Moreover, we consider protocols where the sequence of operations is fixed, but the analog parameters of the pulses (e.g., their amplitudes) are subject to optimization. The interaction Hamiltonian is modulated by a collective pulse envelope, $\mathcal{E}_{int}(t) = \sum_{p} f_{p}(t)$, which activates a bare XY-interaction \cite{a45a}:
\begin{equation}
\hat{H}_i(t) = \mathcal{E}_{int}(t) \sum_{k=1}^{N-1} g_{0,k} \left( \hat{\sigma}_{+,k} \hat{\sigma}_{-,k+1} + \text{h.c.} \right).
\label{eq:ham_int_sequential_new}
\end{equation}
The drive Hamiltonian, $\hat{H}_d(t) = \sum_k \hat{H}_{d,k}(t)$, performs local single-qubit rotations using drives of a specific polarization within this fixed sequence. Note that the $\hat{H}_i(t)$ remains the same for both the linear and circularly polarized pulse protocols.

\subsubsection{Linearly Polarized Drives.}
Linearly polarized drives are a fundamental building block for synthesizing arbitrary single-qubit gates. The polarized pulses are generated by passing unpolarized light through a linear polarizer. A polarizer works by only allowing light waves that oscillate in a specific direction to pass through, while blocking waves that oscillate in other directions. These are widely used in experimental setups because they represent the simplest and most direct way to apply an oscillating field to a qubit. By applying the field along a single, fixed axis, these pulses provide a straightforward method for inducing controlled rotations of the qubit's state in the rotating frame \cite{a45}. The Hamiltonians for drives on qubit $k$ are:
\begin{align}
\hat{H}_{d,k}^{(X)}(t) &= \Omega_{0,k} \mathcal{E}_{d,k}(t) \cos(\omega_{d,k} t) \, \hat{\sigma}_{x,k}, \label{eq:drive_x_single_new} \\
\hat{H}_{d,k}^{(Y)}(t) &= \Omega_{0,k} \mathcal{E}_{d,k}(t) \sin(\omega_{d,k} t) \, \hat{\sigma}_{y,k}, \label{eq:drive_y_single_new} \\
\hat{H}_{d,k}^{(Z)}(t) &= \Omega_{z,0,k} \mathcal{E}_{d,k}(t) \cos(\omega_{d,k} t) \, \hat{\sigma}_{z,k}, \label{eq:drive_z_single_new}
\end{align}
where $\Omega_{0,k}$ and $\Omega_{z,0,k}$ are peak drive amplitudes and $\mathcal{E}_{d,k}(t)$ is the local envelope.

\subsubsection*{RL framework with Polarized Pulses}
\label{sec:pwc_control_polarized}
We formulate the RL problem as a contextual, single-shot parameter-setting task to find the optimal set of pulse amplitudes for a three-window Piecewise-Constant (PWC) Control protocol for the case of linear-polarized sequential protocol.

\paragraph{Action:}
The agent selects a single three-dimensional action vector $\mathbf{a} = (a_1, a_2, a_3)$ at the beginning of the episode based on the initial state $s_0 = \hat{\rho}(t=0)$. Each component $a_k \in [-1, 1]$ defines the constant pulse amplitude during the $k$-th predefined time window. These windows correspond to pulses with fixed linear polarizations in the X, Y, and Z directions, respectively. The policy $\pi_\theta(\mathbf{a} | s_0)$ learns to map the initial state to the optimal set of amplitudes for the entire evolution.

\paragraph{Reward:}
The agent receives an instantaneous reward at each discrete time step $t$, which is based on the change in the entanglement measure $C_t$. This approach guides the learning process towards maximizing the sum of rewards over the trajectory.

For entanglement preservation, the instantaneous reward at step $t$ is:
\begin{equation}
R_t = (C_{t+1} - C_t) \times 10.
\label{eq:reward_pwc_preservation_new_simple}
\end{equation}

For entanglement generation, the instantaneous reward at step $t$ is also given by Eq. \ref{eq:reward_pwc_preservation_new_simple}. Additionally, a large bonus is applied at the end of the episode if a high  entanglement measure is achieved:
\begin{equation}
R_{\text{bonus}} =
\begin{cases}
10.0 & \text{if } C_{\text{final}} > 0.8, \\
0 & \text{otherwise}.
\end{cases}
\end{equation}

Moreover, the reward for each step, $R_t$, is calculated as a direct measure of progress, specifically the change in  entanglement measure between time steps, scaled by a factor of 10:
$R_t = 10 \times (C_{t+1} - C_t)$. For the entanglement generation task, there are two hyperparameters that define a terminal bonus. A bonus of $w_{\text{bonus}} = 10.0$ is awarded if $C_{\text{final}}$ exceeds a threshold of $C_{\text{thresh}} = 0.8$. The total return for the episode is the sum of all instantaneous rewards, including the potential terminal bonus. This formulation turns the optimization into a contextual bandit-like problem, where the agent learns to select the best set of fixed amplitudes for a given initial state to maximize the cumulative reward over the trajectory.

\subsubsection{Circularly Polarized Drives.}
Circularly polarized drives are used to selectively address specific transitions and are efficient for state preparation or engineering chiral interactions \cite{a46}.  To create circularly polarized pulses, one must align the quarter-wave plate so that the incident linearly polarized pulse is at a $45^o$ angle to the fast and slow axes. This ensures the two components have equal amplitude. The $90^o$ phase shift then combines these two orthogonal, equal-amplitude waves to produce a spiraling electric field, resulting in a circularly polarized pulse. The Hamiltonians for left-circularly polarized (LCP) and right-circularly polarized (RCP)  drives on qubit $k$ are:
\begin{align}
\hat{H}_{d,k}^{(\text{LCP})}(t) &= \Omega_{0,k} \mathcal{E}_{d,k}(t) \left[ \cos(\omega_{d,k} t) \hat{\sigma}_{x,k} + \sin(\omega_{d,k} t) \hat{\sigma}_{y,k} \right], \label{eq:drive_lcp_single_new} \\
\hat{H}_{d,k}^{(\text{RCP})}(t) &= \Omega_{0,k} \mathcal{E}_{d,k}(t) \left[ \cos(\omega_{d,k} t) \hat{\sigma}_{x,k} - \sin(\omega_{d,k} t) \hat{\sigma}_{y,k} \right]. \label{eq:drive_rcp_single_new}
\end{align}

\subsubsection*{RL framework with Circularly Polarized Pulses}
\label{sec:control_circular}
We formulate the RL problem as a step-wise parameter-setting task where the agent selects the pulse amplitudes for a two-window pulse protocol at each time step. The polarization of each pulse is fixed as either LCP or RCP, as specified in the code. The agent's goal is to control the dynamics of a quantum state $\hat{\rho}$ in an $N$-qubit system.

\paragraph{Action:}
At each time step $t$, the agent selects a two-dimensional action vector $\mathbf{a}_t = (a_{1,t}, a_{2,t})$ based on the current state $s_t = \hat{\rho}(t)$. Each component $a_{k,t} \in [-1, 1]$ defines the pulse amplitude during the current time step for the $k$-th predefined time window. These windows correspond to pulses with fixed LCP and RCP polarizations, respectively. The policy $\pi_\theta(\mathbf{a}_t | s_t)$ learns to map the current state to the optimal set of amplitudes.

\paragraph{Reward:}
The agent receives a reward at each time step, providing a dense signal for training. This reward is based on the change in the entanglement measure $C$, stability, and an action penalty.

For entanglement preservation, the step-wise reward is:
\begin{equation}
\begin{aligned}
R_t ={}& -w_{\text{dev}} (C_{initial} - C_{t+1})^2 - w_{\text{stab}} (C_{t+1} - C_t)^2 \\
       & - w_{\text{cost}} \sum_{k=1}^{2} |a_{k,t}| - w_{\text{smooth}} \sum_{k=1}^{2} |a_{k,t} - a_{k,t-1}|.
\label{eq:reward_stepwise_preservation}
\end{aligned}
\end{equation}

For entanglement generation, the step-wise reward is:
\begin{equation}
\begin{aligned}
R_t ={}& w_{\text{prog}} (C_{t+1} - C_t)(1+2C_{t+1}^2) + w_{\text{int}} C_{t+1}^2 \\
       & - w_{\text{cost}} \sum_{k=1}^{2} |a_{k,t}| - w_{\text{smooth}} \sum_{k=1}^{2} |a_{k,t} - a_{k,t-1}|.
\label{eq:reward_stepwise_generation}
\end{aligned}
\end{equation}

For the circularly polarized control protocol, regarding the entanglement preservation, the reward is defined by a $w_{\text{dev}}=75.0$, which heavily penalizes the system for any deviation of its  entanglement measure from the initial value. In addition, we set $w_{\text{stab}}=10.0$, which discourages rapid fluctuations in entanglement. For entanglement generation, a more complex reward structure is used with $w_{\text{shape}}= 15.0$, rewarding the agent for making positive progress towards a higher entanglement state, while the $w_{\text{int}}=0.5$ rewards the overall level of entanglement maintained over time. A bonus is awarded if the $C_{\text{final}}> 0.95$ with $w_{\text{bonus}}=75.0$. Conversely, a strong penalty $w_{\text{penalty}}= -30.0$	 is applied for low final  entanglement measure. Additionally, a cost of 0.05 is applied to the magnitude of the control actions ($\|\mathbf{a}\|_1$), and a smoothness penalty of 0.02 is applied to the change in the control action ($\|\mathbf{a}_t - \mathbf{a}_{t-1}\|_1$) to promote a smoother control signal. This continuous reward formulation guides the agent to a more finely-tuned policy compared to the single-shot approach.
\begin{figure}[ht]
\includegraphics[scale=0.29]{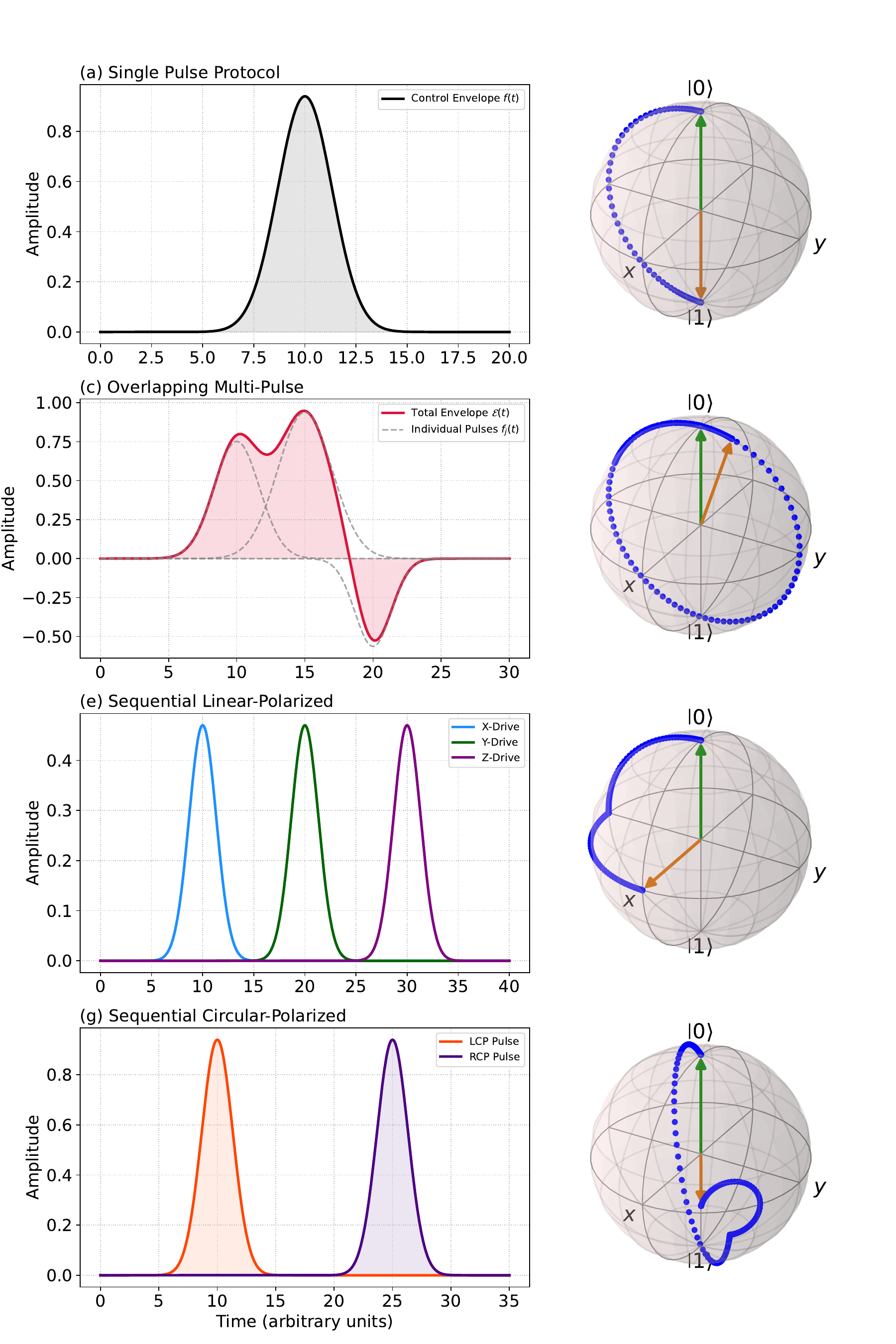}
\caption{Visualizing the general architectures of the above designed protocolsand related impact (as an example, we considered Gaussian pulse).}\label{fig01}
 \end{figure}
 
In Fig. \ref{fig01}, each row corresponds to a distinct control strategy, showing the applied pulse sequence (left column) and the resulting trajectory of a single qubit state on the Bloch sphere, starting from $\ket{0}$ (right column). Fig. \ref{fig01} (a, b) shows the single pulse protocol where a simple Gaussian $\pi$-pulse along the X-axis (black) results in a direct, smooth arc rotating the state from $\ket{0}$ to $\ket{1}$. Fig. \ref{fig01}(c, d) shows the overlapping multi-pulse protocol where a complex total envelope (red), formed by the superposition of three individual pulses (dashed gray), is applied along the X-axis. The resulting trajectory is non-trivial, with its intricate path directly reflecting the positive and negative phases of the driving field.
Fig. \ref{fig01}(e, f) presents the sequential linear-polarized protocol where a time-ordered sequence of three orthogonal $\pi/2$ pulses steers the state. The trajectory shows distinct kicks and sharp turns, with each colored segment corresponding to the action of its respective drive.
Fig. \ref{fig01}(g, h) denotes the sequential circular-polarized protocol with a sequence of an LCP pulse followed by an RCP pulse drives the state. The resulting trajectory consists of two smooth, consecutive spiral paths, demonstrating the precise rotational control achievable with phase-modulated drives. Note that the pulse and protocols are primarily here designed to regulate entanglement, however, to prove that the pulses in a specific control protocol are impactful and differently affecting the configuration, we presented a general trajectory dynamics.


\section{Benchmarking RL Against Established Control Protocols}
In this section, we are interested in comparing RL optimization efficacy to preserve quantum features with protocols which are already established (see Fig. \ref{t-model}). In the following, when incorporated with the pulses, we investigate specific control protocols (such as DD, Floquet engineering, parameter optimization methods, and quantum walk) remain resourceful  in comparison to RL-based optimization. Previously ML has been tested widely in a range of protocols, including averaging gate-errors \cite{a47}, phase controls \cite{a48}, enhancing superconducting circuits \cite{a49} etc. However, a clear representation of the comparison between the standard quantum controls and RL-based controls are lacking. In this regard, according to the complexity of the following protocols, we only focus on the maximally entangled state. As these protocols are already found to be impactful under certain conditions, we assume strict and highly realistic environmental conditions, including pulse errors, initial state errors, and detuning errors, to provide a rigorous comparison.

\subsection{RL against DD for Quantum State Preservation}\label{sec:rl_dd}
To combat decoherence in an N-qubit system, we develop a hybrid strategy that compares RL, DD, and QEC
approaches. The objective is to maximize long-term entanglement preservation, with results demonstrating superior mitigation of environmental noise compared to conventional techniques. To investigate the preservation of entanglement in an open N-qubit system under environmental noise, we employ an RL approach to optimize a DD protocol. The DD scheme is a quantum control technique used to protect qubits from environmental noise by applying carefully timed pulses \cite{a50}. We present a comprehensive methodology, including the quantum system Hamiltonian, noise model, application of DD sequencing, and RL environment. The performance is evaluated by comparing the RL-optimized protocol against standard Dynamical Corrective Gate-2 (DCG-2) \citep{a51}, advanced Uhrig Dynamical Decoupling-16 (UDD-16) \cite{a52}, and hybrid QEC-DD architecture \cite{a53}, using an entanglement measure. The goal is to avoid the complete decay of entanglement in N-qubit state over an extended time, mitigating decoherence effects through tailored control pulses.

For this protocol, the total Hamiltonian, representing a specific instantiation of our general framework for N-qubit dynamical decoupling, is composed of a system and a control part, $\hat{H}(t) = \hat{H}_{\text{sys}}(t) +\hat{H}_{c}(t)$ with $\hat{H}_{c}(t)=\hat{H}_{i}(t)+\hat{H}_{d}(t)$. The system Hamiltonian is defined as given in Eq. \eqref{eq:H_static_N} with bare frequency $\omega_q$ and $\delta\omega(t)$ colored noise affecting the first qubit, modeled as a sum of Lorentzian modes.  In the context of DD sequences we choose only YY-coupling compared to the Eq. \eqref{eq:H_int_base_N} because it aligns with the control pulses’ rotation axes, simplifying the dynamics for RL optimization along with DD-sequences, and effectively counters z-axis noise while preserving entanglement. Therefore, we assume the interaction as:
\begin{equation}
\hat{H}_i(t)=g(t) \sum_{\langle j,k \rangle} \hat{\sigma}_y^j \otimes \hat{\sigma}_y^k,
\end{equation}
where $g(t)$ is the interaction strength. The sum $\sum_{\langle j,k \rangle}$ denotes a sum over interacting pairs. The interaction strength $g(t)$ is modulated during pulse applications:
\begin{equation}
g(t) = 
\begin{cases} 
g \cdot f(t - t_c) & \text{during a pulse centered at } t_c, \\
g & \text{otherwise},
\end{cases}
\end{equation}
where $f(t - t_c)$ is one of the defined pulse envelope functions with duration $\tau_p$.

The drive Hamiltonian, $\hat{H}_{d}(t)$, governs the application of global time-dependent pulses and includes a systematic detuning error:
\begin{equation}
\hat{H}_{d}(t) = \frac{\Omega(t)}{2} \left[ \cos(\phi) \bigotimes_{k=1}^N \hat{\sigma}_x^k + \sin(\phi) \bigotimes_{k=1}^N \hat{\sigma}_y^k \right] + \delta_{\text{det}} \hat{\sigma}_z^1,
\label{eq:control_hamiltonian_N_dd}
\end{equation}
where $\delta_{\text{det}}$ is a detuning error on the first qubit. The pulse amplitude is defined by its nominal peak amplitude factor $\theta$ and envelope function $f(t)$ over a duration $\tau_p$:
\begin{equation}
\Omega(t) = \theta' f(t - t_c),
\label{eq:pulse_amplitude_N_dd}
\end{equation}

where \( \theta' = \theta \cdot \xi \) is the effective pulse amplitude, with \( \xi \sim \mathcal{N}(1, 0.05^2) \) representing a stochastic error drawn from a normal distribution with mean 1 and standard deviation 0.05, corresponding to a 5\% relative error. Here, \( \theta \) is the nominal pulse amplitude factor and \( \phi \) is the phase determining the rotation axis. Because the envelope integral is not normalized to unity, this induces a rotation whose effective angle is proportional to \( \theta' \) and dependent on the pulse shape, allowing shape-dependent variations in control strength.

The system is initialized in a perturbed generalized maximally entangled state:
\begin{equation}
\ket{\psi_0} = \mathcal{N} \left( \frac{\ket{0}^{\otimes N} + \ket{1}^{\otimes N}}{\sqrt{2}} + \epsilon \ket{\delta} \right),
\label{eq:initial_state_N_dd}
\end{equation}
where $\epsilon$ controls the magnitude of the random complex perturbation vector $\ket{\delta}$, and $\mathcal{N}$ is a normalization constant. The time evolution follows the Lindblad master equation \eqref{eq:lindblad_compact}.

\paragraph{Benchmark DD Protocols:}
For comparison, we implement benchmark protocols using global N-qubit pulses, applied within the dynamics governed by the Lindblad master Eq.  \eqref{eq:lindblad_compact} with the considered Hamiltonian. Each protocol applies $\pi$-pulses with an envelope function $f(t)$ over duration $\tau_p$, modulating both the coupling and drive terms. The total evolution time is $T_{\text{total}}$. Pulse amplitudes include a stochastic error \cite{a54}, and dynamics are numerically integrated using a Runge-Kutta 4(5) method. Note that for all the protocols, $\hat{H}_c(t)$ is accordingly utilized and modulated.

\paragraph{DCG-2 Protocol:}
This protocol applies two $\pi$-pulses to suppress low-frequency noise \cite{a51}. The pulses occur at $t_1 = T_{\text{total}}/3$ about the X-axis ($\phi_1=0$) and at $t_2=2T_{\text{total}}/3$ about the Y-axis ($\phi_2=\pi/2$). During each pulse, the drive Hamiltonian is :
\begin{equation}
\hat{H}_{d}(t) = \eta(t) \left[ \cos(\phi_k) \bigotimes_{k=1}^N \hat{\sigma}_x^k + \sin(\phi_k) \bigotimes_{k=1}^N \hat{\sigma}_y^k \right] + \delta_{\text{det}} \hat{\sigma}_z^1,
\label{eq:dcg2_drive}
\end{equation}
with $\eta(t)=\frac{\pi f(t - t_k)}{\tau_p} \frac{1}{2}$. Outside pulse durations, the drive term vanishes except for the detuning error.

\paragraph{UDD-16 Protocol:}
UDD protocol uses $N_p=16$ $\pi$-pulses with non-uniform timings $t_k$ to suppress high-order noise \cite{a52}:
\begin{equation}
t_k = T_\text{total} \sin^2\left(\frac{\pi k}{2(N_p+1)}\right), \quad k=1, \dots, 16,
\label{eq:udd_timings}
\end{equation}
alternating phases ($\phi_k=0$ for odd $k$, $\phi_k=\pi/2$ for even $k$). The drive term during each pulse is identical in form to Eq.~\eqref{eq:dcg2_drive}.

\paragraph{Hybrid QEC-DD Protocol:}
This protocol combines DD and QEC over cycles. Each cycle includes a Carr-Purcell DD sequence with a single $\pi$-pulse about the Y-axis ($\phi=\pi/2$) \cite{a55}.  The CP sequence combats decoherence from slow environmental noise by using a series of equally spaced $\pi$ pulses. These pulses periodically invert the quantum system's evolution, causing accumulated phase errors to cancel out. This refocusing action effectively protects the quantum state, extending its coherence time.  A QEC step is combined with the Carr-Purcell DD sequence. Unlike classical error correction, which can simply copy information, QEC cannot due to the no-cloning theorem. Instead, it encodes a single logical qubit into a highly entangled state of multiple physical qubits. By measuring correlations among these qubits without directly observing their state, QEC can detect and correct errors. Therefore, at each Carr-Purcell cycle’s end, a QEC step is applied with probability $p_{\text{QEC}}$, projecting towards the ideal maximally entangled state (let's say $\varphi \in \{ \text{Bell~or~GHZ}\})$ states, then $\rho_{\text{ideal}} = \ket{\psi_{\varphi}}\bra{\psi_{\varphi}}$:
\begin{equation}
\rho' = \frac{(1 - s) \rho + s \rho_{\text{ideal}}}{\tr((1 - s) \rho + s \rho_{\text{ideal}})},
\label{eq:qec_correction}
\end{equation}
where $s$ is a correction strength dependent on the state fidelity. This combines DD’s suppression of coherent noise with QEC’s correction of stochastic errors.

\subsubsection{RL-Framework for DD Sequence Optimization}\label{sec:rl_dd_framework}
To optimize the DD protocol, we employ the PPO algorithm. The RL environment applies control pulses via the global control Hamiltonian (Eq.~\eqref{eq:control_hamiltonian_N_dd}) within the system dynamics governed by the Lindblad master equation. The environment is defined as follows:

\paragraph{Action Space:} The action space is continuous, defined by an action $a = (\theta, \phi)$, where the pulse amplitude factor $\theta \in [0, 2\pi]$ and the phase $\phi \in [-\pi, \pi]$ specify the parameters for the control Hamiltonian (Eq.~\eqref{eq:control_hamiltonian_N_dd}). A pulse is applied only if $\theta > 0$.

\paragraph{Observation Space:} The observation space is $\mathbb{R}^{4^N}$, comprising the real parts of the $4^N-1$ expectation values $\langle \hat{P}_i \rangle = \tr(\rho \hat{P}_i)$ for non-identity generalized Pauli operators $\hat{P}_i$, augmented with the normalized episode time $t/T_{\text{total}}$.

\paragraph{Reward Function:} The reward function maximizes the final-state  entanglement measure $C_{\text{final}}$ while encouraging sustained high entanglement. At each intermediate step $t$, the reward is $R_t = R_{\text{sustain}} - C_{\text{action}}$, where $R_{\text{sustain}}$ is a small bonus awarded if the  entanglement measure is above a target threshold, and $C_{\text{action}}$ is a small penalty for applying a pulse. At the final step, the reward is 
\begin{equation}
R_{\text{final}} = R_f \cdot C_{\text{final}},~\text{with} ~R_f \gg 1.
\end{equation}

The final reward factor  is set to $R_f=10.0$, which multiplies the final state  entanglement measure $C_{\text{final}}$ to provide a substantial terminal reward. At intermediate steps, the reward function includes a sustain bonus $R_{\text{sustain}}= 0.01$, awarded if the $C_{\text{final}} >0.9$. Additionally, a small penalty for applying a pulse, the action cost $C_{\text{action}}=0.001.$
    \paragraph{Episode Structure:} Each episode consists of $N_{\text{steps}}$ discrete time steps. At each step, a pulse of duration $\tau_p$ is potentially applied, followed by a period of free evolution. The total time for one step is $\tau_{\text{step}} = \tau_p \times m_{\text{free}}$, where $m_{\text{free}}$ is a free evolution multiplier.


\subsection{Floquet Engineering Against RL Enforcement}
\label{sec:floquet_protocol_main}

In this section, we investigatethe  initial maximal entanglement state preservation using Floquet engineering \cite{a56}. The technique employs periodic driving to synthesize a desired effective Hamiltonian \cite{a57, a58}. We design and benchmark control strategies, comparing a non-adaptive protocol with optimized fixed parameters against an adaptive protocol optimized in real-time by an RL agent. The objective is to sustain  a maximally entangled N-qubit state in the presence of a realistic noise models and state/pulse errors.

The system dynamics are modeled in the lab frame, governed by the Hamiltonian $\hat{H}(t) = \hat{H}_{\text{sys}}(t) + \hat{H}_{c}(t) + \hat{H}_{\text{err}}$ with 
\begin{equation}
    \hat{H}_{\text{sys}}(t) = \sum_{j=1}^{N} \frac{\omega_q}{2} \hat{\sigma}_{z,j}.
\end{equation}
While the $\hat{H}_{c}(t)=\hat{H}_{i}(t)+\hat{H}_{d}(t)$  where the first sub part of interaction can be written as
\begin{equation}
\hat{H}_{i}(t) =g \sum_{j=1}^{N-1} \hat{\sigma}_{y,j}\hat{\sigma}_{y,j+1}.
\label{eq:H_sys_floquet_lab}
\end{equation}
The drive Hamiltonian $\hat{H}_{d}(t)$ consists of a multi-harmonic drive modulated by a periodic envelope function $f(t')$:
\begin{equation}
\begin{aligned}
\hat{H}_{c}(t) = \frac{1}{2} A_{\text{pulse}} f\left(t' - \frac{T_\Omega}{2}\right) \sum_{i=1}^{N_h} A'_{i} \cos(\nu_{i} \Omega t) \times \\ \left[ \cos(\phi_{i}) \sum_{j=1}^{N-1} \hat{O}_{j,j+1}^{(x)} + \sin(\phi_{i}) \sum_{j=1}^{N-1} \hat{O}_{j,j+1}^{(y)} \right],
\end{aligned}
\label{eq:H_drive_floquet_lab}
\end{equation}
where $t' = t \pmod{T_\Omega}$, $\hat{O}_{jk}^{(\alpha)} = \hat{\sigma}_{\alpha,j} \otimes \hat{\sigma}_{\alpha,k}$, and $A_{\text{pulse}}$ is a global amplitude factor specific to the envelope shape $f$. The per-harmonic amplitude $A'_{i} = A_{i}(1+\epsilon_A)$ includes the control parameter $A_i$ and a systematic amplitude error $\epsilon_A$. The system's open evolution is governed by the Lindblad master equation with this total Hamiltonian $\hat{H}(t)$. 

Finally, to assume a realistic and loosely corrected environmental setup such that to find the resourcefulness of our Floquet engineering scheme, we also assume an error accompanying term in the Hamiltonian, including classical noise and a systematic detuning on the first qubit:
\begin{equation}
\hat{H}_{\text{err}}(t) = \frac{\delta\omega_1(t)}{2} \hat{\sigma}_{z,1} + \frac{\Delta_{\text{err}}}{2} \hat{\sigma}_{z,1},
\label{eq:H_err_floquet_lab}
\end{equation}
where the second term ($\frac{\Delta_{\text{err}}}{2} \hat{\sigma}_{z,1}$) represents a qubit frequency detuning. This static or slowly varying offset from the nominal operating frequency is a persistent systematic error that arises from imperfect calibration or environmental drifts \cite{a59}. It is a critical error to model as it cannot be eliminated by simple passive protection and requires active correction, such as that provided by a Floquet drive. In the following, we introduce various methods to proceed Floquet engineering.

\paragraph{Reference Drive}

The Reference Drive protocol applies a periodic drive with fixed frequency $\Omega$ and pulse envelope $f(t')$ to the N-qubit system. The Hamiltonian follows \eqref{eq:H_drive_floquet_lab} with a single harmonic ($N_h = 1$) and fixed parameters ($A'_i$, $\phi_i$, $\nu_i = 1$). The envelope $f(t')$ (e.g., Gaussian) is periodic with period $T_\Omega = 2\pi / \Omega$, and the amplitude is scaled by $A_{\text{pulse}}$. Implemented using QuTiP’s \texttt{mesolve}, the system evolves from an N-qubit entangled state with optional perturbations, incorporating Lorentzian noise and dissipation. Finally, entanglement measure is employed  to assess entanglement preservation \cite{a59a}.

\paragraph{Optimized Fixed}

The Optimized Fixed protocol employs a multi-harmonic Floquet drive with fixed parameters optimized to maximize final concurrence. The drive Hamiltonian is given by \eqref{eq:H_drive_floquet_lab} with $N_h = 3$. Parameters ($A_i \in [0.5A_{\text{pulse}}, 2.0A_{\text{pulse}}]$, $\phi_i \in [-\pi, \pi]$, $\nu_i \in [0.5, 1.5]$) are sampled over 50 trials to find the optimal set. The system evolves using a JAX-based ODE solver (\texttt{odeint}) with precomputed superoperators, accounting for noise $\delta\omega_1(t)$ and detuning $\Delta_{\text{err}}$. The envelope $f(t')$ is scaled by $A_{\text{pulse}}$, and multipartite entanglement is tracked using a measure \cite{a59b}.

\paragraph{Lyapunov Control}
The Lyapunov Control protocol is a feedback control method used to stabilize a system's dynamics by ensuring a specific Lyapunov function continually decreases over time \cite{a59c}. The core idea is to find a function, similar to an energy function, that has a minimum value at the desired target state. The control protocol then uses the system's current state to calculate a control signal that actively drives the system towards this minimum, guaranteeing its stability. Here, we adjust the Lyapunov Control protocol dynamically adjusting the drive phase $\phi(t)$ to minimize a Lyapunov function for the N-qubit system. The drive Hamiltonian is:
\begin{equation}
\begin{aligned}
\hat{H}_{d}(t) =& \frac{1}{2} A_{\text{pulse}} f\Big(t' - \frac{T_\Omega}{2}\Big) \cos(\Omega t) \Big[ \cos(\phi(t))\\& \sum_{j=1}^{N-1} \hat{O}_{j,j+1}^{(x)} + \sin(\phi(t)) \sum_{j=1}^{N-1} \hat{O}_{j,j+1}^{(y)} \Big],
\end{aligned}
\end{equation}
where $\phi(t) = \arctan2\Big(-\tr[\mathcal{D}],-\tr[\mathcal{D}]\Big)$ with $\mathcal{D}=M \sum_{j=1}^{N-1} \hat{O}_{j,j+1}^{(\eta)}$,  $\eta \in \{x,y\}$, and $M = i (\rho \rho_{\text{target}} - \rho_{\text{target}} \rho)$ is the commutator with the target N-qubit entangled state. The amplitude is fixed at the maximum, scaled by $A_{\text{pulse}}$, using one harmonic ($\nu_1 = 1$). In the lyapunov control, the protocol computes the gradient of $M$ per step, evolving with the JAX-based solver, including noise and dissipation.
\subsubsection{Reinforcement Learning Framework}
For the case of Floquet engineering, the adaptive protocol is optimized by formulating the control task as a MDP and solving for the optimal policy $\pi^*: \mathcal{S} \to \mathcal{A}$ using the PPO algorithm.

\paragraph{Action Space:} The action space is a continuous, bounded subset of $\mathbb{R}^{3N_h}$. At each discrete time step $k$, the agent's action $\mathbf{a}_k \in \mathcal{A}$ is a vector of the control parameters for all $N_h$ harmonics:
    \begin{equation}
    \mathbf{a}_k = (A_{k,1}, \phi_{k,1}, \nu_{k,1}, \dots, A_{k,N_h}, \phi_{k,N_h}, \nu_{k,N_h}).
    \end{equation}
    These are mapped to physical ranges: amplitude factor $A_{k,i} \in [0, 2\pi]$, phase $\phi_{k,i} \in [-\pi, \pi]$, and frequency multiplier $\nu_{k,i} \in [0.5, 1.5]$.

\paragraph{State Space:}
To enable the RL agent to optimize DD pulses for preserving entanglement in a two-qubit system, the state space \( \mathbf{s}_k \in \mathcal{S} \) at step \( k \) is designed as a feature vector that encapsulates the quantum state, environmental noise, control context, and performance metrics. The observation is defined as:
\begin{equation}
\mathbf{s}_k = \mathbf{v}_{\text{Pauli}} \oplus \mathbf{v}_{\text{context}} \oplus \mathbf{v}_{\text{history}} \oplus \mathbf{v}_{\text{performance}},
\end{equation}
where the components are:
\begin{itemize}
    \item \( \mathbf{v}_{\text{Pauli}} \): Expectation values \( \tr(\rho_k \hat{P}_i) \) for a basis of two-qubit Pauli operators \( \{\hat{P}_i\} \) (excluding the identity), capturing the quantum state’s evolution under the Hamiltonian \( \hat{H}(t)\). This provides the agent with the system’s current configuration to inform pulse decisions.
    \item \( \mathbf{v}_{\text{context}} \): The normalized time \( k / N_{\text{steps}} \), a sample of the colored noise \( \delta\omega(t),$
and the pulse envelope value \( f(t - t_c) \). These reflect the temporal context, environmental noise affecting \( \hat{\sigma}_z^1 \), and the current control strength, enabling the agent to adapt to time-varying dynamics and coupling modulation in \( \hat{H}_c(t)\).
    \item \( \mathbf{v}_{\text{history}} \): The agent’s previous two actions, \( \mathbf{a}_{k-1} \) and \( \mathbf{a}_{k-2} \) (pulse amplitude and phase), providing memory of recent control decisions to ensure coherent pulse sequences.
    \item \( \mathbf{v}_{\text{performance}} \): The current entanglement measure \( C(\rho_k) \), while the previous \( C(\rho_{k-1}) \), and their difference, quantifying entanglement and its trend. This guides the agent toward maximizing entanglement preservation, the primary objective.
\end{itemize}
This state space ensures the agent has comprehensive information to optimize control pulses against decoherence and the coherent interactions induced by the control term \( \hat{H}_c(t) \), aligning with the system’s dynamics and RL objective.

\paragraph{Reward Function:}
The reward function \( R_k \) is designed to guide the RL agent in optimizing pulses to maximize entanglement in a two-qubit system, measured by \( C(\rho_{k+1}) \). At each step \( k \), the reward is calculated based on the resulting state \( \rho_{k+1} \) and the action \( \mathbf{a}_k = (\theta_k, \phi_k) \), representing the pulse amplitude and phase:
\begin{equation}
R_k = R_{\text{ent}}(\rho_{k+1}) - P_{\text{action}}(\mathbf{a}_k) + R_{\text{terminal}}.
\label{eq:reward_function}
\end{equation}
The components are:
\begin{itemize}
    \item{Entanglement Reward:} A positive reward to encourage high entanglement: $ R_{\text{ent}}(\rho_{k+1}) = c_e \cdot C(\rho_{k+1}),$
    where \( c_e = 0.01 \) (sustain bonus) is applied when \( C(\rho_{k+1}) > C_{\text{target}} = 0.9 \). This incentivizes maintaining near-maximal entanglement throughout the evolution.
    
    \item{Action Penalty:} A penalty for applying control pulses to minimize unnecessary actions:
$ P_{\text{action}}(\mathbf{a}_k) = c_p \cdot \mathbb{1}_{\{\theta_k > 0\}},$ where \( c_p = 0.001 \) is applied if a pulse is active (\( \theta_k > 0 \)), promoting sparse control sequences to reduce energy costs and noise amplification.

    \item{Terminal Reward}: A large reward at the final step (\( k = N_{\text{steps}} \)):
$R_{\text{terminal}} = c_t \cdot C(\rho_{N_{\text{steps}}}),$   where \( c_t = 10.0 \) amplifies the final  entanglement measure to prioritize long-term entanglement preservation.
\end{itemize}

The reward function for the RL framework in Floquet engineering is configured to maximize final-state  entanglement measure while penalizing undesirable control actions and low entanglement. The terminal reward is a bonus of  $R_{\text{bonus}}=500$  if the final  entanglement measure is greater than the target threshold  $C_{\text{target}}=0.95$ , and a penalty of  $P_{\text{term}}=100$  otherwise. At each intermediate step, the agent receives a reward scaled by a factor of  $c_{1}=20$ , proportional to the  entanglement measure to the power of  $p=3.0$ . A penalty of  $c_{2}=5$  is also included, calculated as $5 \times (1-C(\rho_k))^2$, along with a cost for action changes, quantified by a penalty factor of  $c_{\Delta a}=0.05$ . The RL agent is a PPO model trained for  $200,000$  timesteps, utilizing a policy network with  three hidden layers of 256 neurons  each. The learning rate is set at  $\alpha=1 \times 10^{-4}$ , with a discount factor of  $\gamma=0.9999$ . The agent controls  $N_{h}=3$  Floquet harmonics with pre-defined weights of  $w_{1}=1.0$ ,  $w_{2}=0.5$ , and  $w_{3}=0.3$ , applied over  $N_{\text{steps}}=100$  discrete time steps. Therefore, this reward structure balances the need to sustain high  entanglement measure against the coherent interactions induced by \( \hat{H}_c(t)\) and environmental noise, and errors, while encouraging efficient control strategies.

\subsection{Benchmarking RL against standard Entanglement Generation protocol}

To extend our RL framework from preserving entanglement to generating it, we now focus on creating a maximally entangled state from an initial separable state \cite{a60, a61}, under the presence of environmental noise and coherent interactions driven by the control pulses. With our framework for an N-qubit state, we implement and benchmark comparing the RL-optimized protocol against standard, Heuristic \cite{a50}, and Knill-DD \cite{a62} using an entanglement metric. Note that to illustrate how much the existing Hamiltonian in original form is capable of generating entanglement, we provide a No-Control scheme which would clarify how much our employed schemes are resourceful.

The system evolves under a total Hamiltonian that includes intrinsic dynamics, environmental noise, and external control. The system Hamiltonian $\hat{H}_{\text{sys}}$, we consider is defined in Eq. \eqref{total Ham} with $\delta\omega_j(t)$, the classical noise is applied only to the first qubit, $j=1$.

As entanglement generation is crucial and different process than preservation, therefore we consider a different YY coupling Hamiltonian over a single-qubit Hamiltonian, such as $\hat{H}_i(t) = g (\hat{\sigma}_y^1 + \hat{\sigma}_y^2)$. Because the two-qubit interaction $\hat{\sigma}_y^1 \otimes \hat{\sigma}_y^2$ directly generates entanglement by correlating the qubits' states in the $y$-$y$ plane, enabling transitions from $\ket{00}$ to superpositions like $\ket{00} + \ket{11}$ \cite{a63}. Hence, the interaction term is assumed as:
\begin{equation}
\hat{H}_{i}(t) = g \sum_{j=1}^{N-1} \hat{\sigma}_{y,j} \otimes \hat{\sigma}_{y,j+1},
\end{equation}

Next, the Floquet control is performed by applying a sequence of shaped pulses locally to the first qubit. During a pulse of duration $T_p$ at step $k$, the interaction strength is modulated by the pulse envelope $f(t)$, and a control-drive term is added, and we get $f(t)\hat{H}_{i}(t) $. In the explicit form, the total Hamiltonian for $t \in [t_k, t_k + T_p]$ is:
\begin{equation}
\begin{aligned}
\hat{H}(t) = &\hat{H}_{\text{sys, free}}(t) + g f(t-t_k) \sum_{j=1}^{N-1} \hat{\sigma}_{y,j} \otimes \hat{\sigma}_{y,j+1} \\&+ \hat{H}_{c}(t; \mathbf{a}_k) + \hat{H}_{\text{err}}(t).
\end{aligned}
\end{equation}
where $\hat{H}_{\text{sys, free}}$ contains the non-interacting terms of $\hat{H}_{\text{sys}}$. The local control-drive Hamiltonian is parameterized by an action vector $\mathbf{a}_k = (\theta_k, \phi_k)$:
\begin{equation}
\hat{H}_{c}(t; \mathbf{a}_k) = \frac{\Omega(t; \theta_k)}{2} \left[ \cos(\phi_k) \hat{\sigma}_{x}^{(1)} + \sin(\phi_k) \hat{\sigma}_{y}^{(1)} \right].
\label{eq:H_c_entanglement}
\end{equation}

To enhance the realism of our simulations, we explicitly include pulse imperfections. These are described by an error Hamiltonian term, $\hat{H}_{\text{err}}(t)$, which accounts for the most common sources of experimental error \cite{a64}. This term includes a systematic detuning and a pulse amplitude error:
\begin{equation}
\begin{aligned}
\hat{H}_{\text{err}}(t) = \frac{\Delta_{\text{det}}}{2} \hat{\sigma}_{z,1} + \frac{\delta\Omega(t)}{2} \left[ \cos(\phi_k) \hat{\sigma}_{x,1} + \sin(\phi_k) \hat{\sigma}_{y,1} \right],
\end{aligned}
\end{equation}
where $\Delta_{\text{det}}$ is a constant detuning error on the first qubit. The pulse amplitude error, $\delta\Omega(t)$, is a stochastic term drawn from a normal distribution, representing random shot-to-shot fluctuations. This term is applied at the same time as the control pulse and on the same axis. Outside of pulse durations, the system evolves under $\hat{H}_{\text{sys}}(t)$ with constant interaction strength $g$.

In the following, we define the protocols employed to find out the Floquet engineering evaluation.
\paragraph{No-Control Protocol.}
This protocol serves as the fundamental baseline. No external control is applied, corresponding to setting the action vector to zero for all time steps, $\mathbf{a}_k = (0, 0) \; \forall k$. The state evolves solely under the native system Hamiltonian and environmental noise. One should note that without applying any specific protocol, the pulses here are impacting the system.

\paragraph{Heuristic Protocol.}
This protocol implements a simple, pre-determined open-loop strategy consisting of alternating pulse areas along a fixed axis \cite{a50}. The action at each step $k$ is:
\begin{equation}
\mathbf{a}_k = \left( (-1)^k \frac{\pi}{2}, 0 \right).
\end{equation}
This applies alternating $+\pi/2$ and $-\pi/2$ pulses along the $\hat{\sigma}_x$ axis of the first qubit. The sequence is fixed and applies a pulse at every time step.

\paragraph{Knill DD Protocol}
This protocol adapts a standard DD sequence, using sparse, high-power pulses \cite{a65}. The sequence consists of $N_p=12$ strong $\pi$-pulses applied at evenly spaced intervals. The action $\mathbf{a}_k$ is zero for most steps, but for the $N_p$ pulse steps, it is:
\begin{equation}
\mathbf{a}_k = (\pi, \phi_{\text{DD}}),
\end{equation}
where the phase $\phi_{\text{DD}}$ cycles through the sequence $\{0, \pi/2, 0, \pi/2, \dots\}$ to counteract noise from different axes.

\subsubsection{RL-framework for Entanglement generation}
\label{sec:rl_ent_gen}

In contrast to  fixed benchmarks, this protocol uses a trained PPO agent to implement a fully adaptive control strategy and generate entanglement. The MDP is defined as follows:

\paragraph{Action Space:} At each step $k$, the agent selects a continuous action $\mathbf{a}_k = (\theta_k, \phi_k)$, where the pulse area $\theta_k \in [0, \pi]$ and phase $\phi_k \in [-\pi, \pi]$ define the local pulse on the first qubit via Eq.~\eqref{eq:H_c_entanglement}.

\paragraph{Observation Space:} The observation $\mathbf{o}_k$ is the vector of expectation values of the $4^N-1$ non-identity N-qubit Pauli operators, $\tr(\rho_k \hat{P}_i)$, concatenated with the normalized time step.

\paragraph{Reward Function:} To incentivize both the generation followed by preservation of entanglement, we employ a shaped reward function. The reward at step $k$ is based on the \textit{change} in entanglement:
    \begin{equation}
    R_k = w_{\text{c}} \left( C_k - C_{k-1} \right) - w_{\text{pen}} \left( \frac{\theta_k}{\pi} \right)^2.
    \label{eq:reward_shaped}
    \end{equation}
    Here, $C_k$ is the entanglement measurement of the state $\rho_k$. The first term, with weight $w_{\text{c}}=50$, rewards improvement in entanglement. The second term is a small penalty, weighted by $w_{\text{pen}}=0.005$, on the squared normalized pulse area to promote resource-efficient solutions. This structure motivates the agent to not only reach a high-entanglement state but to actively maintain or improve it.
 The agent is trained using the same hyperparameters and pulse shapes as the benchmark protocols to facilitate a direct comparison under identical, realistic conditions.

\subsection{Adaptive Quantum Walk Strategy}
\label{sec:quantum_walk_strategy}

This protocol implements an adaptive control strategy for a discrete-time quantum walk (DTQW) on a one-dimensional lattice, as described in \cite{a66}. DTQW is a quantum-mechanical counterpart to a classical random walk. It describes the evolution of a particle on a lattice, governed by quantum principles \cite{a67}. Unlike its classical counterpart that moves in a random direction at each step, a DTQW's evolution is governed by two discrete, sequential operations: a quantum coin operation that puts the particle's internal state into a superposition and a conditional shift operation that moves the particle based on its internal state \cite{a68}. Moreover, due to quantum interference, a DTQW's probability distribution spreads much faster and exhibits distinct, non-smooth peaks, unlike the smooth, diffusive spread of a classical random walk. This unique behavior makes DTQW a valuable model for designing quantum algorithms and a compelling subject for studying quantum control strategies, especially in the presence of noise.

The DTQW is defined on a lattice with $N_s = 11$ sites, labeled $\{-5, \dots, 5\}$. The Hilbert space is $\mathcal{H} = \mathcal{H}_c \otimes \mathcal{H}_p$, where $\mathcal{H}_c = \mathbb{C}^2$ is the coin space and $\mathcal{H}_p = \mathbb{C}^{N_s}$ is the position space \cite{a66, a67}. The system state at discrete step $n$ is represented by the density matrix $\rho_n$. The walk is initialized in a pure state:
\begin{equation}
\ket{\psi_0} = \frac{1}{\sqrt{2}}(\ket{0} + i\ket{1})_c \otimes \ket{0}_p,
\end{equation}
localizing the walker at the lattice center with the coin in a superposition to induce an asymmetric walk, which promotes directional movement toward the target site.

The protocol evolves for $N_{\text{steps}} = 30$ steps, with each step duration $\Delta t = \frac{2 \cdot T}{N_{\text{steps}}}$, where $T$ is the characteristic pulse time. This choice of step number and duration balances the need for sufficient evolution time to explore the lattice with computational efficiency, ensuring the walker can approach the target site within a reasonable number of steps.

Each step consists of two operations: an adaptive coin operation governed by a time-dependent Hamiltonian, followed by a conditional shift. The shift operator is unitary, defined as:
\begin{equation}
S = (\ket{0}\bra{0})_c \otimes S_R + (\ket{1}\bra{1})_c \otimes S_L,
\end{equation}
where $S_R = \sum_{x=-5}^{4} \ket{x+1}\bra{x}$ and $S_L = \sum_{x=-4}^{5} \ket{x-1}\bra{x}$ are the right and left shift operators on the lattice. If the state after the coin operation is $\rho'_n$, the subsequent state is $\rho_{n+1} = S \rho'_n S^\dagger$. This shift operation directs the walker's movement based on the coin state, enabling controlled transport across the lattice.

The coin operation at step $n$ is driven by a Hamiltonian parameterized by the RL agent's action, $\mathbf{a}_n = (a_{\text{amp},n}, a_{\phi,n}, a_{\text{timing},n}) \in [-1, 1]^3$. The action maps to
physical parameters as follows: the pulse amplitude $\theta_n = \frac{a_{\text{amp},n} + 1}{2} \cdot 3\pi$, the phase $\phi_n = a_{\phi,n} \cdot \pi$, and the pulse timing $t_{c,n} = t_{\text{center}} + a_{\text{timing},n} \cdot 3 \cdot \Delta t$, where $t_{\text{center}} = 0.5 \cdot N_{\text{steps}} \cdot \Delta t$. The Hamiltonian is:
\begin{equation}
\hat{H}_c(t; \mathbf{a}_n) = \Omega_0 \cdot \frac{\theta_n}{\pi} \cdot f(t - t_{c,n}) \cdot \hat{H}_{\text{axis}}(\phi_n),
\label{eq:qw_hamiltonian_actual}
\end{equation}
where $\Omega_0$ is the pulse strength, and the control axis is:
\begin{equation}
\hat{H}_{\text{axis}}(\phi_n) = \cos\phi_n \hat{\sigma}_{x,c} \otimes \hat{I}_p + \sin\phi_n \hat{\sigma}_{y,c} \otimes \hat{I}_p
\end{equation}
where $\hat{I}_p$ represents the identity operator. 
The envelope $f(t - t_{c,n})$ shapes the temporal profile of the control pulse, with six possible types. These pulse shapes provide a range of control strategies, from smooth and symmetric profiles to oscillatory or asymmetric modulations, enabling the RL agent to adapt the coin operation to various environmental conditions and optimize the walker's dynamics.

The system evolves under a stochastic master equation to account for environmental decoherence and disorder given in Eq.~\eqref{eq:lindblad_compact} with $\hat{H}_{\text{total}}(t) = \hat{H}_c(t) + \hat{H}_{\text{noise}}(t) + \hat{H}_{\text{disorder}}$. In the context of the quantum walk, the stochastic master equation is critical because it models the open quantum system dynamics, capturing the effects of environmental decoherence (e.g., loss of quantum coherence due to interactions with an external bath) and static disorder (e.g., site-dependent energy variations). Decoherence disrupts the walker's superposition, reducing the efficiency of quantum interference, while disorder introduces randomness in the lattice potential, challenging the walker's ability to localize at the target site $x_{\text{target}} = -3$ \cite{a69}. In addition, the  colored noise Hamiltonian is written as:
\begin{equation}
\hat{H}_{\text{noise}}(t) = \delta\omega_z(t) \cdot (\hat{\sigma}_{z,c} \otimes \hat{I}_p),
\end{equation}
with $\delta\omega_z(t) $ is defined in Eq. \eqref{lorent} .


In an ideal, perfectly uniform lattice, a quantum walker's evolution is purely coherent, leading to ballistic spreading and distinct probability peaks. However, in any real-world physical implementation (e.g., in photonic lattices, cold atoms, or ion traps), there are imperfections. These imperfections, such as local variations in potential energy or fabrication defects, create a non-uniform landscape that the walker moves through \cite{a69}. The disorder Hamiltonian introduces static spatial randomness:
\begin{equation}
\hat{H}_{\text{disorder}} = \hat{I}_c \otimes \sum_{x=-5}^{5} d_x \ket{x}\bra{x},
\end{equation}
with $d_x \sim \mathcal{U}[-0.01 \cdot \Omega_0, 0.01 \cdot \Omega_0]$. This term represents site-specific energy variations, which can impede the walker's coherent propagation and necessitate adaptive control to achieve the target localization.  The disorder Hamiltonian, $\hat{H}_{\text{disorder}}$ introduces a random energy offset, $d_x$, at each site $x$. This random potential breaks the translational symmetry of the ideal system, which has two main effects. By including this disorder term, the model becomes a more accurate representation of a real experiment. It transforms the problem from a simple demonstration of quantum mechanics into a genuine quantum control challenge: how to use adaptive control (like a trained RL agent) to counteract the disorder and guide the walker to a specific target site, even when the environment is actively working against its coherent propagation.

\subsubsection{RL Framework for Quantum Walk}
\label{sec:rl_qwalk}

The RL framework employs a PPO agent with a recurrent neural network (Recurrent PPO) featuring an LSTM-based policy with hidden size 128, trained for 300,000 timesteps across up to 8 parallel environments. The agent optimizes the quantum walk to maximize the probability at the target site $x_{\text{target}} = -3$ under decoherence and disorder, using optimized hyperparameters: learning rate $10^{-4}$, batch size 128, 8 epochs, discount factor $\gamma = 0.995$, GAE $\lambda = 0.98$, clip range 0.15, and entropy coefficient 0.02.

\paragraph{Action Space:}
The action 
\begin{equation}
\mathbf{a}_n = (a_{\text{amp},n}, a_{\phi,n}, a_{\text{timing},n}) \in [-1, 1]^3
\end{equation}
parameterizes the Hamiltonian in Eq.~\eqref{eq:qw_hamiltonian_actual}, controlling the pulse amplitude, phase, and timing for the coin operation at step $n$.

\paragraph{Observation Space}
The observation is a vector of length $2(2N_s)^2 + 6 = 974$, comprising the real and imaginary parts of the flattened density matrix $\rho_n$ (dimension $2N_s \times 2N_s = 22 \times 22$), the normalized step number ($2 \cdot n / N_{\text{steps}} - 1$), the normalized target site ($2 \cdot x_{\text{target}} / N_s - 1 = -6/11$), a tanh-transformed step progress ($\tanh((n - N_{\text{steps}}/2)/(N_{\text{steps}}/4))$), the current target site probability ($10 \cdot p(x_{\text{target}}, n)$), the probability progress ($20 \cdot (p(x_{\text{target}}, n) - p(x_{\text{target}}, 0))$), and the short-term probability trend ($10 \cdot \text{mean}(p(x_{\text{target}}, n-2:n))$).

\paragraph{Reward Function}
The reward at step $n$, denoted by $r_n$, is a sophisticated combination of terms designed to guide the agent and maximize the target site probability (TSP). It is defined as:

\begin{equation}
\begin{aligned}
r_n &= 50 \cdot P_n + 100 \cdot e^{5 P_n} \cdot \mathbb{1}_{P_n > 0.3} \\
&\quad - 2 \cdot \sum_{x=-5}^{5} p(x, n) |x - x_{\text{target}}| \\
&\quad + 20 \cdot \max(0, P_n - \bar{p}_{\text{hist}}) + \delta_{n, N_{\text{steps}}} \cdot 200 \cdot P_n^2 \\
&\quad + 10 \cdot P_n \cdot \frac{n}{N_{\text{steps}}} + 50 \cdot (P_n - P_{n-1}) \\
&\quad - 10 \cdot \mathbb{1}_{\text{var}(p(x_{\text{target}}, n-4:n)) < 10^{-6}, p(x_{\text{target}}, n) < 0.1}
\end{aligned}
\end{equation}

where the terms are defined as follows: $P_n = p(x_{\text{target}}, n)$ is the probability of the particle being at the target site at step $n$. $p(x, n)$ represents the probability distribution of the particle's position and $\bar{p}_{\text{hist}}$ is the mean of the last five target probabilities. The final two terms reward immediate improvements and penalize stagnation at low probabilities, respectively.

 The reward function, $r_n$, is  crafted to aggressively guide the agent toward a high TSP, denoted as $P_n$. This multi-component function rewards momentum and penalizes stagnation. The agent receives a substantial base reward of $50 \cdot P_n$, directly proportional to the TSP. A significant exponential bonus of $100 \cdot e^{5 P_n}$ is awarded if the TSP surpasses a threshold of 0.3, strongly incentivizing the agent to achieve high-probability states. To encourage a concentrated probability distribution at the target, a spatial penalty is applied, calculated as $-2 \cdot \sum_{x} p(x, n) |x - x_{\text{target}}|$. A momentum term of $20 \cdot \max(0, P_n - \bar{p}_{\text{hist}})$ rewards positive changes in the TSP compared to the average of the last five steps ($\bar{p}_{\text{hist}}$). A large final-step bonus of $200 \cdot P_n^2$ is provided at the end of the walk, rewarding the final performance. The reward also includes a linear progress term of $10 \cdot P_n \cdot n/N_{\text{steps}}$, which encourages sustained improvement throughout the walk. Additionally, an immediate improvement reward of $50 \cdot (P_n - P_{n-1})$ is given at each step to reward any increase in TSP. Finally, to prevent the agent from getting stuck in low-performing states, a stagnation penalty of $-10$ is applied if the TSP shows minimal variance and remains below 0.1 over the last five steps. This sophisticated reward structure provides detailed and continuous feedback, enabling the agent to learn complex strategies that go beyond simple maximization.

\paragraph{Curriculum Learning}
The agent employs a curriculum learning strategy, progressively increasing the difficulty of the training environment. Specifically, the target site distance is gradually increased, up to a maximum of $|x_{\text{target}}| = 3$, and the disorder strength is adjusted, reaching a maximum of $0.01 \cdot \Omega_0$. These parameters are updated every 200 episodes based on the agent's training progress, enabling it to generalize to increasingly challenging tasks.

\paragraph{Performance Evaluation}
The performance of the RL agent is rigorously evaluated across 5 distinct realizations. For each run, we compute the mean and standard deviation of three key metrics to account for statistical variability:\\
Target Site Probability (TSP): The probability of the particle reaching the designated target site written as  $p(x, n) = \langle x | \mathrm{Tr}_c(\rho_n) | x \rangle$ \cite{a70}.\\ Entanglement Entropy (EE): A measure of the quantum entanglement between the particle's position and the coin state \cite{a71}. It is defined as $EE = -\Tr(\rho_{\text{pos}} \log \rho_{\text{pos}})$, where $\rho_{\text{pos}} = \Tr_c(\rho)$ is the reduced density matrix of the position subspace.\\
Mutual Information (MI): The total correlation between the coin and position degrees of freedom, calculated as $MI = S_{\text{coin}} + S_{\text{pos}} - S_{\text{total}}$ \cite{a72}. Here, $S_i = -\Tr(\rho_i \log \rho_i)$ represents the von Neumann entropy for a given subspace, with $\rho_{\text{coin}} = \Tr_p(\rho)$ and $S_{\text{total}} = -\Tr(\rho \log \rho)$ representing the reduced density matrices of the coin, position, and total system, respectively.

\section{Numerical Results and Analysis}\label{results}
In this section, we present numerical results for the quantum control protocols developed in the theoretical framework given in Fig. \ref{t-model}. While the formalism was introduced for a general N-qubit system, simulating the dynamics of such systems is computationally demanding due to the exponential growth of the Hilbert space. Therefore, to provide a clear and tractable demonstration of our methods, we now specialize our analysis to the foundational case of a two-qubit system ($N=2$). All Hamiltonians, control operators, and decoherence models used in the following simulations are direct specializations of the general N-qubit equations presented in Sec. \ref{sec:physical_model} with $N=2$ (except the model for the quantum walk in Sec. \ref{sec:quantum_walk_strategy}). Note that in most of the protocols, we utilize concurrence as measure of entanglement defined in Eq. \eqref{con} Sec. \ref{appendix}.

\subsection{Deterministic  against RL protocols initiated with various initial
states}
In this section, we consider various initial states including separable and entangled, see Appendix Sec. Sec. \ref{initial states} to test the efficacy of the protocols under the Gaussian-derived pulse shapes developed in Sec. \ref{sec:control_protoocls} against the RL-framework.

\begin{figure*}[ht]
\centering
 \begin{overpic}[height=8cm,width=16cm]{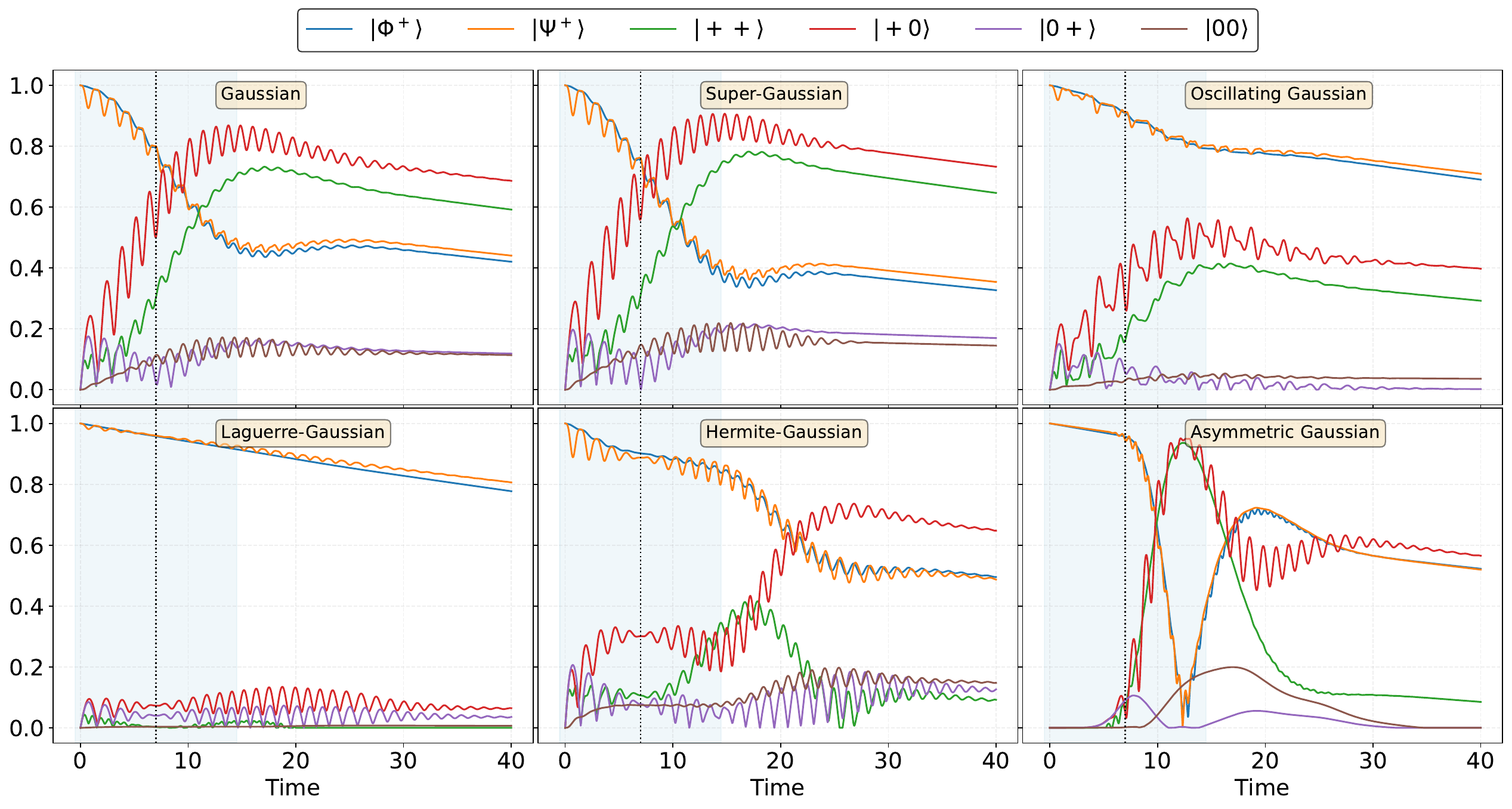}
 \put(4, 40){(a)}
\put(36, 40){ (b)}
\put(68, 40){(c)}
\put(4, 17){(d)}
\put(36, 17){(e)}
\put(68, 17){(f)}
\end{overpic}
\caption{Concurrence versus time for six initial states across six pulse types. Each subplot shows the concurrence with the pulse active region shaded. The parameters are set as follows: inter-qubit coupling $g_{0,1}=0.5$, $\omega_{q,A}=4.8$, $\lambda_{d,A}=1.0$, $\omega_{d,A}=5.0$, with $\phi_{d,A}=0.0$. The pulse is centered at $t_c=7.0$ with a characteristic duration of $T=15.0$ with $\gamma_{1A, 1B}=0.001$, $\gamma_{\phi_A, \phi_B}=0.001$.The Lorentzian colored noise is employed with parameters: $M=1000$, $\Gamma=0.5$, $\tau_c=5.0$, $S_0=0.2$ and $\{ \Omega_{min},\Omega_{max}\}$=$\{-5,5\}$.} \label{fig1}
\end{figure*}
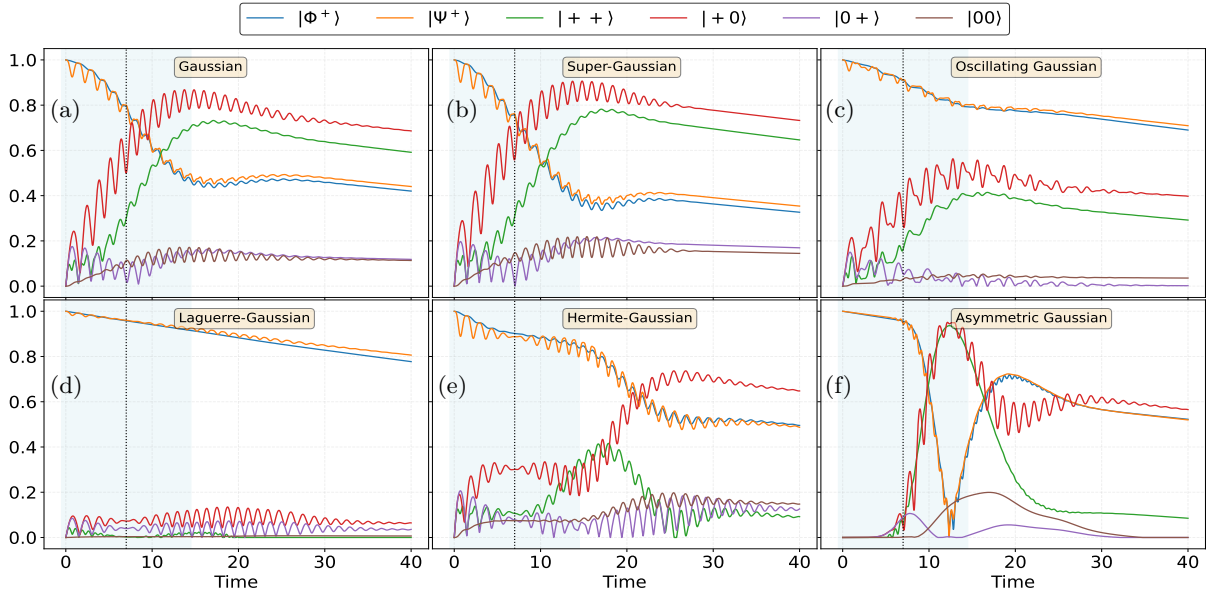 
Fig.~\ref{fig1} provides a comprehensive analysis of entanglement dynamics under a pulse-driven Hamiltonian, illustrating both the preservation of initially entangled states and the generation of entanglement from separable ones. The shaded areas in the figure highlight the time intervals where the external control pulse is active. A key observation is the stark difference in dynamics inside versus outside the pulse-active region. Within this region, the system undergoes complex, highly oscillatory, and non-monotonic evolution, demonstrating the powerful influence of the external drive. After the pulse ceases, the dynamics transitions to a smoother, more predictable decay, characteristic of the natural decoherence channels of the system. When analyzing the preservation of initial Bell states, a clear hierarchy emerges. The \PSI~state demonstrates greater robustness to decoherence compared to the \PHI state across most of the pulse types, indicating that its specific symmetry offers greater protection against the noise model. The choice of pulse envelope is also critical for preservation. The \OGP~(c) and \LGP~(d) pulses are the most effective, inducing gentle oscillations and resulting in the highest final concurrence for the Bell states. In contrast, the more complex pulse \AGP~(f) causes significant, transient entanglement degradation during the driving phase, but the state recovers entanglement after the sudden death. The other pulses also support entanglement preservation in Bell states, but with a lower final concurrence. For entanglement generation from separable states, the \PZ~state followed by the \PP~state are the most favorable for generation. This is a testament to show its specific superposition is readily driven into an entangled subspace by the driven Hamiltonian. The efficacy of this generation is, once again, critically dependent on the pulse. The \GP~(a), \SGP~(b), \OGP~(c), \HGP~(e), and \AGP~(f) pulses act as powerful engines for entanglement, driving the concurrence of the initially separable states to levels that are, at times, comparable to the decaying Bell states. Notably, for \GP~(a), \SGP~(b), \HGP~(e), and \AGP~(f), the generated entanglement exceeds the final concurrence of the Bell states, which is a significant result. This suggests that the complex spectral content of these aggressive, asymmetric pulses is particularly well-suited to resonantly exciting the system into the desired entangled states. The other separable states, such as \ZP~and \ZZ, show much weaker generation and remain less entangled. Our results demonstrate a crucial duality in the function of shaped pulses: such as the pulses like \OGP~ (c) and \LGP~(d) appear optimal for preserving existing quantum correlations, while the remaining pulses are more effective for actively generating entanglement from scratch. Finally, the greater robustness of the \PSI~state is attributed to its specific symmetry, which may leave it in a decoherence-free or decoherence-resistant subspace of the Hamiltonian, making it less susceptible to the primary noise channels. While the system transitions from an initially separable product state to a entangled superposition state because the pulse-driven interaction Hamiltonian provides the necessary energy and symmetry to mix the bare states.

\begin{figure}[ht]
    \centering
    \begin{overpic}[height=6cm,width=8cm]{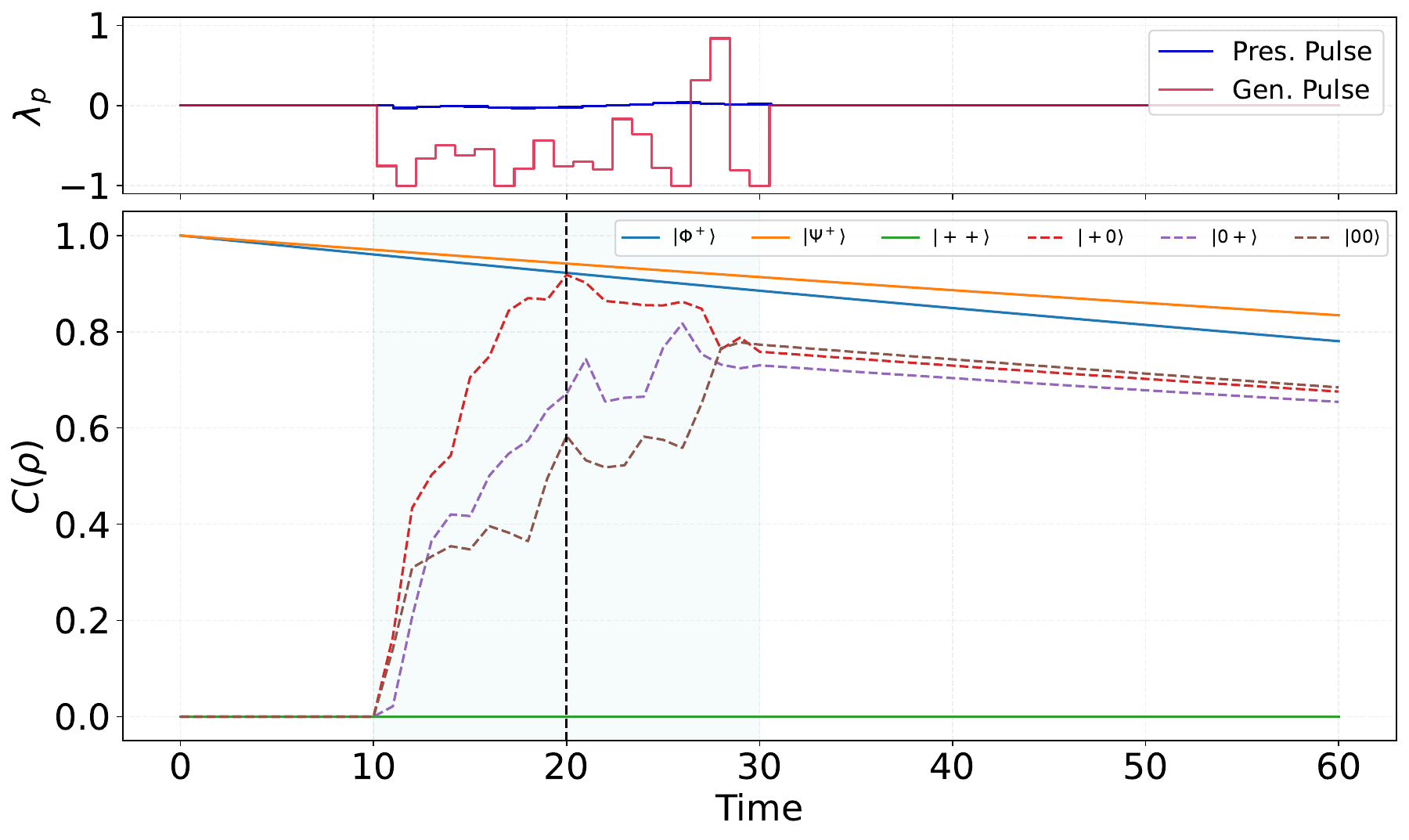}
    \put(10, 68){(a)}
    \put(10, 40){(b)}
    \end{overpic}
    \caption{Top: RL optimized pulse amplitudes ($\lambda_p$) for entanglement preservation and generation tasks, active within the control window $t_{\text{CW}}=[10, 30]$. Bottom: Evolution of concurrence over time under the influence of the RL-optimized pulses, similar to Fig.~\ref{fig1}. The agent is trained using the PPO algorithm. The policy is a Multilayer Perceptron (MlpPolicy) with a shared two-layer network for the actor and critic. Both the preservation and generation tasks are trained for a total of $3 \times 10^5$ timesteps using a rollout buffer size ($n_{steps}$) of $2048$. For the generation task, the learning rate is fixed at $1 \times 10^{-4}$, while the preservation task uses a fixed learning rate of $3 \times 10^{-4}$.}
    \label{fig2}
\end{figure}
Fig.~\ref{fig2} illustrates the impact of applying an RL framework to the entanglement dynamics problem presented in Fig.~\ref{fig1}. Recognizing that the goals of preserving existing entanglement and generating it from scratch are fundamentally different control challenges, two distinct RL agents were trained with specialized reward functions, resulting in the two unique pulse sequences shown in Fig.~\ref{fig2}(a). Therefore, it can be counted an advantage of the deterministic protocols which both can adhere to preserve as well as generate entanglement, while shaping the reward function of the RL may remain largely less effective and challenging in comparison. Therefore, this task-specific approach stands in contrast to using a universal, pre-defined pulse as shown in Fig.~\ref{fig1} and is key to the observed success. Fig.~\ref{fig2}(a) reveals the non-intuitive and structured nature of the learned control policies. The preservation pulse is characterized by lower-amplitude, corrective actions, seemingly designed to gently nudge the system and counteract decoherence without introducing excess noise. Conversely, the generation pulse is far more aggressive, employing strong, alternating positive and negative amplitudes in a bang-bang-like sequence. This policy acts as a powerful, resonant drive, designed to rapidly pump the system from a separable state into an entangled one. The results of these tailored policies are shown in Fig.~\ref{fig2}(b). For the preservation task, the RL agent achieves some stabilization of the dynamics. Comparing this to the chaotic oscillations seen in Fig.~\ref{fig1}, the RL-optimized pulse almost completely suppresses the violent fluctuations, guiding the Bell states (\PHI~and \PSI) along a much smoother decay trajectory. While entanglement loss is still present due to the open nature of the system, the control actively places the system in a state that is maximally robust to decoherence. The inherent robustness of the \PSI~state over the \PHI~state is maintained, suggesting the RL agent optimizes within the physical constraints of the system's decoherence channels. For the generation task, the aggressive pulse from panel (a) proves highly effective. Within the control window, all initially separable states are rapidly driven to significant levels of concurrence however under the active window regoin of the pulse. Notably, the RL agent discovers that the \PZ,~ \ZP,~ and~ \ZZ~ states are particularly amenable to entanglement generation, creating and sustaining a high degree of concurrence that persists long after the control pulse is turned off. The most significant advantage over the protocols in Fig.~\ref{fig1} is the reliability and stability of the generated entanglement. Instead of transient, unpredictable spikes, the RL agent creates a robustly entangled state that then decays smoothly. Therefore, a tailored RL approach can discover sophisticated, non-trivial control sequences that qualitatively transform the system's dynamics from chaotic and inefficient to stable and highly effective for specific quantum control objectives.
 
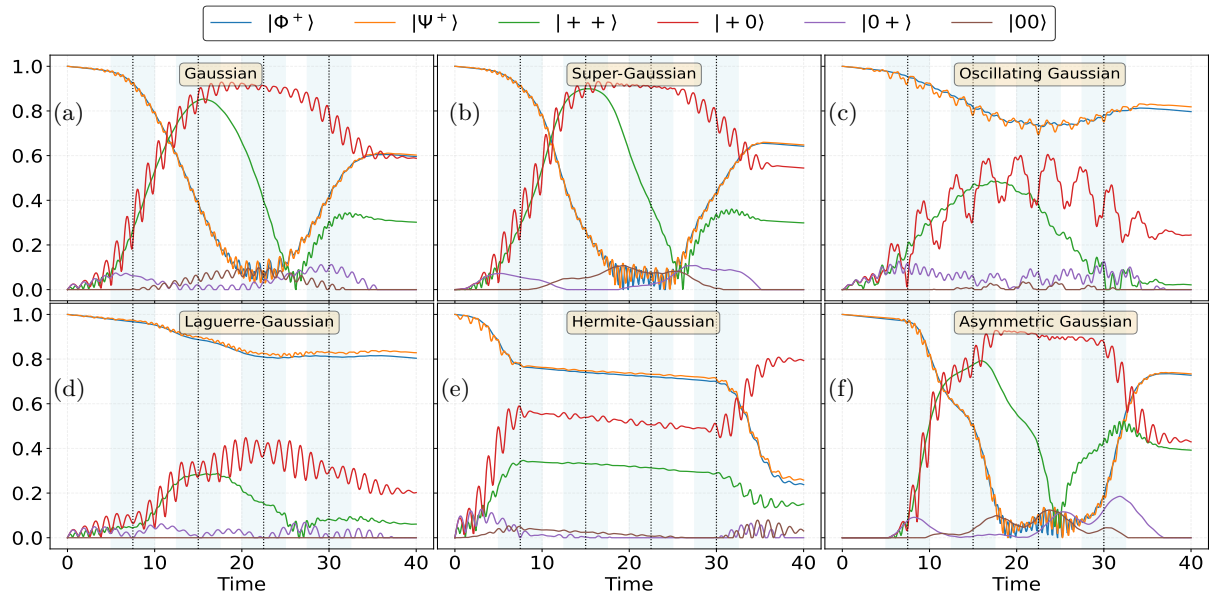
\begin{figure*}[ht]
\centering
\begin{overpic}[height=8cm,width=16cm]{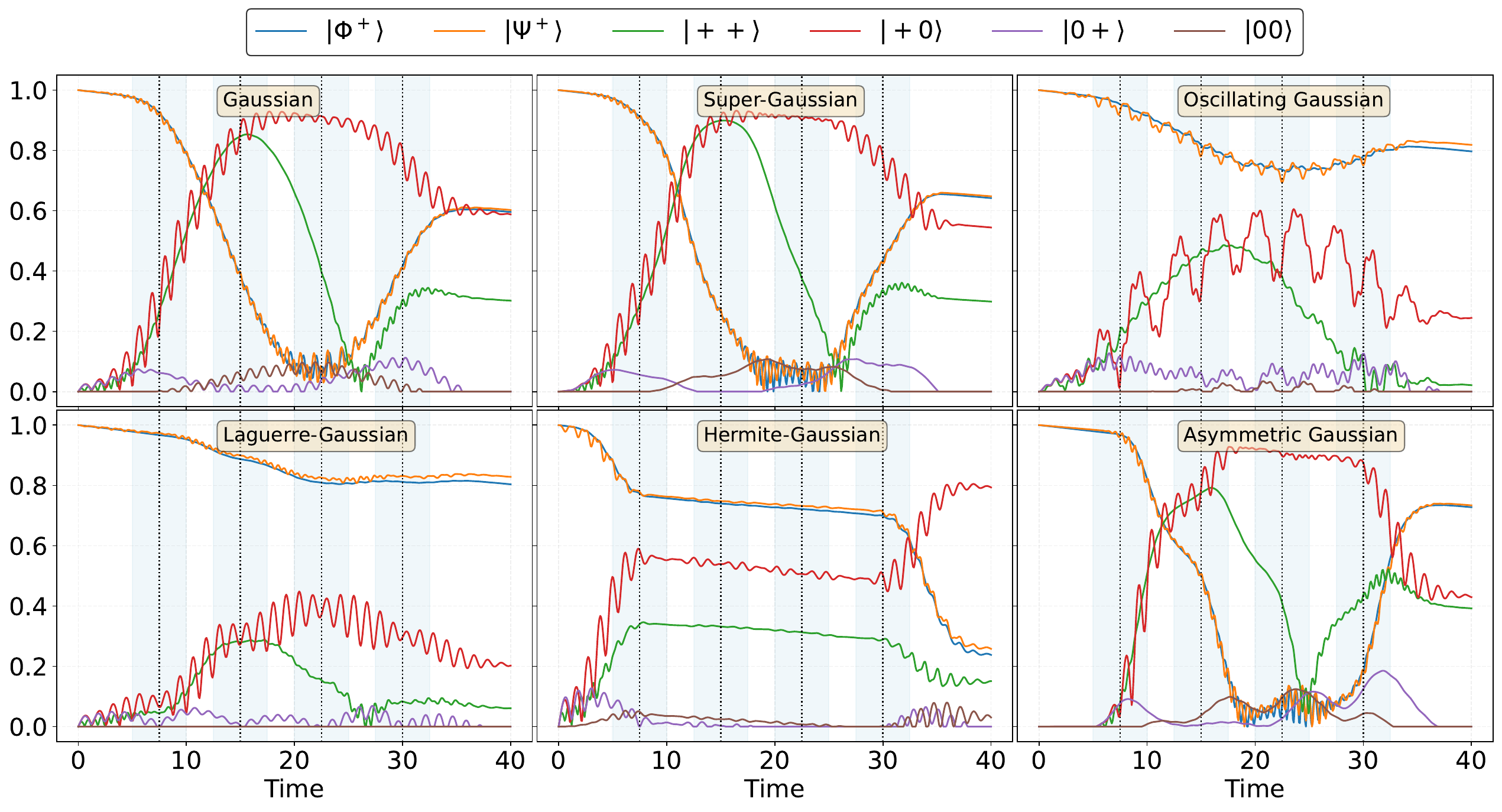}
\put(4, 40){(a)}
\put(36, 40){ (b)}
\put(68, 40){(c)}
\put(4, 17){(d)}
\put(36, 17){(e)}
 \put(68, 17){(f)}
\end{overpic}
\caption{Concurrence versus time under a Hamiltonian driven by four sequential pulses, with six initial states across six pulse types. Each subplot shows the concurrence with the pulse active regions ($t_c + k \cdot \Delta t \pm T/2$, where $k=0,1,2,3$) shaded. The parameters are: a coupling strength of $g_{0,1}=0.5$, a shared qubit Zeeman energy of $\omega_{q,A}=\omega_{q,B}=4.8$, $\lambda_{d,A}=1.0$, $\omega_{d,A}=5.0$, and $\phi_{d,A}=0.0$. The characteristic pulse duration is $T=5.0$ while the protocol uses $N_p=4$ pulses, with the first pulse centered at $t_c=7.5$ and subsequent pulses separated by $\Delta t=7.5$ (a factor of $s=1.5 \times T$). Decoherence rates are set to $\gamma_{1A, 1B}=0.001$ for amplitude damping and $\gamma_{\phi_A, \phi_B}=0.001$ for dephasing. The Lorentzian colored noise is employed with parameters: $M=1000$, $\Gamma=0.1$, $\tau_c=5.0$, $S_0=0.2$ and $\{ \Omega_{min},\Omega_{max}\}$=$\{-5,5\}$.}
\label{fig3}
\end{figure*}

Fig.~\ref{fig3} investigates the effect of a periodic multi-pulse control strategy, where the system is subjected to four sequential pulse applications. The primary goal is to understand how repeated interactions with the control field alter the entanglement dynamics. A clear and immediate consequence of this protocol is the significant enhancement of complex, chaotic behavior during the active pulse regions, especially when compared to the single-pulse case. Under the \OGP~(c) and \LGP~(d) pulses, the Bell states face less decay, and even better enhancement than the single pulse cases, both the single-pulse deterministic case in Fig. \ref{fig1} and the RL-optimized case Fig. \ref{fig2}.  Entanglement starts to improve after the projection of the third pulse envelope, which aligns with the pulse behavior observed in Fig.~\ref{fig1}. It is important to note that the multi-pulse case induces higher preservation and generation of entanglement in selective cases. A counter-intuitive phenomenon occurs under the \GP~(a), \SGP~(b), and \AGP~(f) pulses: for the initially entangled Bell \PSI~and \PHI~states, the repeated, intense driving first destroys the initial entanglement, driving concurrence to nearly zero. The subsequent pulses, however, regenerate a significant amount of partial entanglement, in some cases reaching a non-zero final value. Even more impressive is the effect on the separable \PZ~state, followed by the \PP~state. All the pulses act as a powerful entanglement engine, driving concurrence to near-maximal levels, as seen with the \GP~(a), \SGP~(b), \HGP~(e), and \AGP~(f) pulses. This demonstrates that the multi-pulse protocol with these aggressive envelopes acts less as a preservation tool and more as a robust state-generation and state-conversion protocol. The average entanglement of the initially separable states is significantly higher. The sequential nature of the protocol appears to have a beneficial, corrective effect, leading to a higher final concurrence for the \PSI~and \PHI~states compared to the single-pulse case, while still maintaining relative stability. A particularly noteworthy result is the behavior under the \HGP~(e) pulse. In the single-pulse case, this pulse induced highly chaotic and unstable dynamics. In stark contrast, the multi-pulse \HGP~protocol (e), despite initial volatility, ultimately guides the system to the most stable final entanglement configuration across both initially entangled and separable states. This complete reversal of character suggests that the sequence of \HGP~pulses (e) synergistically creates a robust, decoherence-resilient subspace. The multi-pulse protocol is a double-edged sword whose function is dictated by the pulse shape. It can amplify generation to near-perfect levels at the cost of transiently destroying initial entanglement, or it can be used to refine and improve preservation. After the final pulse, the dynamics universally transition to a stable decay without revivals. As before, the \ZP~and \ZZ~states show only insignificant entanglement generation.

\begin{figure}[ht]
    \centering
    \begin{overpic}[height=6cm,width=8cm]{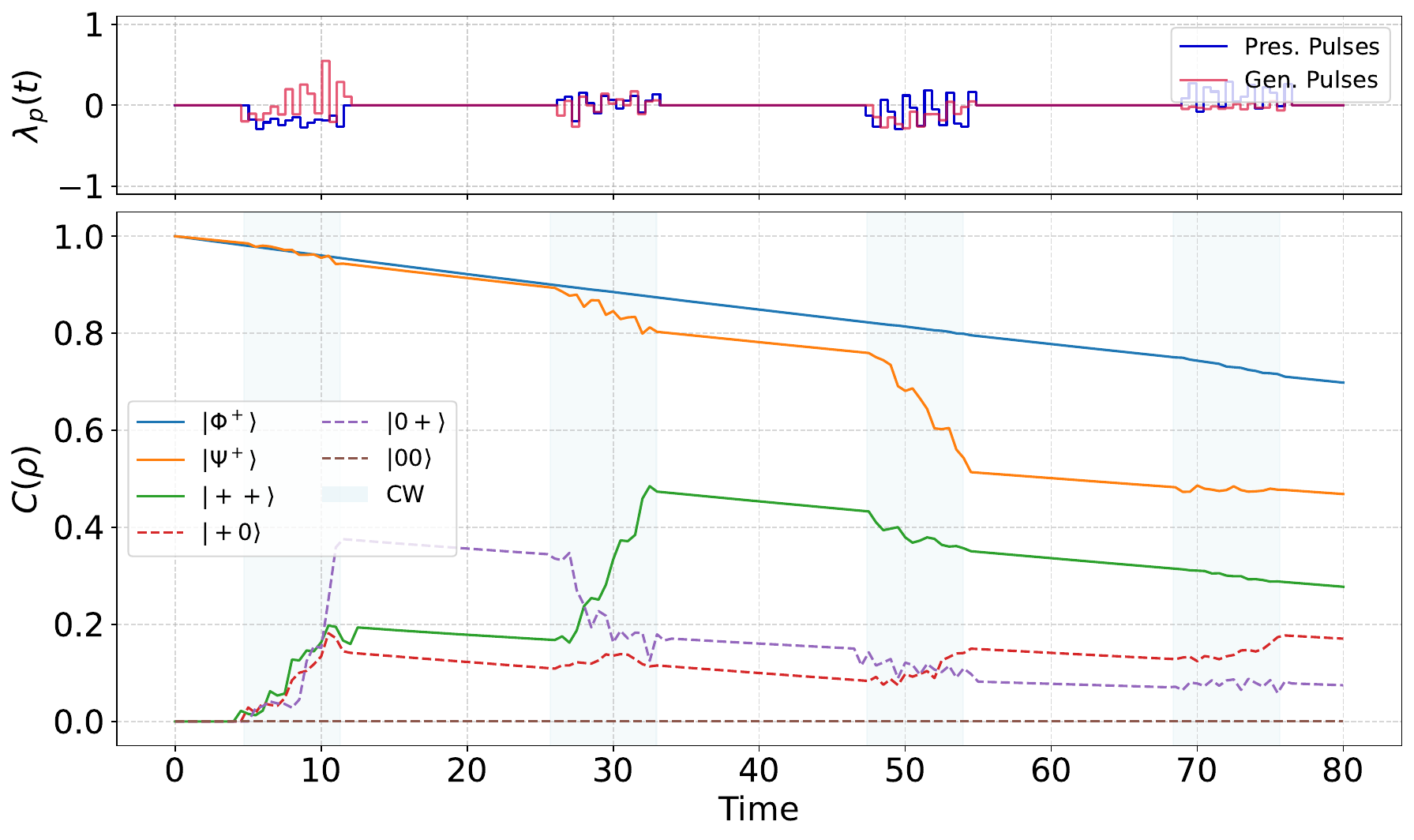}
    \put(10, 68){(a)}
    \put(10, 40){(b)}
    \end{overpic}
    \caption{Top: RL-optimized pulse amplitudes ($\lambda_p$) for simultaneous multi-pulse protocols for entanglement preservation and generation with configuration settings as shown in Fig. \ref{fig3}. The pulses are active in a series of control windows, as indicated by the shaded regions. Bottom: The corresponding evolution of concurrence under these optimized pulses. The agent is trained using the PPO algorithm with a MlpPolicy architecture. Key hyperparameters for the entanglement preservation task are: a total of $2 \times 10^5$ timesteps, a rollout buffer size ($n_{steps}$) of $2048$, and a fixed learning rate of $3 \times 10^{-4}$. For the generation task, the agent uses an $n_{steps}$ of $4096$, a linear learning rate schedule from $3.0 \times 10^{-4}$ to $1 \times 10^{-5}$, a batch size of $128$, and $10$ epochs per rollout.}
    \label{fig4}
\end{figure}

Fig.~\ref{fig4} extends the analysis of entanglement dynamics by applying the RL framework to a multi-pulse control scenario, analogous to that shown in Fig.~\ref{fig3}. Panel (a) displays the RL-optimized pulse sequences designed for both entanglement preservation and generation. These policies are markedly different from the single, broad pulse of Fig.~\ref{fig2}, exhibiting discrete, well-separated periods, reflecting the multi-pulse nature of the control. Fig.~\ref{fig4}(b) shows that for the preservation task, the RL agent enhances the stability of the initially entangled Bell states. The \PHI state maintains a high level of concurrence throughout the entire evolution, showing minimal decay. The \PSI~ state in contrast, while exhibiting some oscillations within the pulse regions, demonstrates significantly lesser preserved entanglement. This highlights the RL agent's ability to not only stabilize the dynamics but also effectively counteract cumulative decoherence effects over extended periods for the Bell states however, lesser than the deterministic protocol we designed. For entanglement generation, the RL agent adapts its strategy to the multi-pulse format where except the \ZZ~state, all the initially separable states are  slowly driven into slightly less entangled states within the first pulse window.  While maintaining a constant level entanglement, with the second pulse application region, the \PP~state begins to show strong entanglement generation while other undergoing decay.  The third and fourth active pulse windows fail to generate entanglement and show decay if the existed generated entanglement. This results in a stair-step generation/decay pattern for these states, where entanglement plateaus are maintained during the silent periods and then either generate (first two windows)  or decay (last two windows) during the active pulse regimes. However, this is also true that although with lesser entangelement, the adaptive, repeated-pulse strategy allows for the sustained presence of non-zero entanglement from initially separable states except the \ZZ~ state. In contrast to the multi-pulse case in Fig. \ref{fig3}, the \PZ~and \ZZ~states show higher entanglement generation, confirming their persistent resistance to entanglement under this Hamiltonian, even beating the adaptive multi-pulse control. Overall, Fig.~\ref{fig4} provides evidence that RL-optimized multi-pulse protocols offer moderate performance for both stable entanglement preservation and sustained entanglement generation, demonstrating the ability of adaptive control to leverage complex pulse sequences for robust quantum state engineering under environmental noise. However, one should also note that compared to the case of a single pulse optimized by RL ( Fig.~\ref{fig2}), the current case induces lesser entanglement.

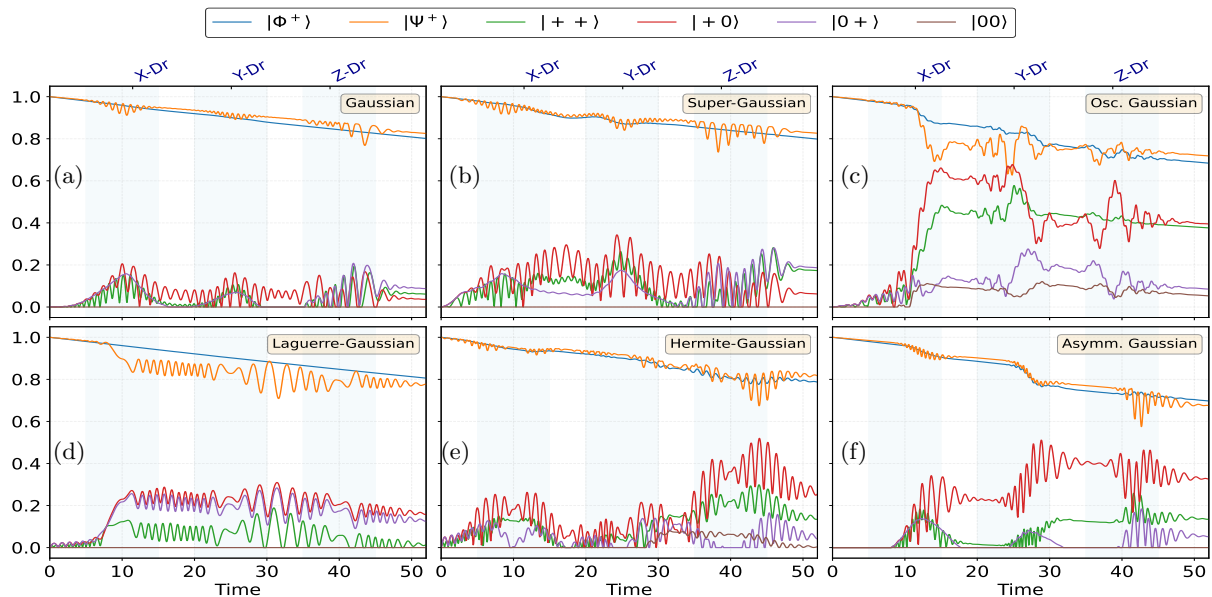
\begin{figure*}[ht]
\centering
 \begin{overpic}[height=8cm,width=16cm]{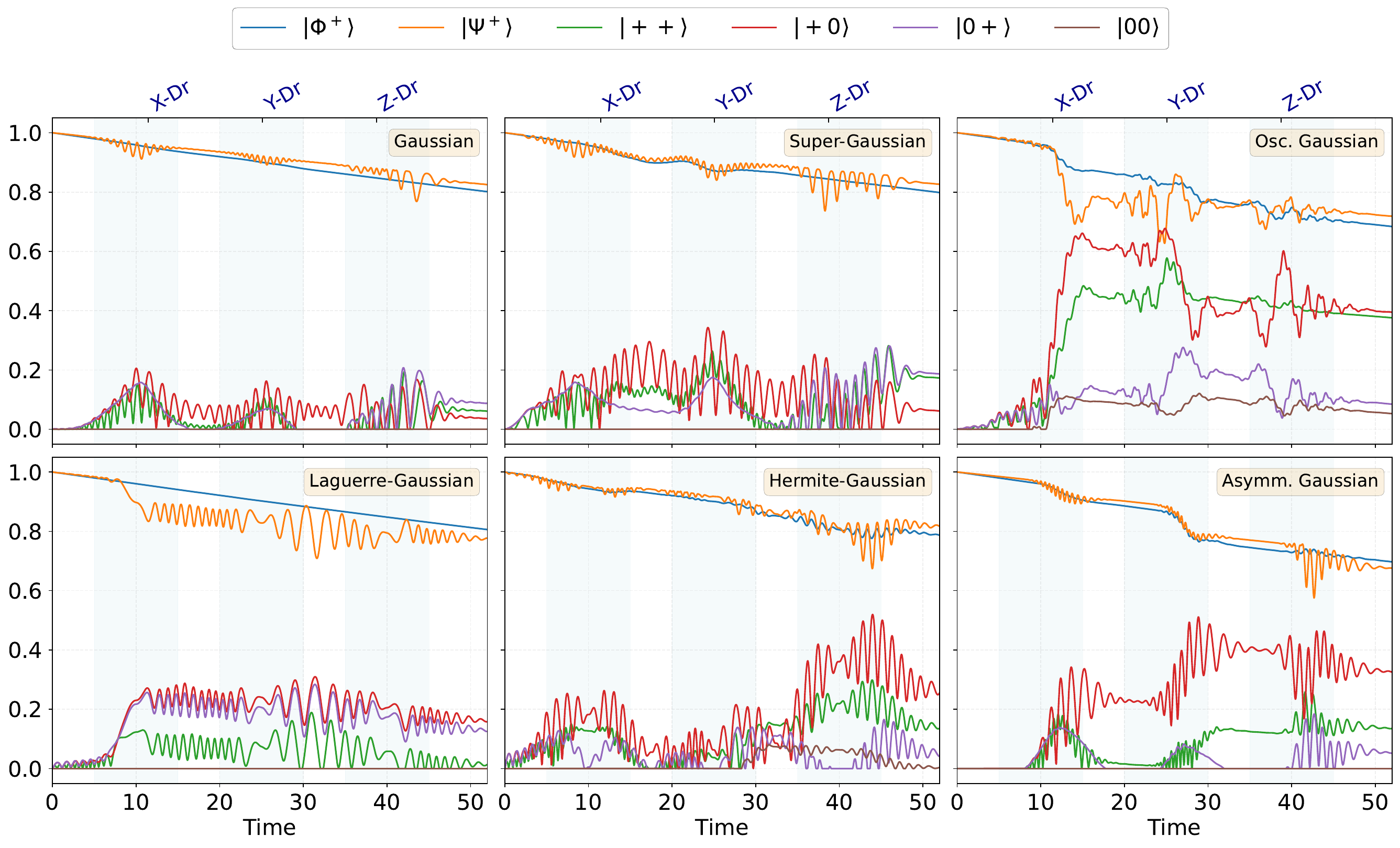}
 \put(4, 35){(a)}
 \put(36, 35){ (b)}
 \put(69, 35){(c)}
 \put(4, 12){(d)}
 \put(36, 12){(e)}
 \put(69, 12){(f)}
\end{overpic}
\caption{Concurrence versus time under pulse-driven Hamiltonians with six initial states across six pulse types, for different linear (X-, Y-. Z-) drive polarizations. Each subplot shows the concurrence, with the pulse active regions shaded. The parameters are: $g_{0,1}=0.5$, $\omega_{q,A}=\omega_{q,B}=4.8$, and $\omega_{d,A}=5.0$. The drive amplitudes for the X, Y, and Z-polarizations are $\Omega_{0,k}^{(xy)}=1.0$ and $\Omega_{z,0,k}=1.0$. The protocol uses $N_p=3$ pulses, each with a characteristic duration of $T=15$, with the first pulse centered at $t_c=10$ and subsequent pulse centers are separated by $\Delta t=5.0$. The decoherence rates are $\gamma_{1A, 1B}=0.001$ for amplitude damping and $\gamma_{\phi_A, \phi_B}=0.001$ for dephasing. The Lorentzian colored noise is employed with parameters: $M=1000$, $\Gamma=0.5$, $\tau_c=5.0$, $S_0=0.2$, $\Omega_c=0.5$, and $\{ \Omega_{min},\Omega_{max}\}$=$\{-5,5\}$.}
\label{fig5}
\end{figure*}
In Fig.~\ref{fig5}, we explore a sequential multi-pulse protocol where six distinct pulses are applied in a time-ordered sequence. Each pulse has a specific linear polarization: X-Dr, Y-Dr, and then Z-Dr with application in sequential  order. Their active durations are highlighted by the colored-shaded regions in each subplot. This protocol is designed to emulate concatenated quantum gates and leverage pulse shaping for superior entanglement control. By systematically applying different polarizations, we can deconstruct their individual and collective contributions to the evolution of both initially maximally entangled and separable states. This sequential protocol generally improves entanglement preservation compared to the previously considered schemes including both RL and non-RL protocols. The observed dynamical patterns are rich and highly dependent on the pulse shape. For the maximally entangled initial states, this linear polarization sequence is generally more effective at mitigating decoherence. Across all the pulse shapes (a-f), the \PSI~ and \PHI~ states consistently maintain a higher final concurrence. A notable observation is an inversion of robustness under the \OGP~(b) and \LGP~(d) pulses, where the \PSI~state retains a higher average entanglement than the \PHI~ state. This underscores the non-trivial interplay between pulse shape and state preservation. It is interesting to note that the \PSI~state experiences higher revivals compared to the \PHI~ state, suggesting that the pulse interacts more strongly with the subspace occupied by the former. Furthermore, the protocol's ability to generate entanglement from initially separable states is strongly contingent on the pulse type. For the pulse shapes \GP~(a), \SGP~(b), and \LGP~(d), the generated concurrence for the unentangled states is relatively low and often transient. However, the \OGP~(c) pulse generates a higher degree of entanglement, followed by the \HGP~(e) and \AGP~(f) pulses in relative performance. In comparison, the current linear sequence is best for entanglement preservation within a sequential gating framework. Conversely, we find this sequential framework to be less efficient for entanglement generation compared to the single-pulse (Fig.~\ref{fig1}) and multi-pulse (Fig.~\ref{fig2}) cases.

\begin{figure}[ht]
    \centering
    \begin{overpic}[height=6cm,width=8cm]{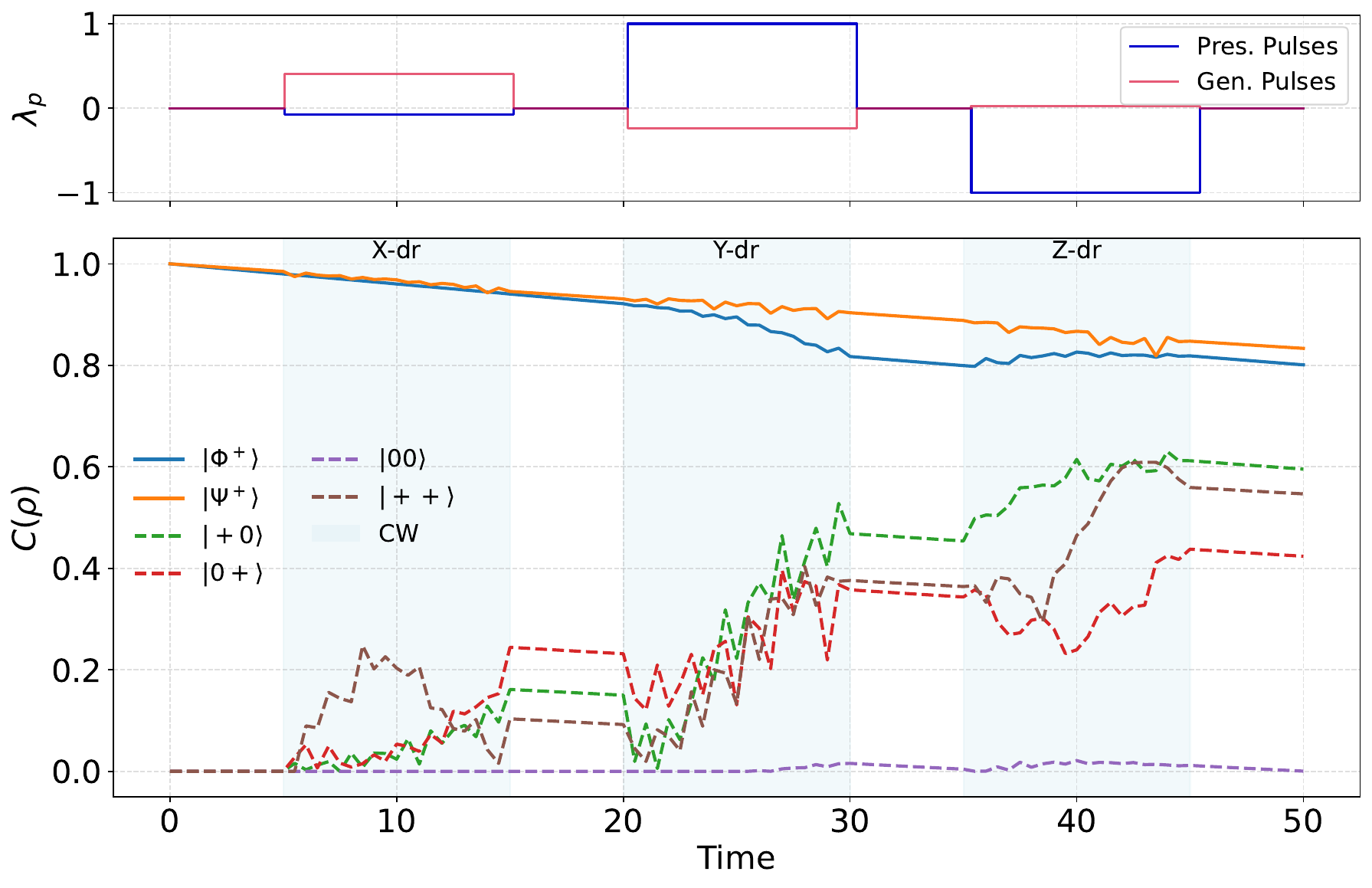}
        \put(10, 68){(a)}
        \put(10, 40){(b)}
    \end{overpic}
    \caption{RL-optimized pulse amplitudes ($\lambda_p$) for sequential linear multi-pulse protocols with configuration parameters likewise Fig. \ref{fig5}. The agent is trained using the PPO algorithm to achieve the dual goals of generating and preserving entanglement. The policy is a MlpPolicy with a shared neural network for both the actor and critic, consisting of two layers, each with 64 neurons. The RL environment is configured with a total simulation time of $50$ time units, divided into $100$ discrete steps. The agent's control consists of three distinct, sequential pulses: an X-Dr pulse, a Y-Dr pulse, and a Z-Dr pulse. Training is conducted for a total of $2 \times 10^{5}$ timesteps, with a rollout buffer size ($n_{steps}$) of $2048$ for both the entanglement preservation and generation tasks. The learning rate is fixed at $3 \times 10^{-4}$ for the preservation task, and at $1 \times 10^{-4}$ for the generation task.}
    \label{fig6}
\end{figure}

Fig.~\ref{fig6} demonstrates the performance of a sequential multi-pulse protocol where the pulse amplitudes are determined by the RL framework. The objective is to simultaneously preserve existing initial entanglement and generate entanglement from separable states using a sequence of linearly polarized X-, Y-, and Z-drives. Fig.~\ref{fig6}(a) displays the distinct, optimized pulse sequences discovered by the RL agent for these two tasks: a preservation and a generation pulse. As seen in Fig.~\ref{fig6}(b), both the entangled states exhibit high entanglement during the X- and Y- and Z-Dr regimes, with only minimal decay. This suggests the RL agent has successfully identified drive parameters that counteract the dominant decoherence mechanisms in these subspaces. However, the application of the Y-Dr introduces a significant disturbance, particularly for the \PHI~state, which experiences a decrease in concurrence, however, regaining it at the exposure of z-Dr. This may indicate that the polarized Y-Dr excites a resonance or pushes the \PHI~state into a region of the Hilbert space that is more susceptible to the specific noise channels of the system. In contrast, the \PSI~state remains more robust. For the task of entanglement generation, the dynamics is markedly different. The RL-optimized protocol slowly generates entanglement in all unentangled states except \ZZ~state. Its concurrence remains negligible during all the drive regimes. All the other separable states continue to gain entanglement, most prominently under Y-Dr. Finally, in the Z-Dr regime, the named separable states again generate entanglement and finally stabilizing at non-zero entanglement values. This implies that the Y- and Z-drives, in sequence, function as a highly effective entangling gate for this specific initial state.  In comparing the current results and that in Fig. \ref{fig4} for multi-pulse case to the single-pulse RL protocol shown previously in Fig.~\ref{fig2}, a key conclusion emerges. While this sequential protocol demonstrates the potential for task-specific optimization (i.e., step-wise generating high entanglement in separable states), the single optimized pulse appears to offer a more robust and efficient solution for achieving both high preservation and generation across a broader range of states. This suggests that the increased complexity of a sequential protocol does not necessarily guarantee superior performance over a single, well-optimized pulse.

\begin{figure*}[ht]
 \centering
\begin{overpic}[height=10cm,width=18cm]{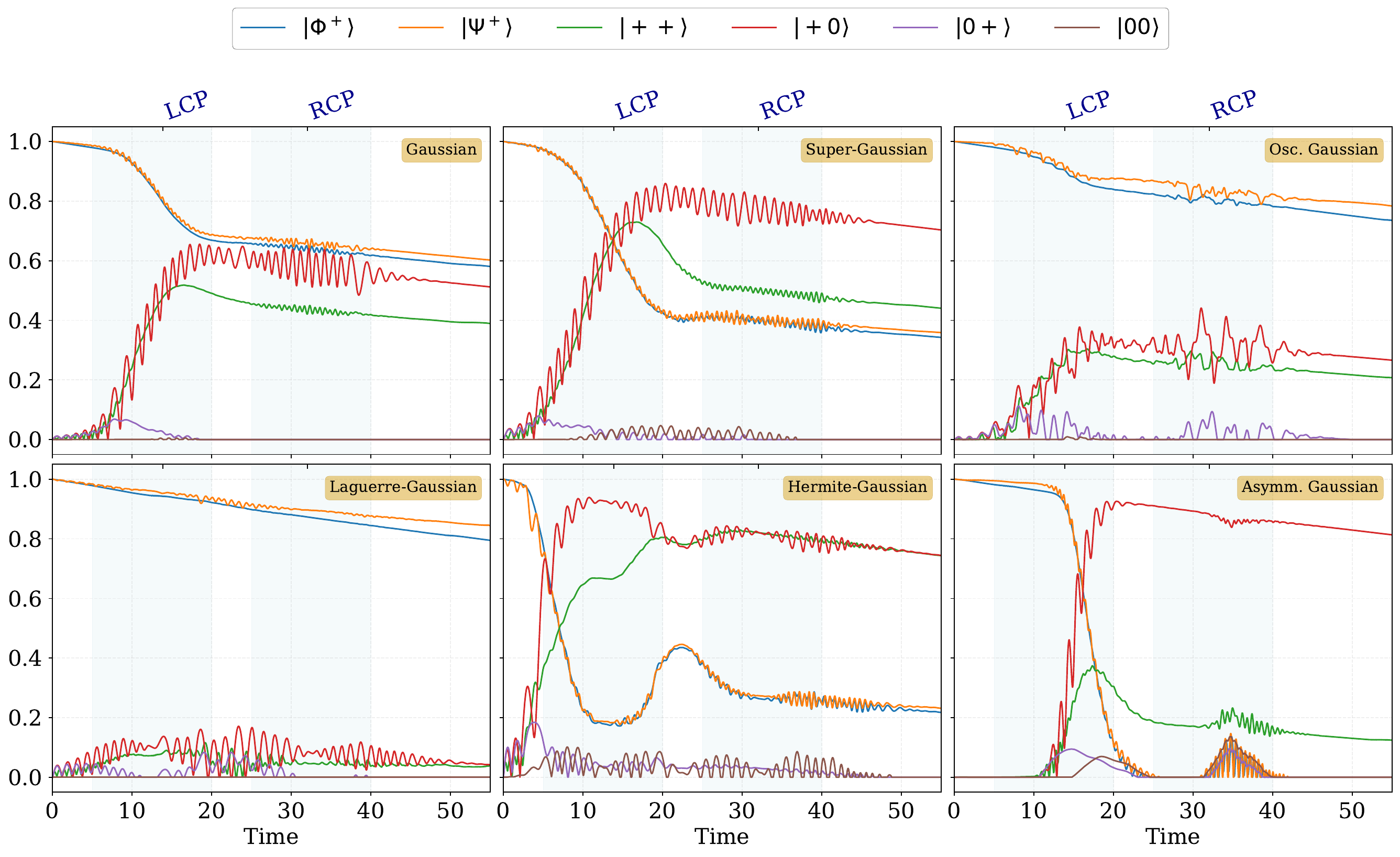}
\put(4, 35){(a)}
\put(36, 35){ (b)}
\put(69, 35){(c)}
\put(4, 12){(d)}
\put(36, 12){(e)}
\put(69, 12){(f)}
 \end{overpic}
\caption{Concurrence versus time for sequential multi-pulse protocols, specifically illustrating two circularly polarized pulses: Left-circularly Polarized (LCP-Dr) and Right-circularly Polarized (RCP). The parameters are the same as in Fig. \ref{fig5}.}
\label{fig7}
\end{figure*}

Fig.~\ref{fig7} extends our analysis to a sequential protocol that uses two circularly polarized pulses with opposite helicity: an LCP drive followed by an RCP drive. This framework allows us to explore how the chirality of the driving field, when switched mid-protocol, influences entanglement preservation and generation. This provides a crucial point of comparison to the linear drive sequence discussed in Fig.~\ref{fig5}. For entanglement preservation, the linear protocol generally offers more robust protection for the \PSI~state than the \PHI~state, except the \HGP~(e), and \AGP~(f) pulse, where the entanglement equally stabilize at similar levels. Consistent with Figs.~\ref{fig1}, \ref{fig3} and \ref{fig5}, the \OGP~(c) and \LGP~(d) followed by \GP~(a), and \SGP~(b) pulses are highly effective at preserving entanglement in the Bell states. However, the circular protocol introduces highly pulse-dependent dynamics. The most obvious example is the \AGP~(f) pulse, which induces a near-total collapse of concurrence in the \PSI~ and \PHI~ states. This strong selectivity suggests that the time-ordered LCP-RCP sequence creates an effective Hamiltonian that is nearly commutative with the subspace of the separable states. For entanglement generation, the sequential circular protocol demonstrates dominance over all the linear polarization schemes discussed in Fig.~\ref{fig5}. It is particularly effective at generating a high level of entanglement in the initially separable \PZ~ and \PP~ states under the \GP~(a), \SGP~(b), \HGP~(e), and \AGP~(f) pulses. In comparison, the circularly polarized protocol remains highly effective under \HGP~(e), and \AGP~(f) pulses. This result was not as prominently seen in the linearly polarized sequential along with single and multi-pulse protocols. In particular, the generated entanglement in the \PZ~ and \PP~ states under pulses (e, f) is nearly comparable to the maximal entanglement. This suggests that the sequential application of drives with opposite helicity provides an efficient pathway for generating entanglement, likely by using the LCP pulse to create a superposition and the RCP pulse to rotate it into an entangled configuration. The dynamics under this protocol are visibly more complex, featuring rapid, large-amplitude jumps under the LCP pulse, followed by a gain in stability under the RCP pulse. This behavior likely arises from the rotating nature of the drive exciting multiple, interfering transition pathways (Floquet sidebands). While complete entanglement sudden death for the Bell states is not prevalent except with \AGP~(f), the system exhibits instances of entanglement suppression and revival under LCP and then maintaining the entanglement under RCP regime, underscoring the rich and highly controllable dynamics accessible through structured, circularly polarized pulse sequences.

\begin{figure}[ht]
\centering
\begin{overpic}[height=6cm,width=8cm]{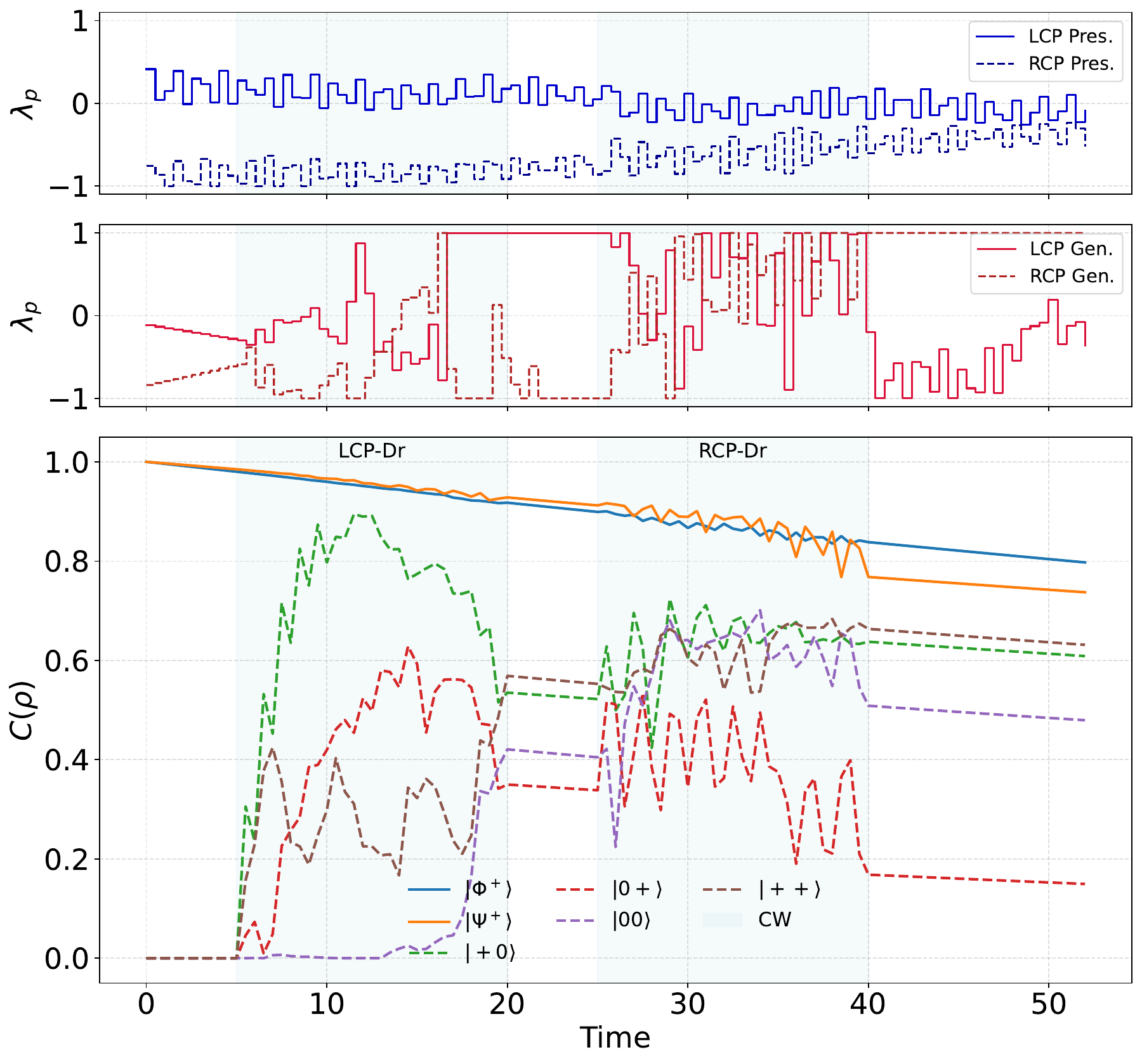}
\put(10, 70){(a)}
\put(9, 56){ (b)}
\put(9, 35){ (b)}
\end{overpic}
\caption{Evolution of concurrence under RL-optimized, sequential circularly polarized (LCP followed by RCP) multi-pulse protocols. (a) Top: Learned pulse amplitudes for entanglement preservation (LCP-Pres., RCP-Pres.). (b) Same as (a) but for entanglement generation. (c) Bottom: Corresponding concurrence dynamics for various initial states. The RL agent, utilizing a PPO algorithm with an `MlpPolicy`, was trained for $2\times 10^5$ timesteps for preservation and $2.5\times 10^5$ for generation, with a rollout buffer size of $2048$.  The RL framework used a fixed learning rate of $1\times 10^{-4}$ for the generation task, while the preservation task used $3\times 10^{-4}$. The tasks were specialized with distinct reward functions for preservation and generation. The RL agent was optimized to counteract decoherence and generate entanglement efficiently for specific states, respectively.}
\label{fig8}
\end{figure}

Fig.~\ref{fig8} presents the results of a sequential LCP-RCP protocol where the pulse amplitudes are optimized by RL-framework. As shown in panel (a, b), the RL agent discovers distinct, piecewise-constant control sequences for the separate tasks of entanglement preservation and generation.  The RL agent designed a pulse with a relatively low amplitude and corrective, oscillating nature. It avoids large, sudden changes and remains within a narrow range of amplitudes. This gentle, wiggling action is designed to counteract the system's decoherence and noise without introducing additional destructive oscillations. The two components of the pulse, LCP and RCP, are distinct but similarly low-amplitude. In contrast, the generation pulse is highly aggressive, characterized by large, abrupt changes between positive and negative amplitudes. This bang-bang type of control is designed to rapidly pump the system's energy to facilitate the generation of entanglement. The performance of the preservation protocol, shown in panel (c), is highly state-dependent. The \PHI~state is well-preserved, with its concurrence remaining high throughout both the LCP and RCP drive phases. In contrast, the \PSI~state experiences decoherence during the RCP drive, followed by a revival decay of entanglement during the subsequent RCP drive. This  revival dynamic suggests that the RL agent has learned a sequence where the second pulse fails against the detrimental evolution induced by the environment under RCP-pulse exposure. For entanglement generation, the RL-optimized protocol is largely successful. Unlike the previous cases, here all the separable states undergo  entanglement generation during the LCP phase while \PZ~state remaining more favorable under LCP pulse. Interestingly under the RCP pulse, the curves for the separable states undergo chaotic revival behavior, while \PP~states achieving the highest concurrence followed by the \PZ~ and \ZZ~states. Finally, the generated entanglement stabilize at a meaningful high level after the RCP phase. This indicates that the learned generation policy, likely optimized for a short-term reward, highly succeeds to produce persistent entanglement. Drawing a comparison with the single-pulse RL protocol (Fig.~\ref{fig2}), it is clear that the simpler approach is superior. The single-pulse protocol achieves more robust preservation and far more effective generation of stable entanglement. However, the current RL-protocol for LCP, and RCP pulses has the advantage to induce entanglement in all separable states where the previous RL-single, and RL-multi-pulse protocols completely failed to do. This result provides a key insight: for this system, increasing the complexity of the control protocol from a single pulse to a sequential one does not yield better performance and can even be counterproductive. The RL-framework itself guides us to the conclusion that a simpler control strategy is more effective.

\subsection{Advanced and Adaptive control protocols with maximally entangled state}

In this section, we provide comparative results for the protocols described in Sec. \ref{sec:rl_dd}-\ref{sec:quantum_walk_strategy} with RL-frame work. The protocols in the following are complex to handle along with their comparable RL strategies, therefore, we choose a single state inclusion in the following. On should note that, the current physical model does not only contains dephasing and Lorentzian noise effects, but we also consider initial state errors, pulse errors and detuning errors with certain strengths to impose strict decaying and noisy realistic environments. The motive behind this is to find out entanglement preservation under realistic conditions, as the experimental frames may include such types of errors.

\begin{figure*}[ht]
\centering
\begin{overpic}[height=10cm,width=18cm]{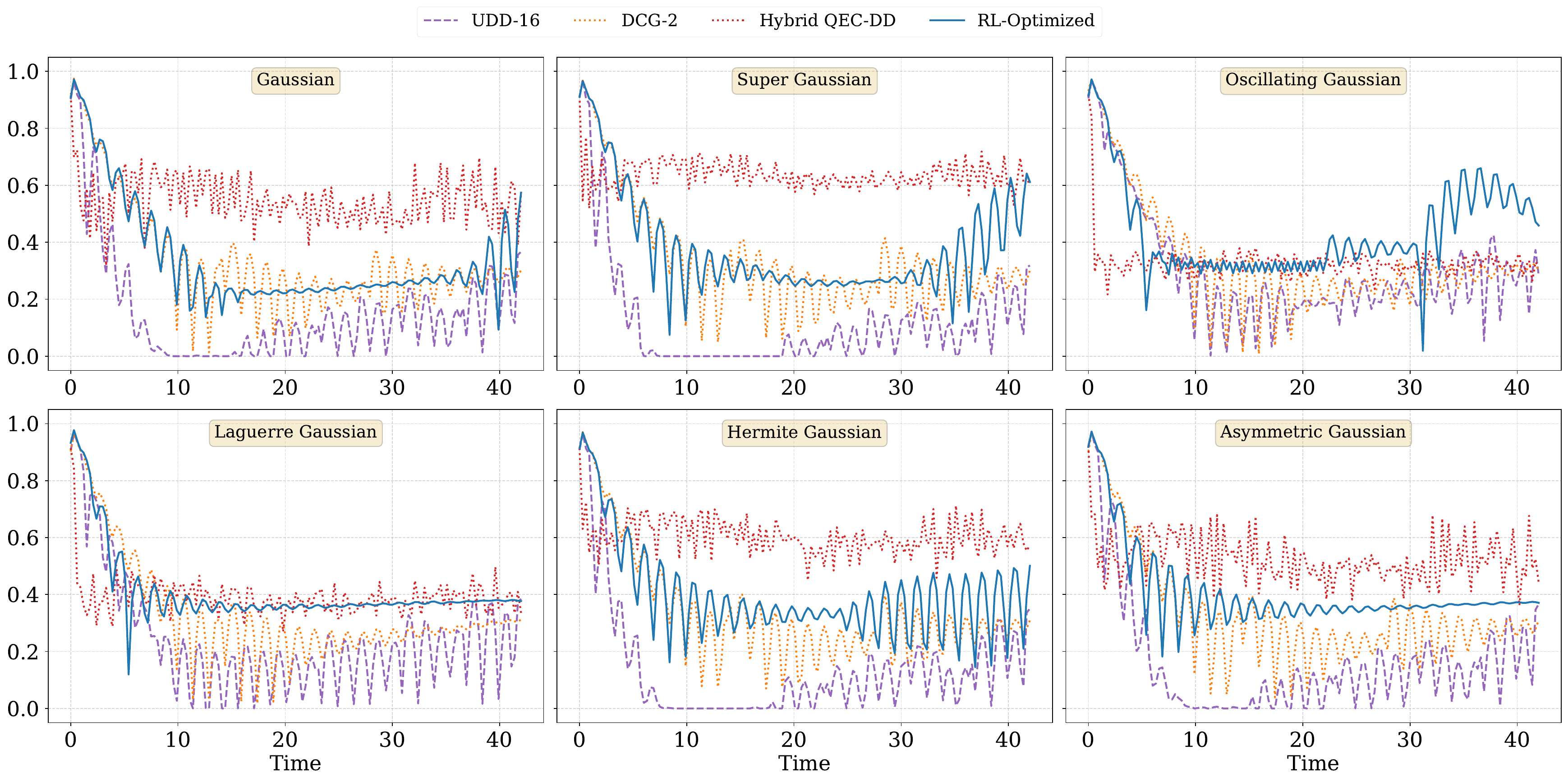}
\put(5, 28){(a)}
\put(37, 28){ (b)}
\put(69, 28){(c)}
\put(5, 07){(d)}
\put(39, 07){(e)}
\put(71, 07){(f)}
\end{overpic}
\caption{Time evolution of concurrence for various protocols showing the performance of UDD-16, DCG-2, Hybrid QEC-DD, and an RL-Optimized protocol using six different pulse envelopes. The system parameters are set as follows: interaction strength $g=0.5$, $\omega_q=2.0$, $\gamma_{1k}=0.05$, $\gamma_{\phi k}=0.05$, $M=1000$ $S_0=0.2$, $\Gamma=0.5$,  $\{\Omega_{\min},\Omega_{\max}\}=\{-5.0, 5.0\}$, $T=0.15$, free evolution multiplier of  $\mathcal{F}_m=2.0$, $t=42.0$ over $N_t=140$ steps. The RL-Optimized protocol uses the PPO algorithm to find an optimal control sequence. The RL agent's policy is a two-layer neural network with 128 neurons per layer. Training was conducted for $2 \times 10^5$ timesteps  with a rollout buffer size ($n_{steps}$) of $2048$, a batch size of $64$, and a discount factor ($\gamma$) of $0.99$. The reward function includes a final reward factor of $10.0$, a sustain bonus of $0.01$, and a minor penalty for each action taken ($0.001$). For the Hybrid QEC-DD protocol, the correction probability is $p_{\text{corr}}=0.8$, with a maximum correction strength of $s_{\text{max}}=0.3$ and a dead time of $0.01$.}
\label{fig9}
\end{figure*}

Fig.~\ref{fig9} presents a  comparison of quantum control protocols under a physically realistic and challenging simulation environment. The underlying model, governed by a Lindblad master equation, incorporates not only standard Markovian decoherence channels (amplitude damping and dephasing) but also a crucial non-Markovian component: structured colored noise, simulated as a superposition of a thousand Lorentzian oscillators. This structured noise, coupled with realistic imperfections such as pulse amplitude errors, detuning errors, and initial state preparation infidelity, creates a complex control landscape that serves as a stringent test for each protocol. The protocols employed demonstrated varying degrees of proficiency. The standard UDD-16 protocol was the least effective, though it still prevented the complete decay of entanglement. Its performance is heavily dependent on the type of pulse used. For instance, with the \OGP~(c), the UDD-16 protocol's performance is comparable to more advanced hybrid QEC-DD protocols. Similarly, a noticeable amount of entanglement was preserved with the \LGP~(d). The DCG-2 protocol consistently outperformed the UDD-16, as it did not exhibit the sudden "entanglement death and rebirth" phenomena seen with UDD-16 across all pulse types. The relatively weaker performance of both UDD-16 and DCG-2 can be attributed to their open-loop nature. These protocols apply a pre-calculated sequence of pulses with timings and phases optimized for simpler, often idealized noise models. They are fundamentally mismatched to the specific spectral density of the simulated colored noise and lack the adaptability to correct for stochastic pulse errors or state drift. Consequently, they are unable to prevent the rapid decay of concurrence. On the other hand, the Hybrid QEC-DD protocol demonstrates a significant leap in performance due to its combined strategy. It operates in discrete cycles, applying a simple Carr-Purcell (CP) DD sequence for passive error suppression within each cycle. Critically, at the end of each cycle, it performs an active error correction step. This step probabilistically nudges the system's density matrix back towards the ideal Bell state, with a strength proportional to the fidelity loss. This periodic, feedback-like correction mechanism effectively resets a portion of the accumulated error, allowing the protocol to maintain a stable, high level of concurrence. This explains the characteristic flat-line like behavior (on average) of the red dotted curve, representing a robust, showing an optimal error management strategy. In contrast, the RL-Optimized protocol, while learning a highly active and non-trivial control policy, ultimately falls short in comparison to the QEC-DD protocol. The large oscillations of the blue curve indicate that the agent has successfully learned to dynamically manipulate the system to sustain a high average level of entanglement throughout the evolution. However, it fails to translate this dynamic control into a maximized final state concurrence overall. This suggests a potential shortcoming in the agent's learned strategy. Despite a reward function that heavily incentivizes the final outcome, the agent may have converged to a local optimum: a juggling policy that is effective at continuous error mitigation but lacks the specific sequence required to perfectly phase the system's oscillations for a peak concurrence at the precise final moment. The agent's behavior is shaped entirely by its reward function, which is dominated by a large bonus for the final concurrence. This design incentivizes the discovery of a non-trivial, long-term strategy. The pronounced oscillations in the blue curve are therefore not random fluctuations but the signature of this learned policy in action: a dynamic sequence of carefully timed pulses designed to proactively steer the quantum state along an optimal trajectory through Hilbert space, actively counteracting the deleterious effects of the environment to maximize the final entanglement. While the Hybrid protocol periodically corrects errors, the RL agent learns to prevent and pre-empt them, leading to the highest overall performance across all tested noise profiles. In the case of pulse-oriented results, the Hybrid QEC-DD protocol, as shown by the dotted red line, maintains the highest level of entanglement under \GP~(a), \SGP~(b), \HGP~(c), and \AGP~(f) compared to any of the other protocols. The \SGP~(b) and \LGP~(d) pulses provide a less chaotic and smoother evolution for the QEC-DD protocol, which is indicative of a more stable interaction between the pulse envelope and the QEC-DD's feedback mechanism. Conversely, the RL-Optimized protocol, performs exceptionally well, often surpassing the other protocols in specific scenarios. It demonstrates the highest final concurrence under the \OGP~(c) pulse, even outperforming the QEC-DD protocol. This is followed by strong performance with the \SGP~(b), \GP~(a), and \HGP~(e) pulses. The most stable and non-oscillatory entanglement dynamics are learned by the RL-agent under the \LGP~(d) and \AGP~(f) pulses, where the protocol effectively suppresses fluctuations and guides the system along a smooth, high-concurrence trajectory. The DCG-2 protocol, exhibits more chaotic dynamics with a similar average baseline of entanglement across most pulses. However, the \GP~(a) pulse appears to favor this protocol more than the others. Finally, the UDD-16 protocol, consistently performs the worst across all pulse conditions, failing to preserve entanglement effectively. Its highly oscillatory and rapidly decaying concurrence signifies its ineffectiveness in counteracting the specific decoherence channels present in this system. Our analysis demonstrates that while RL can discover sophisticated, active control sequences, it does not inherently guarantee superiority. For the problem and parameters studied here, the Hybrid QEC-DD protocol's blend of robust, periodic correction and passive decoupling proves to be the more effective and reliable strategy for maximizing final entanglement. This highlights that well-designed, physics-informed hybrid strategies can outperform more complex, black-box optimization methods, especially if the latter face challenges in navigating a complex control landscape to find a true global optimum.

\begin{figure*}[ht]
\centering
\begin{overpic}[height=10cm,width=18cm]{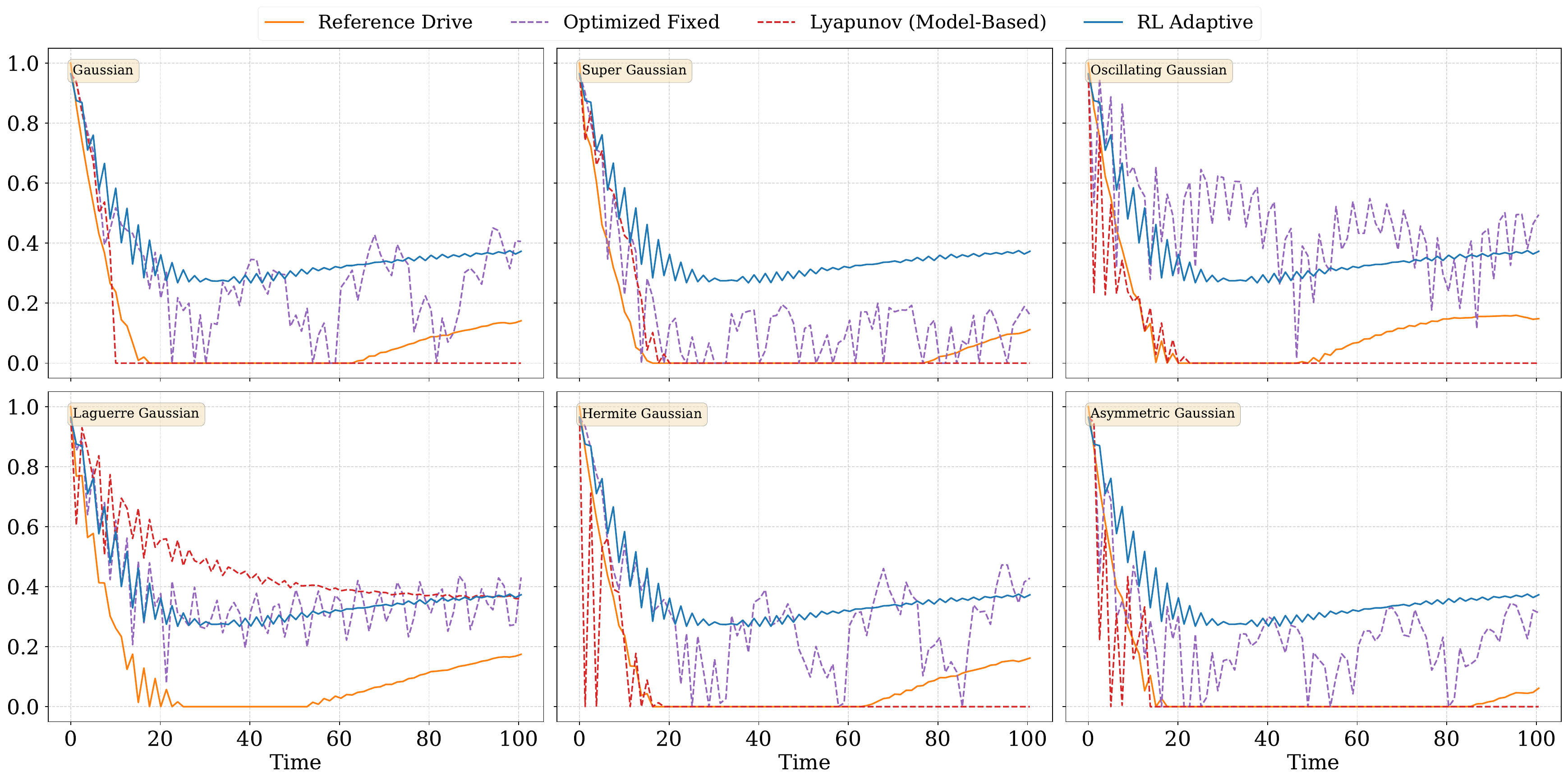}
\put(28, 40){(a)}
\put(61, 40){ (b)}
\put(94, 40){(c)}
\put(28, 19){(d)}
\put(61, 19){(e)}
\put(94, 19){(f)}
\end{overpic}
\caption{Time evolution of concurrence under RL Adaptive, Lyapunov Control (Model-Based), Optimized Fixed, and Reference Drive protocols across six pulse types. Each subplot shows concurrence versus time over $N_{\text{steps}}=100$ steps, with a Floquet period of $T_\Omega \approx 1.2566$ and a drive frequency of $\Omega=5.0$. The total simulation time is $\tau=100.0$. The system parameters are: $g=1.2$, $\omega_a=3.5$, $\gamma_{1k}=0.03$, $\gamma_{\phi k}=0.03$ with $M=1000$, $S_0=0.05$, $\Gamma=0.5$, $\{ \Omega_{min}, \Omega_{max}\} \in [0.001, 10.0]$. Additional parameters include a pulse amplitude error $\epsilon_{A}=0.01$ and a detuning error $\Delta_{\text{err}}/\omega_{a,0}=0.01$. The RL Adaptive protocol uses PPO with a three-layer neural network (256 neurons per layer) trained for $2 \times 10^5$ timesteps. The RL agent optimizes amplitude, phase, and frequency multipliers (range $[0.5, 1.5]$) per harmonic. The reward function includes a target concurrence threshold of $0.95$, a concurrence power of $3.0$, a low-concurrence penalty, and a frequency change penalty of $0.05$. The RL and Optimized Fixed protocols use three harmonics with weights $[1.0, 0.5, 0.3]$.}
\label{fig10}
\end{figure*}
Fig.~\ref{fig10} provides a comparison of four control paradigms to stabilize entanglement in a continuously driven configuration. The simulation environment is designed to be particularly challenging, incorporating not only standard Markovian decoherence but also a high-dimensional (1000-mode) colored noise bath with a Lorentzian spectral density. This creates a complex, non-Markovian control problem where the environment possesses memory, a scenario that critically tests the assumptions underlying each control strategy. The performance of the baseline Reference Drive protocol is, as expected, poor. It consistently leads to a rapid destruction of entanglement within the first 20 time units, though a minor, late-time resurgence is observable across most pulse profiles, suggesting a coincidental resonance at the end of the evolution. A more nuanced and revealing case is the Lyapunov (Model-Based) protocol. This controller, which relies on a simplified analytical model of the system, fails unequivocally in most scenarios. Its control actions, optimal for its internal model, prove counterproductive in the complex reality of the simulation. However, a striking exception occurs under the \LGP~(d) pulse, where it performs exceptionally well, even surpassing the RL agent. This suggests that for this specific pulse characteristic, the simplified model happens to be a fortuitously accurate approximation of the true dynamics. This isolated success highlights the protocol's extreme sensitivity to the noise and pulse type, underscoring its non-robust and generally unreliable nature. The Optimized Fixed protocol emerges as a strong competitor. As an open-loop strategy derived from an offline optimization, it successfully sustains a significant level of entanglement, preventing the complete decoherence seen in the baseline cases. Its performance is nonetheless erratic, characterized by large oscillations or deaths and rebirths of entanglement. Notably, this protocol proves highly effective under specific conditions, outperforming the RL agent for the \OGP~(c) and achieving a comparable average concurrence to the RL under \LGP~(d). In other scenarios, its performance varies, sometimes exceeding and sometimes falling short of the RL protocol. This demonstrates that while a fixed, optimized drive can be a powerful tool, its efficacy is inconsistent and it lacks the adaptability for robust performance across diverse conditions. Ultimately, the RL Adaptive protocol demonstrates the best overall performance and stability. Its superiority stems from a fundamentally different, model-free, and closed-loop approach. By processing real-time information about the system's state and the surrounding noise, the agent learns a dynamic, state-dependent policy, continuously adjusting multiple drive parameters to actively steer the system. The resulting trajectory, an initial decay followed by a confident recovery to a stable, high-concurrence plateau, is a direct result of this adaptive learning process, guided by a reward function engineered for both high fidelity and stability. While it can be narrowly surpassed by less adaptive protocols under specific, fortuitous conditions, its strength lies in its robustness. The RL agent consistently delivers high performance across the entire suite of diverse and challenging noise environments and pulse types, marking it as the most generally applicable and reliable strategy among those tested.

\begin{figure*}[ht]
\centering
\begin{overpic}[height=10cm,width=18cm]{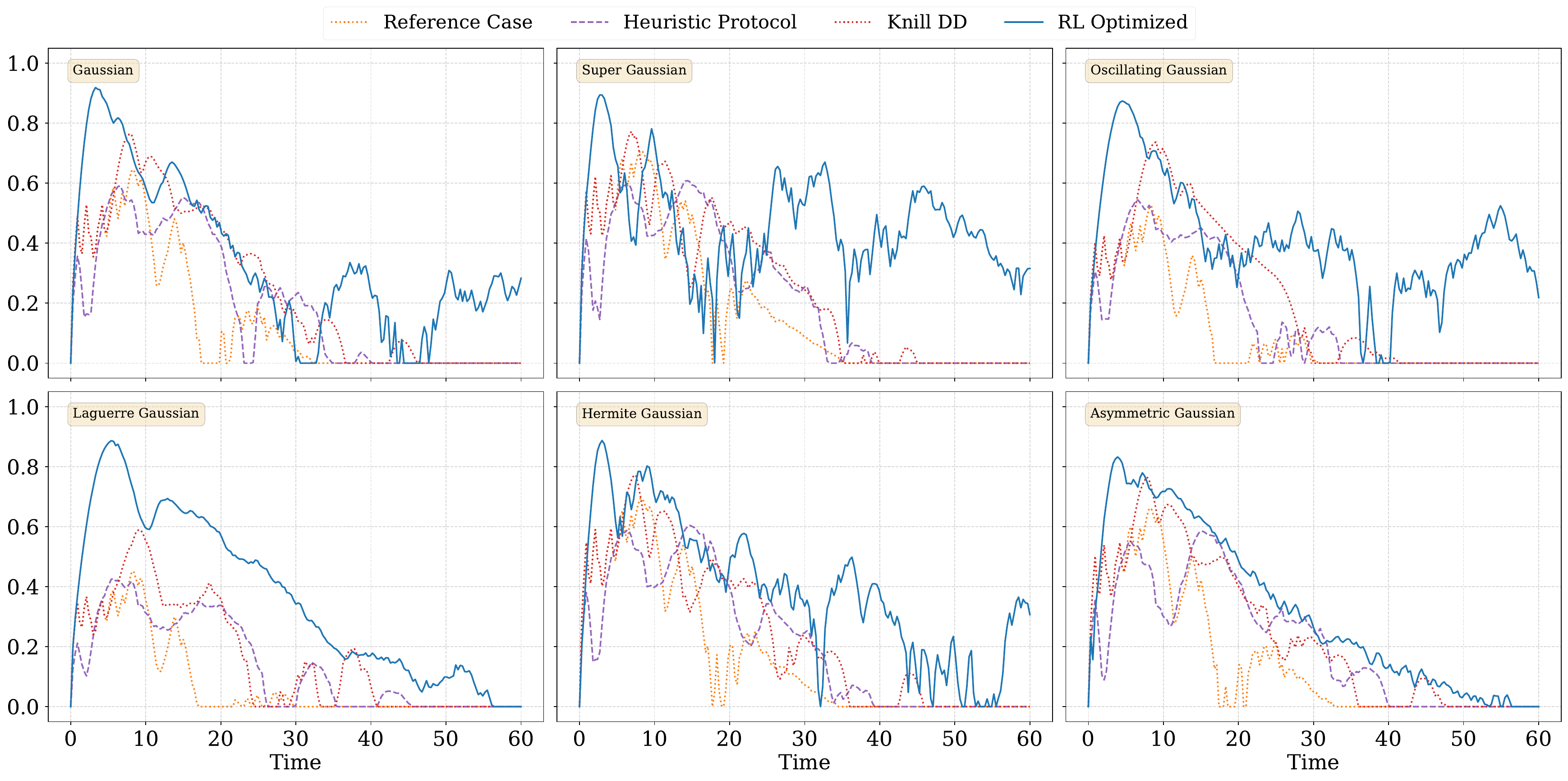}
\put(28, 40){(a)}
\put(61, 40){(b)}
\put(94, 40){(c)}
\put(28, 19){(d)}
\put(61, 19){(e)}
\put(94, 19){(f)}
\end{overpic}
\caption{Comparative analysis of entanglement generation protocols under various pulse envelopes via concurrence dynamics. The simulations start from the separable state $\ket{00}$ under four control protocols: No Control (passive evolution), Heuristic protocol, Knill dynamical decoupling (DD) sequence, and an RL-optimized protocol trained using proximal policy optimization (PPO). Key simulation parameters are: coupling strength $g=0.5$, qubit frequency $\omega_a=2.0$, relaxation rate $\gamma_{1k}=0.01$, dephasing rate $\gamma_{\phi k}=0.01$, pulse duration $T_p=0.05$, total simulation time $\tau=60.0$, amplitude error $\epsilon_\text{amp}=0.01$, and relative detuning error $\delta_\text{det}/\omega_a = 0.01$ with $M=1000$ trajectories, drive amplitude range ${\Omega_{\text{min}}, \Omega_{\text{max}}} = [-10.0, 10.0]$, and initial state preparation error $S_0 = 0.01$. The RL protocol optimizes the pulse angle and phase at each step using the PPO algorithm with the following hyperparameters: learning rate $3\times 10^{-4}$, discount factor $\gamma=0.995$, 1024 steps per environment, batch size of 64, and 10 optimization epochs. The reward function includes a weight of 50.0 for concurrence increase and a penalty of 0.005 for large pulse angles to discourage aggressive control. The RL protocol was trained for $2 \times 10^5$ timesteps.}
\label{fig11}
\end{figure*}

Fig.~\ref{fig11} provides a detailed analysis of a quantum control task: the active generation of entanglement from a separable state ($|00\rangle$) within a complex, non-Markovian noise environment under six pulse profiles. The results reveal a clear hierarchy of performance that is directly attributable to the sophistication and adaptability of the control strategy employed. The core of the dynamics is a competition between the system's inherent entangling interaction $S_Y \otimes S_Y$ type coupling and the decohering effects of dissipation and structured noise. The performance of the open-loop protocols provides critical insight into the limitations of pre-programmed control. A key physical feature of our model is the modulation of the intrinsic qubit-qubit interaction by the control pulse envelope. This induces a time-dependent entangling Hamiltonian, $H_{\text{int}}(t) \propto f(t) \cdot (S_Y \otimes S_Y)$, where $f(t)$ is the specific pulse shape. This dynamic interaction landscape is the critical context in which entanglement must be induced and managed against experimental noise and errors. Since the pulse impacts in Figs. \ref{fig1}, \ref{fig3}, \ref{fig5}, and \ref{fig7} are quite resourceful for entanglement generation. Therefore, for the reason is that the pulses employed are highly favorable for entanglement generation. Therefore, we used pulse-based entanglement generation case called Reference Case, without using a specific protocol where we found it to be a very resourceful method.  The Reference Case, representing passive evolution, shows that the system's natural Hamiltonian is itself entangling, but this process is slow and is quickly overwhelmed by decoherence. The Reference Case successfully generates entanglement, though its performance is highly dependent on the pulse shape. For example, generation is most effective under \GP~(a), \SGP~(b), \HGP~(e), and \AGP~(f); however, the resulting entanglement is quickly destroyed by the strong noise profile. The Knill DD protocol attempts to improve this by applying a standard DD sequence. Its goal is not to generate entanglement itself, but rather to protect the entanglement naturally created by the pulse-modulated $S_Y \otimes S_Y$ interaction. It offers only a marginal improvement, underscoring that the protection scheme is fundamentally mismatched to the complex spectral density of the colored noise. Its performance is relatively consistent across most pulse shapes, with the notable exception of the \LGP~(d) case, where the entanglement generated is quite low. The Heuristic Protocol, which applies a static sequence of alternating $\pi/2$ pulses, represents a different philosophy: active, ``brute force'' entanglement generation. This active approach is more effective than passive protection, achieving a higher peak concurrence. However, its rigid, open-loop nature is its downfall. Applying the same pulse sequence regardless of the system's state or the specific noise realization leads to an inefficient process where control actions become mis-timed, resulting in a swift decay after the peak. While the Knill DD protocol often achieves a higher initial peak, the Heuristic protocol tends to sustain a low, non-zero level of entanglement for a longer duration at later times. Finally, the unequivocal success of the RL Optimized protocol lies in its fundamentally different, closed-loop, and model-free paradigm. The agent learns a dynamic, state-dependent policy that effectively creates a hybrid strategy, deciding at each moment whether to prioritize generation or protection. The crucial insight from its design is the reward function, which is proportional to the ``rate of change'' of concurrence ($dC/dt$). This objective function does not simply reward high entanglement; it rewards the most efficient and rapid pathway to achieving it. This has two profound consequences:
\begin{enumerate}
    \item Optimal Ascent: The RL protocol exhibits a significantly steeper initial rise in concurrence. The agent learns to exploit system resonances and apply pulses at opportune moments to maximize the rate of entanglement generation, effectively ``outrunning'' the onset of decoherence.
    \item Dynamic Stabilization: After reaching a peak, where the potential for rapid generation diminishes, the agent's strategy naturally shifts. To continue receiving positive rewards, it must now actively combat decoherence. It learns to apply corrective pulses---a form of learned dynamical decoupling---to ``juggle'' the state and protect the entanglement it has created. This explains the sustained, high-concurrence plateau absent in all other protocols.
\end{enumerate}
For complex tasks like entanglement generation in realistic environments, the control strategy's adaptability is paramount. The RL agent's ability to learn a state-dependent, hybrid policy optimized for the dynamics of the process allows it to vastly outperform static protocols limited to either naive generation or mismatched protection schemes. In all pulse shapes, the RL Optimized protocol rapidly generates near-maximal entanglement. However, due to the strong noise profiles and errors, the concurrence subsequently decays, with the degree of loss depending on the specific pulse type. In some cases, such as \LGP~(d) and \AGP~(f), the entanglement seems to completely degrade. However, under all other pulse types, the entanglement exhibits revivals after significant decay. This revival is particularly notable under the \GP~(a), \SGP~(b), \OGP~(c), and \HGP~(e) profiles, where the entanglement returns to a significant value. Overall, the RL Optimized protocol is particularly highly effective at generating and then preserving entanglement under the \SGP~(b) pulse profile.

\begin{figure*}[ht]
\centering
\begin{overpic}[height=10cm,width=18cm]{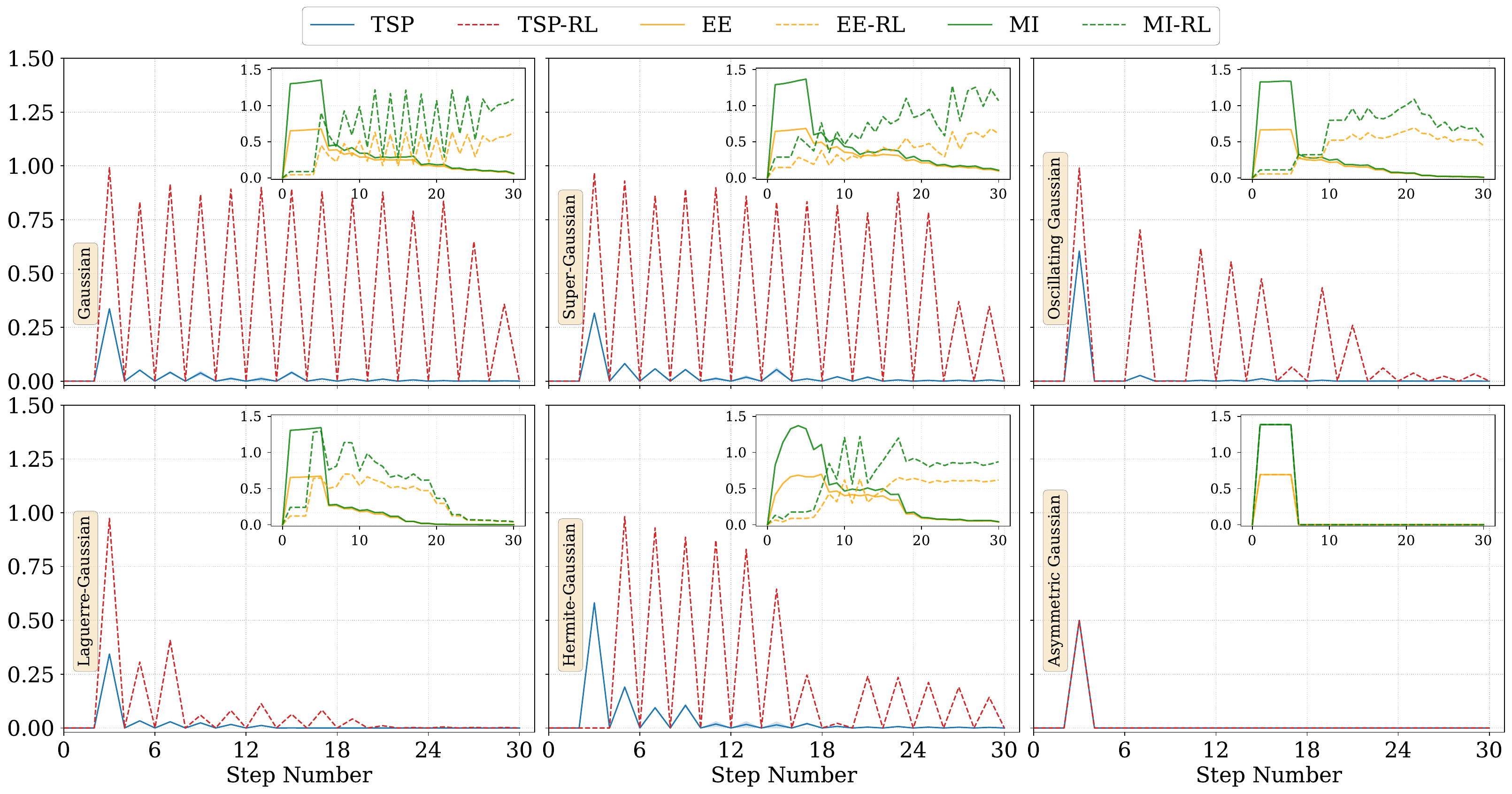}
\put(28, 40){(a)}
\put(61, 40){ (b)}
\put(94, 40){(c)}
\put(28, 19){(d)}
\put(61, 19){(e)}
\put(94, 19){(f)}
\end{overpic}
\caption{Target site probability (TSP), entanglement entropy (EE), and mutual information (MI) for a quantum walk on a $1D$ lattice with $N_s = 11$ sites under baseline and RL-optimized protocols across six pulse types. Each subplot shows the evolution of TSP (at target site $x_{\text{target}} = -3$), EE, or MI versus step number over $N_{\text{steps}} = 30$ steps, with a time per step of $\Delta t = 0.5$ ($\tau=7.5$). The simulation uses a pulse strength of $\Omega_0 = 1.0$ and a pulse center offset of $t_{\text{offset}} = 0.5$, $S_0=0.04$, $\gamma=0.5$, and a disorder strength of $0.01$. The RL-Optimized protocol uses a RecurrentPPO agent with a `MlpLstmPolicy`. The agent was trained for $3 \times 10^5$ total timesteps. Key hyperparameters are: a learning rate of $1 \times 10^{-4}$, $n_{\text{steps}}=512$, a batch size of $128$, $n_{\text{epochs}}=8$, a discount factor $\gamma=0.995$, a GAE $\lambda=0.98$, a clipping range of $0.15$, and an entropy coefficient of $0.02$. The agent optimizes a 3D action space representing the pulse amplitude, phase, and timing shift. The reward function is a sophisticated combination of a heavily weighted target probability ($50 \times \text{TSP}$), an exponential bonus for high probabilities, a penalty for distance from the target, a momentum reward for increasing probability, and a large bonus for a high final target probability.}
\label{fig12}
\end{figure*}
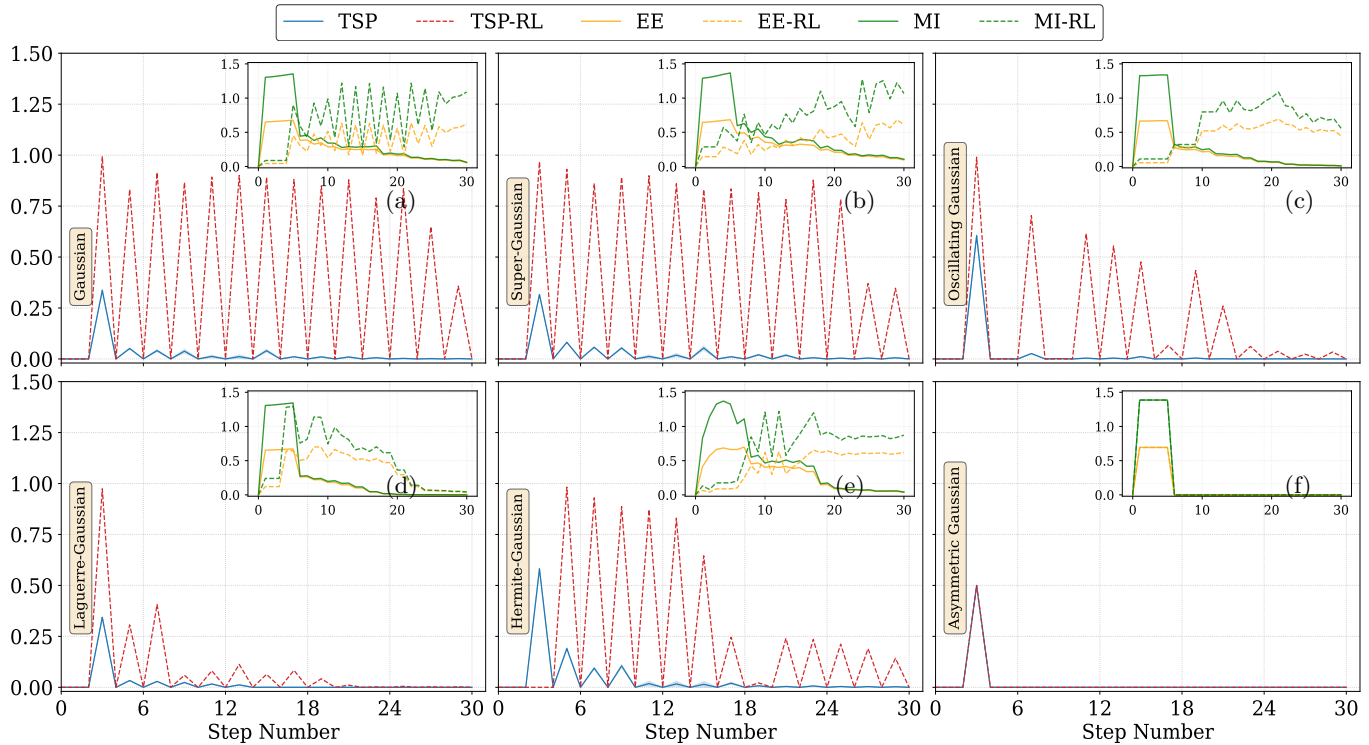
Fig.~\ref{fig12} provides a comprehensive analysis of control strategies for a quantum walk on a disordered 1D lattice. The comparison between a baseline and an RL-optimized protocol, tested across six pulse envelopes, reveals how performance in this complex transport problem is dictated by the interplay between the control strategy's adaptability and the nature of the external drive. The primary metric, TSP, is contextualized by the underlying quantum correlations of EE and MI. The universal failure of the Baseline Protocol, is evident across its TSP, EE, and MI, is instructive. The open-loop strategy is unable to navigate the complex energy landscape created by any of the pulse envelopes, leading to rapid decoherence and delocalization. In stark contrast, RL-Optimized protocol's performance demonstrates a clear dependence on the complexity of the control landscape defined by each pulse. Its success is enabled by a Recurrent PPO architecture with LSTM memory, which allows it to learn time-correlated actions tailored to the specific dynamics induced by each envelope. The agent's most remarkable successes, achieving near-perfect transport (TSP $\approx 1.0$), occur with the \GP~(a), \SGP~(b), and \HGP~(e) pulse envelopes. These pulses provide a structured control field, creating a tractable learning environment for the agent to master a policy of coherent state manipulation. The corresponding inset plots confirm this, showing that the agent successfully engineers a large, sustained bubble of EE and MI, the essential resource for coherent transport. It is also notable that under these successful profiles, the RL-driven EE and MI curves show sustained growth, indicating the agent is not just reaching a static level of correlations but is actively enhancing it over time. The control problem becomes more challenging with the internally complex \OGP~(c) and \LGP~(d) pulses. While the agent still finds a vastly superior solution, its peak TSP is diminished, and the more erratic evolution of EE and MI reflects the difficulty of maintaining coherence with such a drive. The near-total failure under the \AGP~(f) pulse is particularly illuminating. This envelope likely introduces a strong, uncontrolled bias, creating a pathologically difficult control problem that exceeds the capabilities of the learned policy, preventing the generation of necessary quantum correlations. Finally, the results demonstrate that while a memory-enabled RL agent can discover powerful solutions for quantum control, its performance is coupled to the complexity of the control landscape. The pulse envelope effectively sets the ``difficulty level'' for the task of coherence engineering, and the agent's success is a direct measure of its adaptability within that landscape. The RL-optimized protocol's ability to maintain high TSP, EE, and MI is particularly pronounced under the \GP~(a) and \SGP~(b) pulses, providing a stark contrast to the baseline, which fails completely to generate or sustain these essential quantum resources.

\section{Conclusion}
\label{sec:conclusion}
In this comprehensive study, we have systematically investigated and benchmarked a diverse array of quantum control methodologies. Our study encompasses a progression from deterministic pulse sequences to advanced adaptive and hybrid control protocols, evaluated across a suite of demanding quantum information processing tasks. We considered a realistic simulation environment incorporating  colored noise accompanied with Linblad master equation and experimental imperfections. hence, our work elucidates a clear picture of the trade-offs between protocol complexity, adaptability, and relative performance. Our initial exploration of deterministic protocols, including single-pulse, multi-pulse, and sequential polarized schemes, established that while control is achievable, performance is often brittle and highly contingent on the specific initial state, task, and pulse shape. We found that certain pulse envelopes are predisposed to either preserving or generating entanglement, highlighting the limitations of a "one-size-fits-all" approach and underscoring the need for task-specific optimization. These open-loop strategies, while foundational, lack the robustness required for reliable performance in complex, fluctuating environments. Notably, our deterministic protocols proved effective for both entanglement generation and preservation, particularly in selected single- and multi-pulse cases. In contrast, RL-based optimization remained challenging when using a single reward function to address both entanglement generation and preservation simultaneously, performing worse than the deterministic protocols. Furthermore, our deterministic protocols achieved higher average concurrence compared to the corresponding RL-based optimizations. While the RL optimization although with lesser entanglement somehow provided stable entanglement dynamics. RL optimization was more effective in single-pulse scenarios than in multi-pulse cases for entanglement generation and preservation. Even with complex and enhanced reward functions, the RL multi-pulse cases were less efficient than RL single-pulse  optimization. Hence, increasing the complexity of the RL optimization does not always yield fruitful results.

The comparison of more advanced protocols revealed a crucial insight: adaptability is paramount, but its implementation matters. In the task of preserving a maximally entangled state, the physics-informed Hybrid QEC-DD protocol, we designed proved to be the most effective strategy, outperforming even the RL-optimized agent. Its blend of passive error suppression and active, periodic correction established a robust and stable defense against decoherence that the more dynamic, yet potentially suboptimal, RL policy could not consistently surpass. This result powerfully demonstrates that well-designed, physics-informed hybrid strategies can represent the state-of-the-art and serve as critical benchmarks for machine learning methods. However, for tasks requiring the discovery of highly non-trivial, dynamic solutions, such as rapid entanglement generation from a separable state, Floquet engineering or coherent transport in a disordered quantum walk, the model-free RL framework demonstrated its unique power. By leveraging sophisticated reward engineering and recurrent memory architectures, the RL agent was able to learn complex, state-dependent policies that vastly outperformed all other methods. It successfully navigated high-dimensional control landscapes to actively engineer quantum correlations, a feat intractable for pre-programmed or simpler model-based approaches. Nevertheless, the agent's success was not absolute; its performance was still coupled to the complexity of the control landscape defined by the pulse envelope, as evidenced by its failure in the most pathologically difficult scenarios. In synthesis, our work establishes that there is no single "panacea" for quantum control. The optimal strategy is task-dependent. While the robust stability of hybrid methods like QEC-DD makes them ideal for preservation tasks, the unparalleled adaptability of RL is best suited for discovering novel, high-performance solutions to complex dynamical problems. The future of robust quantum control will likely involve a synthesis of these paradigms: leveraging physical insight to structure control problems and design reward functions, while harnessing the power of machine learning to navigate the immense complexity of controlling large-scale quantum systems.

\section{Appendix}\label{appendix}
In this section, we give some detailed information about the previous sections.
\subsection{Derivatives of the Gaussian Pulses}
For the Gaussian pulses in Sec. \ref{pulses}, defining the normalized time $\mathcal{C} = (t-t_c)/T$ and the base Gaussian derivative term $g(\mathcal{C}) = -2\mathcal{C}/T \cdot e^{-\mathcal{C}^2}$, the derivatives are:
\begin{align}
\dot{f}^{\GP}(t) &= 
g(\mathcal{C}) \\
\dot{f}^{\SGP}(t) &= 
-\frac{n\mathcal{C}^{n-1}}{T} e^{-\mathcal{C}^n} \\
\dot{f}^{\OGP}(t) &= 
g(\mathcal{C}) \cos(5\pi \mathcal{C}) - \frac{5\pi}{T} e^{-\mathcal{C}^2} \sin(5\pi \mathcal{C}) \\
\dot{f}^{\LGP}(t) &= 
e^{-\mathcal{C}^2} \left[ \frac{\text{sgn}(\mathcal{C})}{T} - \frac{2|\mathcal{C}|\mathcal{C}}{T} \right] \\
\dot{f}^{\HGP}(t) &= 
\frac{2\sqrt{2}}{T} e^{-\mathcal{C}^2} \left( 1 - 2\mathcal{C}^2 \right) \\
\dot{f}^{\AGP}(t) &= 
g(\mathcal{C}) \left[1 + \mathrm{erf}\left(\frac{\mathcal{C}T}{\tau_a}\right)\right] + \frac{2 e^{-\mathcal{C}^2} e^{-(\frac{\mathcal{C}T}{\tau_a})^2}}{\sqrt{\pi}\tau_a} \label{eq:all_derivs}
\end{align}
These derivatives ensure accurate implementation of control terms involving time-dependent modulation.

\subsection{Initial State consideration}\label{initial states}
All initial states $\ket{\psi}$ are prepared as pure state density matrices $\rho_0 = \ket{\psi}\bra{\psi}$, constructed from the single-qubit basis states $\ket{0}$, $\ket{1}$, and $\ket{+} = (\ket{0} + \ket{1})/\sqrt{2}$. The goal is to create entanglement from both an initially separable and maximally entangled states. The RL agent was trained starting from the ground state $\ket{00}$, with the final policy evaluated on the full set of separable states: \ZZ, \PZ, \ZP, \PP. Besides, we also aim to protect an existing entangled state from decoherence. The initial states for this task are the maximally entangled Bell states:
\begin{subequations}
\label{eq:bell_states}
\begin{align}
\ket{\Phi^+} &= \frac{1}{\sqrt{2}} \left( \ket{00} + \ket{11} \right) \\
\ket{\Psi^+} &= \frac{1}{\sqrt{2}} \left( \ket{01} + \ket{10} \right)
\end{align}
\end{subequations}
Additionally, the separable state $\ket{++} = \ket{+}\otimes\ket{+}$ is included as a non-entangled control case to observe its evolution under the same protocol.

\subsection{Concurrence}
To quantify the degree of entanglement in the two-qubit system, we employ the concurrence $C(\rho)$~\cite{Wooters1998}. For a general two-qubit density matrix $\rho$,first we construct the non-Hermitian matrix $R = \rho \tilde{\rho}$ with $\tilde{\rho} = (\hat{\sigma}_y \otimes \hat{\sigma}_y) \rho^* (\hat{\sigma}_y \otimes \hat{\sigma}_y)$ where $\rho^*$ is the complex conjugate of $\rho$. The $C(\rho)$ is then given by the expression:
\begin{equation}
C(\rho) = \max(0, \lambda_1 - \lambda_2 - \lambda_3 - \lambda_4) \label{con}
\end{equation}
where $\lambda_i$ are the square roots of the eigenvalues of the matrix $R$, sorted in descending order. Throughout our analysis, the $C(\rho)$ serves as the primary figure of merit for evaluating the performance of our quantum control protocols in generating and preserving entanglement.
\section{Acknowledgments}
This work was supported in part by the National Natural Science Foundation of China (NSFC) under the Grants 12475087 and 12235008, and by the University of Chinese Academy of Sciences.

\end{document}